\documentclass[letterpaper,11pt]{article}

\pdfoutput=1

\usepackage{hyperref}
\hypersetup{
    colorlinks=true,       
    linkcolor=blue,          
    citecolor=blue,        
    filecolor=blue,      
    urlcolor=blue           
}

\usepackage{float}
\usepackage{textcomp}               
\usepackage{braket,slashed,bm}
\usepackage{array,multirow}
\usepackage[normalem]{ulem}
\usepackage{xcolor,cancel,youngtab}

\usepackage[T1]{fontenc} 

\usepackage{mathrsfs}
\usepackage{booktabs}
\usepackage{adjustbox}
\usepackage{mathtools}

\usepackage{soul}

\usepackage{tikz}
\usetikzlibrary{arrows,decorations.pathmorphing,backgrounds,positioning,fit,petri,automata,shadows,calendar,mindmap,
decorations.markings,calc}

\definecolor{labelkey}{rgb}{0,0.5,0.0}

\usepackage{xspace}
\usepackage{slashed}
\usepackage{array,multirow}
\usepackage{graphicx}
\usepackage{graphics}
\usepackage{epsfig}
\usepackage{amsfonts}
\usepackage{amsmath}
\usepackage{amssymb}
\usepackage{dsfont}
\usepackage{color}
\usepackage{xcolor}
\usepackage{cite}
\usepackage{pdflscape}
\usepackage{pifont}
\usepackage{pgfplots}
\usepackage{tikz} 
\usepackage{pgfplotstable}
\usepackage{bbold}

\newcommand{\hc}{\mathrm{h.c.}}
\newcommand{\ep}{\epsilon}

\newcommand{\al}{\alpha}

\newcommand{\g}{\gamma}

\usepackage{bm}  %

\newcommand{\beq}{\begin{equation}}
\newcommand{\eeq}{\end{equation}}
\newcommand{\be}{\begin{equation}}
\newcommand{\ee}{\end{equation}}
\newcommand{\bea}{\begin{eqnarray}}
\newcommand{\eea}{\end{eqnarray}}
\newcommand{\ben}{\begin{eqnarray*}}
\newcommand{\een}{\end{eqnarray*}}

\newcommand{\bma}{\begin{pmatrix}}
\newcommand{\ema}{\end{pmatrix}}

\def\lixo#1{}

\def\slashchar#1{\setbox0=\hbox{$#1$}           
  \dimen0=\wd0                                    
  \setbox1=\hbox{/} \dimen1=\wd1                  
  \ifdim\dimen0>\dimen1                           
    \rlap{\hbox to \dimen0{\hfil/\hfil}}            
    #1                                             
  \else                                          
    \rlap{\hbox to \dimen1{\hfil$#1$\hfil}}        
    /                                           
 \fi}                                           %

\newcommand\gsim{\gtrsim}

\newcommand{\dslash}[1]{#1 \llap{/\kern-0.5pt}}
\newcommand{\Dslash}[1]{#1 \llap{/\kern+1.5pt}}
\newcommand{\DDslash}[1]{#1 \llap{/\kern+2.3pt}}
\newcommand{\dslashh}[1]{#1 \llap{/\kern+1pt}}
\newcommand{\abs}[1]{|#1|}

\newcommand{\nn}{\nonumber}

\definecolor{cadmiumgreen}{rgb}{0.0, 0.42, 0.24}
\definecolor{darkpastelgreen}{rgb}{0.01, 0.75, 0.24}
\definecolor{darkspringgreen}{rgb}{0.09, 0.45, 0.27}
\definecolor{forestgreen(web)}{rgb}{0.13, 0.55, 0.13}
\definecolor{forestgreen(traditional)}{rgb}{0.0, 0.27, 0.13}
\definecolor{cobalt}{rgb}{0.0, 0.28, 0.67}
\definecolor{darkblue}{rgb}{0.0, 0.0, 0.75}
\definecolor{darkred}{rgb}{0.55, 0.0, 0.0}
\definecolor{palatinatepurple}{rgb}{0.41, 0.16, 0.38}
\definecolor{burntorange}{rgb}{0.8, 0.33, 0.0}

\begin{document}

\begin{titlepage}
	\begin{flushright}
		APCTP Pre2020-027\\
		BONN-TH-2020-10\\
		RBRC-1328
	\end{flushright}

\newcommand{\AddrBONN}{%
	Bethe Center for Theoretical Physics \& Physikalisches Institut der 
	Universit\"at Bonn,\\ Nu{\ss}allee 12, 
	53115 Bonn, Germany
}
\newcommand{\AddrAMHERST}{%
	Amherst Center for Fundamental Interactions, Department of Physics, \\ University of Massachusetts, Amherst, MA 01003, USA}

\newcommand{\AddrRIKEN}{%
	RIKEN BNL Research Center, Brookhaven National Laboratory, Upton, NY 11973-5000, USA}
\newcommand{\AddrAPCTP}{%
    Asia Pacific Center for Theoretical Physics (APCTP) 
    - Headquarters San 31,\\ Hyoja-dong, Nam-gu, Pohang 790-784, Korea}
\newcommand{\AddrNTHU}{%
	Department of Physics, National Tsing Hua University, Hsinchu 300, Taiwan
}
	
	\begin{center}
		{\LARGE  \bf 
			Long-lived Sterile Neutrinos at the LHC \\
			in Effective Field Theory
		
		}
		\vspace{2cm}

	 Jordy~de~Vries${}^{a,b}$\footnote{jdevries@umass.edu},
	 Herbert~K.~Dreiner${}^{c}$\footnote{dreiner@uni-bonn.de},
	 Julian~Y.~G\"unther${}^{c}$\footnote{guenther@physik.uni-bonn.de},
	 Zeren~Simon~Wang${}^{d,e}$\footnote{wzs@mx.nthu.edu.tw},
	 and Guanghui~Zhou${}^{a}$\footnote{gzhou@umass.edu}\\[0.2cm]

	{\small \textit{ 
			${}^a$\AddrAMHERST\\[0pt] 
	}} 
    	{\small \textit{ 
    		${}^b$\AddrRIKEN\\[0pt] 
    }} 
	
	{\small \textit{ 
			${}^c$\AddrBONN\\[0pt] 
	}} 

    {\small \textit{ 
    		${}^d$\AddrNTHU\\[0pt] 
    }}  

	{\small \textit{ 
			${}^e$\AddrAPCTP\\[0pt] 
	}}

	\begin{abstract}
		
		We study the prospects of a displaced-vertex search of sterile neutrinos at the Large Hadron Collider (LHC) in the framework of the neutrino-extended Standard Model Effective Field Theory ($\nu$SMEFT).
		The production and decay of sterile neutrinos can proceed via the standard active-sterile neutrino mixing in the weak current, as well as through higher-dimensional operators arising from decoupled new physics.
		If sterile neutrinos are long-lived, their decay can lead to displaced vertices which can be reconstructed.
		We investigate the search sensitivities for the \texttt{ATLAS}/\texttt{CMS} detector, the future far-detector experiments: \texttt{AL3X}, \texttt{ANUBIS}, \texttt{CODEX-b}, \texttt{FASER}, \texttt{MATHUSLA}, and \texttt{MoEDAL-MAPP}, and at the proposed fixed-target experiment \texttt{SHiP}.
	We study scenarios where sterile neutrinos are predominantly produced via rare charm and bottom mesons decays through minimal mixing and/or dimension-six operators in the $\nu$SMEFT Lagrangian. We perform simulations to determine the potential reach of high-luminosity LHC experiments in probing the EFT operators, finding that these experiments are very competitive with other searches. 
		\end{abstract}
	
		\end{center}
	\newpage

	\vfill
\end{titlepage}
\tableofcontents
\section{Introduction}
\setcounter{footnote}{0} 

Neutrino oscillations have proven without doubt that neutrinos are massive particles.
The Standard Model of particle physics (SM) contains no right-handed neutrino fields.
This forbids the generation of a neutrino mass via the Higgs mechanism, which generates the masses of the other elementary particles.
This situation can be remedied by adding a sterile neutrino field to the SM \cite{Minkowski:1977sc,Yanagida:1979as,Mohapatra:1979ia,GellMann:1980vs,Schechter:1980gr}. The sterile neutrino, also called heavy neutral lepton (HNL), is a right-handed 
gauge-singlet spin-1/2 field and couples to left-handed neutrinos and the Higgs field through Yukawa interactions. This generates a Dirac neutrino mass after electroweak 
symmetry breaking.

In general, nothing forbids an additional Majorana mass term for the right-handed neutrino field, leading to Majorana mass eigenstates and lepton number violation (LNV).
However, lepton number can be an (approximate) symmetry of extension beyond the SM (BSM), such that low-energy LNV signals, \textit{e.g.} neutrinoless double beta decay ($0\nu\beta\beta$), is suppressed.
Sterile neutrinos may not only account for neutrino masses, but have also been linked to explanations of other problems of the SM.
Light sterile neutrinos can account for dark matter \cite{Drewes:2013gca,Kusenko:2009up,Adhikari:2016bei,Boyarsky:2018tvu}, while sterile neutrinos with a broad range of masses can account for the baryon asymmetry of the Universe through leptogenesis \cite{Davidson:2008bu}.
Sterile neutrinos are thus a well-motivated solution to a number of major outstanding issues in particle physics and cosmology.

While the observation of neutrino masses provides a hint for the existence of sterile neutrinos, it does not specify their mass scale.
They might very well be light and accessible in present-day and near future experiments.
A large number of experimental and theoretical works have gone into the search for sterile neutrinos in so-called minimal scenarios, where sterile neutrinos only interact with SM fields through renormalizable Yukawa interactions (see Refs.~\cite{Deppisch:2015qwa,Bryman:2019bjg} for a review).
Here we take a more general approach.
In broad classes of BSM models, sterile neutrinos appear sterile at lower energies, but interact at higher energies through the exchange of heavy BSM fields.
Examples are left-right symmetric models \cite{Mohapatra:1974gc,Pati:1974yy,Mohapatra:1980yp}, grand unified theories \cite{Bando:1998ww}, Z' models \cite{Chiang:2019ajm}, or leptoquark models \cite{Dorsner:2016wpm}, which contain new fields that are heavy compared to the electroweak scale.
Independent of the details of these models, at low energies the sterile neutrinos can be described in terms of local effective operators in the framework of the neutrino-extended Standard Model effective field theory ($\nu$SMEFT) \cite{delAguila:2008ir,Liao:2016qyd}. 

In this work, we study relatively light GeV-scale sterile neutrinos (see e.g. Refs~\cite{Cottin:2018nms,Cottin:2018kmq,Drewes:2019fou} for LHC searches for somewhat heavier neutrinos).
Such sterile neutrinos can be produced either via direct production with parton collisions, or via rare decays of mesons that are copiously produced at the LHC interaction points \cite{Shrock:1980vy,Shrock:1980ct}.
For sterile neutrino masses below the $B$-meson threshold the primary production mode is through rare decays of mesons with subleading contributions from partonic processes, which we estimated using MadGraph5 3.0.2 \cite{Alwall:2014hca} to be less than 10\%. The latter become more important and even dominant for heavier sterile neutrinos.
In this work we choose to focus on the mass range below about 5 GeV and hence on the rare meson decays, and we leave the direct production channel for future studies.
If the sterile neutrinos are relatively long-lived, their decays lead to displaced vertices that can be reconstructed in LHC detectors.
We consider a broad range of (proposed) LHC experiments: \texttt{ATLAS} \cite{Aad:2008zzm}/\texttt{CMS} \cite{Chatrchyan:2008aa}, \texttt{CODEX-b} \cite{Gligorov:2017nwh}, \texttt{FASER} \cite{Feng:2017uoz,Ariga:2018uku}, \texttt{MATHUSLA} \cite{Chou:2016lxi,Curtin:2018mvb,Alpigiani:2020tva}, \texttt{AL3X} \cite{Gligorov:2018vkc}, \texttt{ANUBIS} \cite{Bauer:2019vqk}, \texttt{MoEDAL-MAPP} \cite{Pinfold:2019nqj,Pinfold:2019zwp}, as well as the proposed CERN SPS experiment \texttt{SHiP} \cite{Calviani:2063300, Anelli:2015pba,SHiP:2018yqc}, and discuss their potential in probing $\nu$SMEFT operators.
We calculate $\nu$SMEFT corrections to sterile neutrino production and decay processes and perform simulations for the various detectors, to estimate their search sensitivities.
Our simulations show that the experimental reach is strong, probing dimension-six operators associated to BSM scales up to a hundred TeV.  In a simple $3+1$ model, adding just one sterile neutrino field, we compare our results to existing $0\nu\beta\beta$ decay limits, showing that these experiments are complementary. 

This paper is organized as follows. In Sec.~\ref{sec:model} we introduce the model framework of $\nu$SMEFT, followed by Secs.~\ref{sec:prod-of-HNLs} and \ref{sec:decays-of-HNLs} detailing the calculation of production cross sections and decay widths of the sterile neutrinos in both the minimal model and from higher-dimensional operators.
Sec.~\ref{sec:scenarios} shows the theoretical scenarios considered by numerical study in this work, and Sec.~\ref{sec:exp} goes through the different experiments we study in detail and briefly introduces the Monte-Carlo simulation procedure.
In Sec.~\ref{sec:flavorbenchmarks} we present the numerical results for both the minimal scenario and a number of flavor benchmarks.
These results are compared with a number of other experimental probes including $0\nu\beta\beta$ decay in Sec.~\ref{sec:comparison}.
In a set of appendices, we present the details of the production and decay rate computations, as well as the physical parameters, decay constants, and form factor input we employ.
We conclude and provide an outlook in Sec.~\ref{sec:conclusion}.

\section{Standard Model Effective Field Theory extended by Sterile Neutrinos}\label{sec:model}

\subsection{The Effective Neutrino Lagrangian}

We are interested in the production and decay of sterile neutrinos at the LHC. In particular, we investigate the production of sterile neutrinos in the 
decay of mesons containing a single $b$ or $c$ quark, as these are copiously produced and are sufficiently massive to produce GeV sterile neutrinos. 
This is an interesting mass range that appears in scenarios of low-scale leptogenesis \cite{Ghiglieri:2017gjz,Hernandez:2016kel,Akhmedov:1998qx,Asaka:2005an,Asaka:2005pn,Shaposhnikov:2006nn,Canetti:2012vf}. 
The sterile neutrinos are assumed to be singlets under the SM gauge group and, at the renormalizable level, only interact with SM fields via a Higgs 
Yukawa coupling to the lepton doublet. The renormalizable part of the Lagrangian is given by
\begin{eqnarray}\label{eq:smeftdim4}
\mathcal L &=&  \mathcal L_{SM} - \left[ \frac{1}{2} \bar \nu^c_{R} \,\bar M_R \nu_{R} +\bar L \tilde H Y_\nu \nu_R + \rm{h.c.}\right]\,.
\end{eqnarray}
$\mathcal L_{SM}$ denotes the SM Lagrangian, $L$ is the lepton doublet, and $H$ is the SM complex Higgs doublet field with $\tilde H = i \tau_2 H^*$.
We work in the unitary gauge 
\begin{equation}
H = \frac{v}{\sqrt{2}} \left(\begin{array}{c}
0 \\
1 + \frac{h}{v}
\end{array} \right)\,,
\end{equation}
where $v=246$ GeV is the Higgs vacuum expectation value of the Higgs real scalar $h$. $\nu_R$ is an $n\times 1$ column vector of $n$ 
right-handed gauge-singlet neutrinos. $Y_\nu$ is a $3\times n$ matrix of Yukawa couplings. $\bar M_R$ is a complex symmetric $n\times n$  
mass matrix. In general we can work in a basis where the charged leptons and all quarks are in their mass eigenstates,  except the $d^i_L$, 
$i=1,2,3$, for which we have $d^i_L = V^{ij} d_L^{j,\,\rm mass}$ with $V$ the CKM matrix. $\Psi^c$ is the charge conjugate field of $\Psi$ with 
$\Psi^c = C \bar \Psi^T$ and $C$ is the charge conjugation matrix, $C=-i\gamma^2\gamma^0$, which satisfies the relation $C = - C^{-1} = -C^T 
= - C^\dagger$. We define $\Psi_{L,R}^c = (\Psi_{L,R})^c =  C \overline{\Psi_{L,R}}^T= P_{R,L} \Psi^c$, in terms of the projectors $P_{R,L}=(1
\pm\gamma_5)/2$. 

The Lagrangian in Eq.~\eqref{eq:smeftdim4} can account for the observed active neutrino masses and mixing angles if $n\geq 2$ (in case of $n=2$ 
the lightest neutrino is massless \cite{Donini:2012tt}). As lepton number is explicitly violated by the Majorana masses, $\bar M_R$, Eq.~\eqref{eq:smeftdim4} 
in general leads to Majorana neutrino mass eigenstates and thus to $0\nu\beta\beta$ decay and other LNV processes, unless additional structure is imposed on the 
matrices $\bar M_R$ and $Y_\nu$. 

In various popular extensions of the SM, right-handed neutrinos appear naturally, but are not completely sterile. For instance, in left-right symmetric 
models right-handed neutrinos are charged under a right-handed $SU(2)_R$ gauge group and interact with right-handed gauge bosons and new 
scalar fields. If such bosons exist,  they must be heavy, with masses well above the electroweak scale, to avoid experimental constraints. The 
right-handed neutrinos on the other hand, can remain light. From this point of view, the scale separation suggests the use of an EFT framework where 
the degrees of freedom are the usual SM fields, as well as a set of $n$ neutrinos, which are singlets under the SM gauge groups. The 
interactions in Eq.~\eqref{eq:smeftdim4} form the dimension-four and lower part of a more general Lagrangian containing higher-dimensional operators 
that is often referred to as the $\nu$SMEFT \cite{delAguila:2008ir, Chala:2020vqp, Dekens:2020ttz,Banerjee:2020jun}. We begin by introducing the higher-dimensional 
operators at a scale $\Lambda \gg v$, where $\Lambda$ denotes the scale where we match the microscopic UV theory to the $\nu$SMEFT.

The first operators have dimension 5
\be
 \mathcal L^{(5)}_{\nu_L} = \ep_{kl}\ep_{mn}(L_k^T\, C^{( 5)}\,CL_m )H_l H_n\,,\qquad  \mathcal L^{(5)}_{\nu_R}=- \bar \nu^c_{R} \,M_R^{(5)} \nu_{R} H^\dagger H\,.
\ee
At lower energies, after electroweak symmetry breaking these operators contribute to, respectively, Majorana mass terms for active and sterile neutrinos. 
The first operator, the famous Weinberg operator \cite{Weinberg:1979sa}, can for instance be induced by integrating out sterile neutrinos with masses of 
$\mathcal O(\Lambda)$ usually referred to as a type-I seesaw mechanism. The second operator, for our purposes, can simply be absorbed in $\bar M_R$ 
in Eq.~\eqref{eq:smeftdim4}. For $n\geq 2$ there appears a dim-5 transition dipole operator, but we will not consider it here.

{\renewcommand{\arraystretch}{1.3}\begin{table}[t!]\small
		\center
		\begin{tabular}{||c|c||c|c||}
			\hline Class $1$& $\psi^2 H^3$  & Class $4$ &  $\psi^4 $\\
			\hline
			$\mathcal{O}^{(6)}_{L\nu H}$ & $(\bar{L}\nu_R)\tilde{H}(H^\dagger H)$ & $\mathcal{O}^{(6)}_{du\nu e}$ & $ (\bar{d}\gamma^\mu u)(\bar{\nu}_R \gamma_\mu e)$  \\ \cline{1-2}
			Class $2$&  $\psi^2 H^2 D$ &  $\mathcal{O}^{(6)}_{Qu\nu L}$ & $(\bar{Q}u)(\bar{\nu}_RL)$  \\ \cline{1-2}
			$\mathcal{O}^{(6)}_{H\nu e}$ & $(\bar{\nu }_R\gamma^\mu e)({\tilde{H}}^\dagger i D_\mu H)$ & $\mathcal{O}^{(6)}_{L\nu Qd}$ & $(\bar{L}\nu_R )\epsilon(\bar{Q}d))$ \\ \cline{1-2}
			Class $3$ & $\psi^2 H^3 D$  & $\mathcal{O}^{(6)}_{LdQ\nu }$ & $(\bar{L}d)\epsilon(\bar{Q}\nu_R )$ \\ \cline{1-2}
			$\mathcal{O}^{(6)}_{\nu W}$ &$(\bar{L}\sigma_{\mu\nu}\nu_R )\tau^I\tilde{H}W^{I\mu\nu}$  & &\\
			\hline
		\end{tabular}
		\caption{$\nu$SMEFT dim-6 operators \cite{Liao:2016qyd} involving one sterile neutrino field.
		} \label{tab:O6R}
\end{table}}

We are mainly interested in the operators that appear at dimension-6 \cite{Grzadkowski:2010es,Liao:2016qyd}.  We focus on operators that involve a 
single right-handed neutrino\footnote{Operators with a left-handed neutrino also contribute to the same observables we discuss here, but the contributions 
are suppressed by small heavy-light neutrino mixing angles.} and limit the set of effective operators to those that lead to hadronic processes at tree level. 
A more general set of interactions is left for future work. The operators are presented in Table \ref{tab:O6R}. For our purposes, the operator $\mathcal{O}^{(6)}_{L\nu H}$ can be absorbed in a shift in $Y_\nu$ in Eq.~\eqref{eq:smeftdim4}. The related Higgs 
phenomenology was discussed in Ref.~\cite{Butterworth:2019iff}. This leaves us with the remaining six  operators that, in general, have 
arbitrary flavor indices although certain couplings can be suppressed if minimal flavor violation is assumed \cite{Caputo:2017pit,Barducci:2020ncz}.

We evolve the operators from $\Lambda$ to the electroweak scale using one-loop QCD anomalous dimensions. The operators $\mathcal{O}^{(6)}_{H\nu e}$, $\mathcal{O}^{(6)}_{\nu W}$, and $\mathcal{O}^{(6)}_{du\nu e}$ do not evolve under QCD at one loop. The remaining three operators evolve simply as \cite{Dekens:2020ttz}
\bea\label{eq:STrge}
\frac{d C^{(6)}_{Qu\nu L}}{d\ln\mu} = \left(\frac{\al_s}{4\pi}\right)\gamma_{S}\,C^{(6)}_{Qu\nu L }\,,\qquad \frac{dC^{(6)}_{S}}{d\ln\mu} = \left(\frac{\al_s}{4\pi}\right)\gamma_{S}\,C^{(6)}_{S }\,,\qquad \frac{dC^{(6)}_{T}}{d\ln\mu} = \left(\frac{\al_s}{4\pi}\right)\gamma_{T}\,C^{(6)}_{T}\,,
\eea
where $C^{(6)}_S$ and $C^{(6)}_T$ are defined as the linear combinations
\bea
C_{S}^{(6)} =-\frac{1}{2} C_{LdQ\nu}^{(6)}+C_{L\nu Qd}^{(6)}\,,\qquad C_{T}^{(6)} =-\frac{1}{8} C_{LdQ\nu}^{(6)}\,,
\eea
and 
\be
\gamma_S = -6 C_F\,, \qquad \g_T = 2C_F\,,
\ee
where $C_F = (N_c^2-1)/(2N_c)$ and $N_c=3$, the number of colors. 

At the electroweak scale, we integrate out the heavy SM fields (W-, Z-, and Higgs-boson and top quark) and match to a $SU(3)_c \times U(1)_
{\mathrm {em}}$-invariant EFT. The tree-level matching relations for $\nu$SMEFT operators up to dimension-7 were given in 
Ref.~\cite{Dekens:2020ttz} and here we use a subset of these results. We obtain 
\begin{eqnarray}\label{DL2lag}
\mathcal L_{} &=&  \mathcal L_{SM}-  \left[\frac{1}{2} \bar \nu^c_{L} \, M_L \nu_{L}   +  \frac{1}{2} \bar \nu^c_{R} \, M_R \nu_{R} +\bar \nu_L M_D\nu_R +\hc \right]\nn \\
	&&+\mathcal L^{(6)}_{\Delta L = 0} +  \mathcal L^{(6)}_{\Delta L = 2} +   \mathcal L^{(7)}_{\Delta L =0 }\,,
	\label{eq:eff-lagrangian-all}
\end{eqnarray}
where $\mathcal L_{SM}$ contains dimension-four and lower operators involving light SM fields. $\mathcal L^{(6)}_{\Delta L = 0}$ 
includes dim-6 operators that conserve lepton number ($\Delta L=0$) and is given by
\bea\label{flavor}
\mathcal L^{(6)}_{\Delta L = 0}& =& \frac{2 G_F}{\sqrt{2}} \Bigg\{ 
  \bar u_L \gamma^\mu d_L \left[  \bar e_{L}  \gamma_\mu c^{(6)}_{\textrm{VL}} \,  \nu_{L}+ \bar e_{R}  \gamma_\mu \bar c^{(6)}_{\textrm{VL}} \,  \nu_{R} \right]+
  \bar u_R \gamma^\mu d_R\,\bar e_{R}\,  \gamma_\mu  \bar c^{(6)}_{\textrm{VR}} \,\nu_{R}\nn \\
& & +
  \bar u_L  d_R \,\bar e_{L}\, \bar c^{(6)}_{ \textrm{SR}}  \nu_{R} + 
  \bar u_R  d_L \, \bar e_{L} \, \bar c^{(6)}_{ \textrm{SL}}    \nu_{R} +  \bar u_L \sigma^{\mu\nu} d_R\,  \bar e_{L}  \sigma_{\mu\nu} \bar c^{(6)}_{ \textrm{T}} \, \nu_{R}
\Bigg\}  +{\rm h.c.} \label{lowenergy6_l0}\,.
\eea
For the operators in Table~\ref{tab:O6R} the Lagrangians $\mathcal L^{(6)}_{\Delta L = 2}$ and   $\mathcal L^{(7)}_{\Delta L =0 }$ only contain a single term each
\bea\label{lowenergy7_l0} 
\mathcal L^{(6)}_{\Delta L = 2}& =& \frac{2 G_F}{\sqrt{2}} \Bigg\{ 
  \bar u_L \gamma^\mu d_L \,  \bar e_{L}  \gamma_\mu \bar C^{(6)}_{\textrm{VL}} \,  \nu^c_{R} \Bigg\}  +{\rm h.c.}\,,\nn\\
    \mathcal L^{(7)}_{\Delta L = 0} &=& \frac{2 G_F}{\sqrt{2} v} \Bigg\{  \bar u_L \gamma^\mu d_L\, \bar e_{L} \, \bar c^{(7)}_{\textrm{VL}} \,  i \overleftrightarrow{D}_\mu  \nu_{R} \Bigg\}  +{\rm h.c.}
  \eea
where $\overleftrightarrow D_\mu = D_\mu - \overleftarrow D_\mu$. In these expressions we have suppressed flavor indices on the 
Wilson coefficients. Each Wilson coefficient carries indices $ijkl$ where $i,j =\{1,2,3\}$ indicate the generation of the involved up-type 
and down-type quarks, respectively, $k=\{1,2,3\}$ the generation of the charged lepton (we will often use the labels $e,\mu,\tau$ instead for clarity) and $l=\{1,2,3\}$ for $ c^{(6)}_{\textrm{VL}}$ the 
generation of the active neutrino, while $l = \{1,\dots ,n\}$ for the remaining Wilson coefficients involving sterile neutrinos.

The explicit matching relations are given by Ref.~\cite{Dekens:2020ttz}
\bea\label{Massmatch}
M_L = -v^2 C^{(5)}\,,\qquad
M_R = \bar M_R + v^2 \bar M_R^{(5)}\,,\qquad
M_D =\frac{v}{\sqrt{2}} \left[Y_\nu -\frac{v^2}{2}C_{L\nu H}^{(6)}\right]\,,
\eea 
for the mass terms in Eq.~(\ref{eq:eff-lagrangian-all}), and 
\bea\label{match6LNC}
c_{\rm VL}^{(6)} &=& -2V \mathbb{1}-\frac{4\sqrt{2}v}{g} C^{(6)}_{\nu W}V M_D^\dagger\,,\nn\\
\bar c_{\rm VL}^{(6)} &=& \left[-v^2C_{H\nu e}^{(6)}V- \frac{4\sqrt{2} v}{g}   \left(C_{\nu W}^{(6)}\right)^\dagger V M_e\right]^\dagger\,,\nn\\
\bar c_{\rm VR}^{(6)} &=& v^2\left(C_{du\nu e}^{(6)}\right)^\dagger\,,\nn\\
\bar c_{\rm SR}^{(6)}&=& -v^2C_{L\nu Qd}^{(6)}+\frac{v^2}{2} C_{LdQ\nu }^{(6)}\,,\nn\\
\bar c_{\rm SL}^{(6)}&=& v^2\left(C_{Qu\nu L}^{(6)}\right)^\dagger V\,,\nn\\
\bar c_{\rm T}^{(6)} &=& \frac{v^2}{8} C_{LdQ\nu }^{(6)}\,,\nn\\
\bar C_{\rm VL}^{(6)} &=&-\frac{4\sqrt{2} v}{g}C_{\nu W}^{(6)}V M_R^\dagger,\nn\\
\bar c_{\rm  VL}^{(7)} &=& \frac{4\sqrt{2} v^2}{g} C_{\nu W}^{(6)}V\,,
\eea
for the remaining operators. Here $V$ denotes the CKM matrix, $\mathbb{1}$ the $3\times3$ identity matrix in lepton flavor space, and $M_e = {\rm diag} (m_e,\, m_\mu,\, m_\tau)$ is the diagonal matrix of charged lepton masses. 

The first term in the expression for $c_{\rm VL}^{(6)}$ denotes the contribution from the SM weak interaction. All other entries arise from the 
dimension-6 operators, in some cases with additional insertions of leptonic mass matrices. The operators with Wilson coefficients $\bar C_
{\rm VL}^{(6)}$ and $\bar c_{\rm  VL}^{(7)}$ are only induced by the dimension-6 operator $\mathcal O_{\nu W}^{(6)}$. This operator involves 
a derivative acting on a charged $W^\pm$ field which, after integrating out the $W^\pm$ bosons, leads to operators involving an explicit 
derivative ($\bar c_{\rm  VL}^{(7)}$) or an insertion of a lepton mass by using the equations of motion. $C_{\nu W}^{(6)}$ is strictly constrained 
because it generates neutrino dipole moments  at one-loop \cite{Butterworth:2019iff,Canas:2015yoa}.  Additionally, if these constraints are 
avoided $N$ would decay relatively fast into  two body final states via $N\rightarrow\nu \gamma$ \cite{Duarte:2015iba,Duarte:2016miz}. To 
ensure that $N$ is long-lived, we suppress $\mathcal O_{\nu W}^{(6)}$ in the following. This effectively implies we do not consider the effects 
of $\bar C_{\rm VL}^{(6)}$ and $\bar c_{\rm  VL}^{(7)}$. 

Besides the charged currents listed in Eq.~\eqref{flavor}, we also include the effects of  the SM weak neutral currents that 
contribute to decay processes of sterile neutrinos \begin{equation}\label{SMNeutralCurrent}
\begin{aligned}
\mathcal L^{(6)}_{\rm neutral}&=\frac{-4 G_F}{\sqrt{2}}\bar \nu^i_L \gamma^\mu \nu^i_L\Bigg\{ 
   \bar e_{L}  \gamma_\mu (-\frac{1}{2}+\sin^2\theta_w)  e_L+ \bar e_{R}  \gamma_\mu (\sin^2\theta_w) e_R \\
 & +  \bar u_L \gamma^\mu (\frac{1}{2}-\frac{2}{3}\sin^2\theta_w)u_L\,+\bar u_{R}\,  \gamma_\mu   (-\frac{2}{3}\sin^2\theta_w)u_R \\
& +  \bar d_L \gamma^\mu (-\frac{1}{2}+\frac{1}{3}\sin^2\theta_w)d_L\,+\bar d_{R}\,  \gamma_\mu   (\frac{1}{3}\sin^2\theta_w)d_R \\
&
+ \frac{1}{4}(2-\delta_{ij})\bar \nu^j_L \gamma^\mu \nu^j_L \Bigg\}\,, 
\end{aligned}
\end{equation}
where $i,j$ are the  flavor indices of  active neutrinos and $\theta_w$ is the Weinberg angle.

\subsection{Rotating to the Neutrino Mass Basis}
After electroweak symmetry breaking the neutrino masses can be written as
\bea
\mathcal L_m = -\frac{1}{2} \bar N^c M_\nu N +{\rm h.c.}\,,\qquad M_\nu = \bma M_L &M_D^*\\M_D^\dagger&M_R^\dagger \ema \,,
\eea 
where $M_\nu$ is a $\bar{n}\times \bar{n}$ symmetric matrix with $\bar{n}=3+n$ and  $N = (\nu_L,\, \nu_R^c)^T$.  We use 
a $\bar{n}\times \bar{n}$ unitary matrix, U, to diagonalize the mass matrix
\bea\label{Mdiag}
U^T M_\nu U =m_\nu \equiv {\rm diag}(m_1,\dots , m_{3+n})\,,
\eea
and define $N=UN_m$. In absence of sterile neutrinos, $U$ is the usual PMNS matrix. We write the Majorana mass eigenstates 
as $\nu\equiv N_m+N^c_m=\nu^c$ that appear in the Lagrangian as
\bea
\mathcal L_\nu = \frac{1}{2} \bar \nu i\slashed \partial \nu -\frac{1}{2} \bar \nu^{ } m_\nu \nu\,.
\eea
We introduce $3\times \bar n$ and $n \times \bar n$ projector matrices 
\be
P = \begin{pmatrix}\mathcal I_{3\times 3} & 0_{3 \times n}  \end{pmatrix}\,,\qquad
P_s = \begin{pmatrix} 0_{n\times 3} & \mathcal I_{n \times n}  \end{pmatrix}\, ,
\ee 
to express the relation between the neutrinos in the flavor and mass basis as 
\bea
\nu_L = P_L(P U) \nu \,,\qquad \nu_L^c =P_R (P U^*) \nu\,,\nn\\
\nu_R =P_R (P_s U^*) \nu \,,\qquad \nu_R^c = P_L(P_s U) \nu\,.
\eea
In the mass basis, the operators in Eqs.~\eqref{lowenergy6_l0}-\eqref{lowenergy7_l0} become
\bea \label{final67}
\mathcal L^{(6,7)}_{\rm mass}& =& \frac{2 G_F}{\sqrt{2}} \Bigg\{ 
  \bar u_L \gamma^\mu d_L \left[  \bar e_{L}  \gamma_\mu C_{\rm VLL}^{(6)} \,  \nu+ \bar e_{R}  \gamma_\mu C_{\rm VLR}^{(6)} \, \nu \right] +  
  \bar u_R \gamma^\mu d_R\,\bar e_{R}\,  \gamma_\mu  C_{\rm VRR}^{(6)}  \,\nu\nn \\
  &&  
  \bar u_L  d_R \,\bar e_{L}\, C_{\rm SRR}^{(6)}   \nu  +   \bar u_R  d_L \, \bar e_{L} \, C_{\rm SLR}^{(6)}  \nu+  \bar u_L \sigma^{\mu\nu} d_R\,  
  \bar e_{L}  \sigma_{\mu\nu}C_{\rm TRR}^{(6)} \nu   \nn\\
  &&
+ \frac{1}{v}   \bar u_L \gamma^\mu d_L\, \bar e_{L} \,C_{\rm VLR}^{(7)} \,  i \overleftrightarrow{D}_\mu  \nu \Bigg\}  +{\rm h.c.}\,,
\eea
where
\begin{align}\label{redefC6}
C_{\rm VLL}^{(6)} &=   c_{\rm VL}^{(6)}P U +  \bar C_{\rm VL}^{(6)}P_sU	  \,,	\qquad 	&C_{\rm VLR}^{(6)} &= 	\bar c_{\rm VL}^{(6)}P_s U^*\,,\nn\\
C_{\rm VRR}^{(6)} &=\bar c_{\rm VR}^{(6)}P_s U^*\,,\qquad  &C_{\rm SRR}^{(6)} &= \bar c_{\rm SR}^{(6)}P_s U^*\,,\nn\\
C_{\rm SLR}^{(6)} &= \bar c_{\rm SL}^{(6)}P_s U^*\,,\qquad  &C_{\rm TRR}^{(6)} &= \bar c_{\rm T}^{(6)}P_s U^*\,,\nn\\
C_{\rm VLR}^{(7)} &= \bar c_{\rm VL}^{(7)}P_s U^*\,.
\end{align}
Each Wilson coefficient again carries four flavor indices $ijkl$ where $i,j,k=\{1,2,3\}$ indicate the generation of the involved up quark, 
down quark, and charged lepton, respectively. $l$ now denotes the particular neutrino mass eigenstate and runs from $\{1,\dots,\bar n \}$. 

Finally, we evolve the operators down to the bottom or charm mass. The vector currents do not evolve while the scalar and tensor dimension-6 and -7 couplings evolve in the same way as the scalar and tensor currents in Eq.\ \eqref{eq:STrge}, with
\bea
C^{(6)}_S = C_{\rm SRR,\,\rm SLR}^{(6)}\,,\qquad C^{(6)}_{\rm  T}=C_{\rm TRR}^{(6)}\,.
\eea

In what follows we only consider the dimension-six terms in Eq.~\eqref{final67} and neglect the dimension-seven operator proportional to  
$C_{\rm VLR}^{(7)}$ whose low-energy effects are suppressed by $m_{b,c}/v$. Furthermore, as discussed below Eq.~\eqref{match6LNC},  
$C_{\rm VLR}^{(7)}$ is only induced by the $\nu$SMEFT operator $C_{\nu W}^{(6)}$ which is strongly constrained by other probes.

\section{Production of Sterile Neutrinos}\label{sec:prod-of-HNLs}

In this section we discuss the production of sterile neutrinos at collider and fixed-target experiments. For concreteness we consider the case
where the sterile neutrinos are Majorana particles. We consider the production through the decay of mesons, produced at the interaction points, containing a single charm or bottom quark. We neglect the subdominant contribution from $B_c$, $J/\Psi$, and $\Upsilon$ mesons although they would allow to probe a larger neutrino mass range.
Production via the decay of pseudoscalar mesons dominates over the contribution from vector mesons of the same 
quark composition, due to the much shorter lifetime of the latter. 
Sterile neutrino production via direct decays of $W$-, $Z$-, and Higgs bosons is subdominant for $\mathcal{O}(\text{GeV})$ 
neutrinos, mainly because of their smaller production cross sections \cite{Helo:2018qej,Curtin:2018mvb, Hirsch:2020klk}. 

\subsection{Sterile Neutrino Production in Minimal Models} 
We begin by discussing sterile neutrino production in minimal models where sterile neutrinos interact with SM fields only via 
mixing. For simplicity, we consider a single sterile neutrino with mass $m_N$ and set $n=1$ ($\overline{n}=4$). The production 
then arises solely from the first term in Eq.~\eqref{final67} with $(C_{\rm VLL}^{(6)})_{ijk4}=-2 V_{ij}\,U_{k4}$, where $V$ is the 
CKM matrix and $U$ the lepton mixing matrix. A broad range of processes are relevant. Naively one might think that leptonic 
meson decays $ M_{ij}^\pm \rightarrow N + l_k^\pm$ would dominate because of phase space suppression associated to semi-leptonic 
decays, but CKM factors and powers of meson/neutrino masses in the amplitude expressions change this picture. To calculate 
the production rate, we require the number of mesons produced at the various experiments and the branching ratio to final 
states including a sterile neutrino. The former is discussed below, while here we calculate the latter. For minimal models, these 
branching ratios have been calculated in the literature, see Refs.~\cite{Bondarenko:2018ptm,Coloma:2020lgy} for recent 
discussions, and here we confirm (most of) these results. 

We consider leptonic and semi-leptonic decays of $D^\pm$,  $D^0$, $D_s$, $B^\pm$, $B^0$, and $B_s$ mesons.
For semi-leptonic decays, we consider final-state pseudoscalar and vector mesons.
The decay rate formulae and the associated decay constants and form factors are given in the Appendix.
In the left and right panels of Fig.~\ref{fig:BRminimal} we depict a selection of branching ratios for decay processes of $D^-$, $D_s$, and $B^-$, $B_s$ mesons, respectively. 
Branching ratios for analogous decays of neutral $D^0$ or $B^0$ are similar and not shown to not clutter the plots too much.
For these examples we considered a final-state electron and set $U_{e4} =1$.
All branching ratios are in excellent agreement with Ref.~\cite{Bondarenko:2018ptm}. From the plots it is clear that, depending on the mass $m_N$, both leptonic (solid lines) and semi-leptonic processes (dashed, dotted, and 
dot-dashed lines) must be included and the latter involve both final-state pseudoscalar and vector mesons.

\begin{figure}[t]
	\centering
	\includegraphics[width=0.49\linewidth]{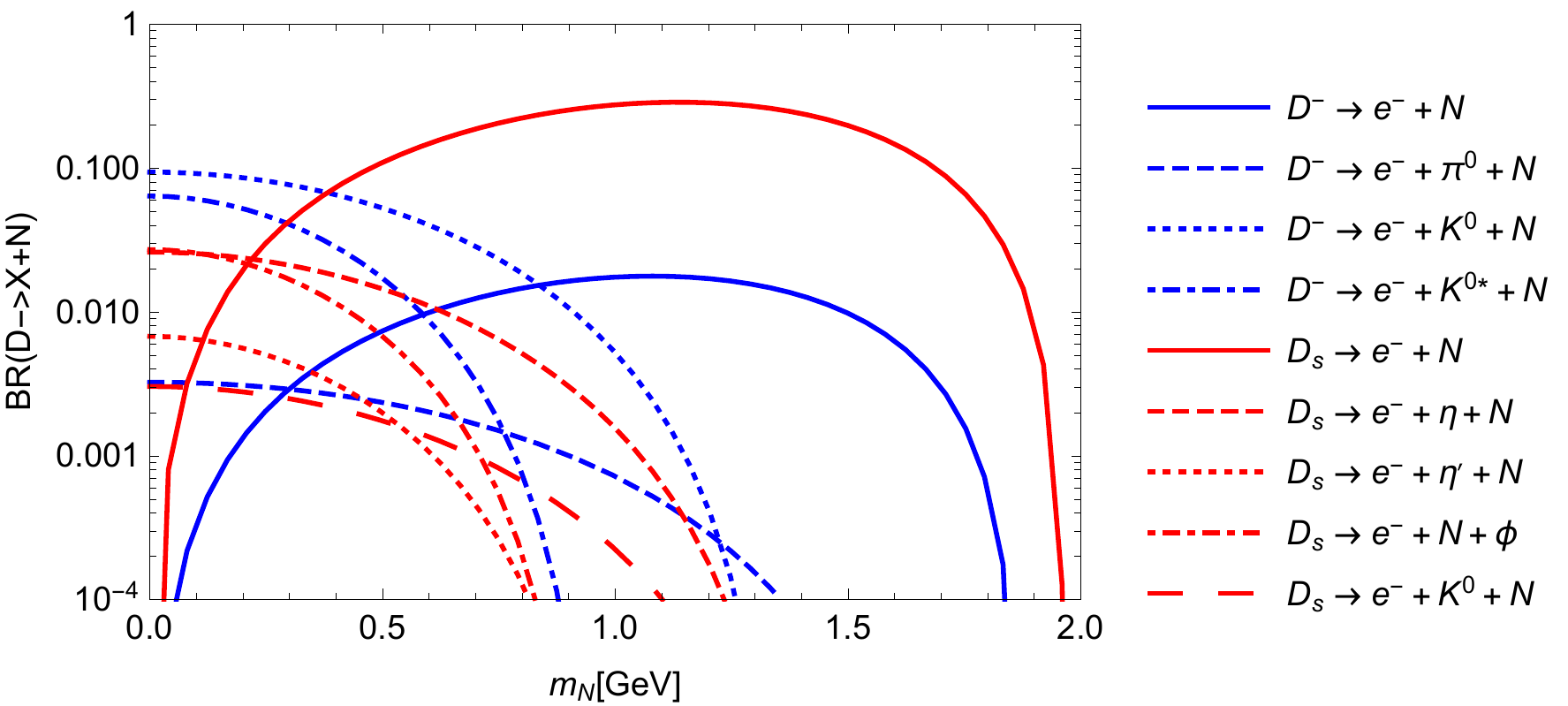}
	\includegraphics[width=0.49\linewidth]{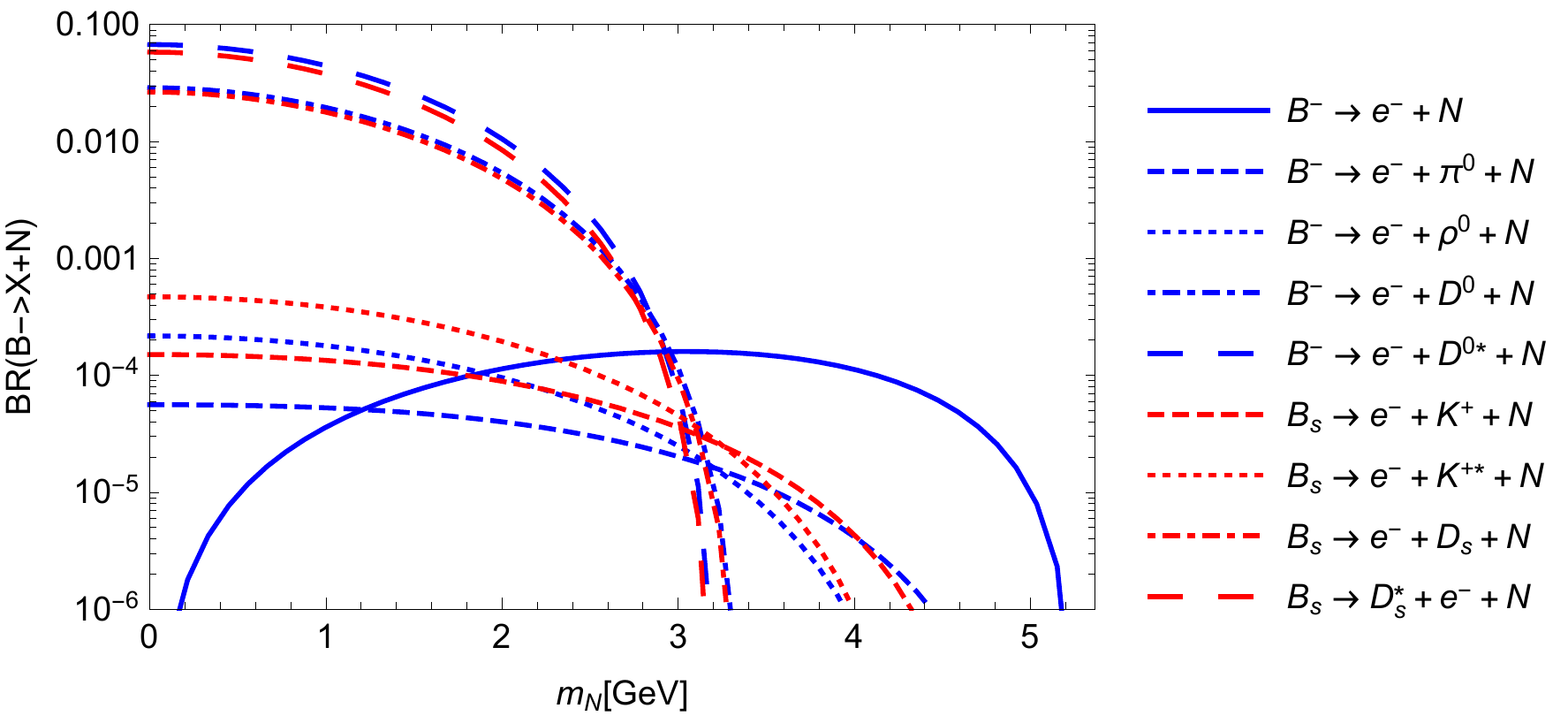}
	\caption{Branching ratios of sterile neutrino production channels through $D$ (left figure) or $B$ (right figure) mesons in the minimal scenario for final-state electrons and $U_{e4} =1$.}
	\label{fig:BRminimal}
\end{figure}

\subsection{Sterile Neutrino Production from Higher-Dimensional Operators} 

\begin{figure}[t]
	\centering
\includegraphics[width=0.49\textwidth]{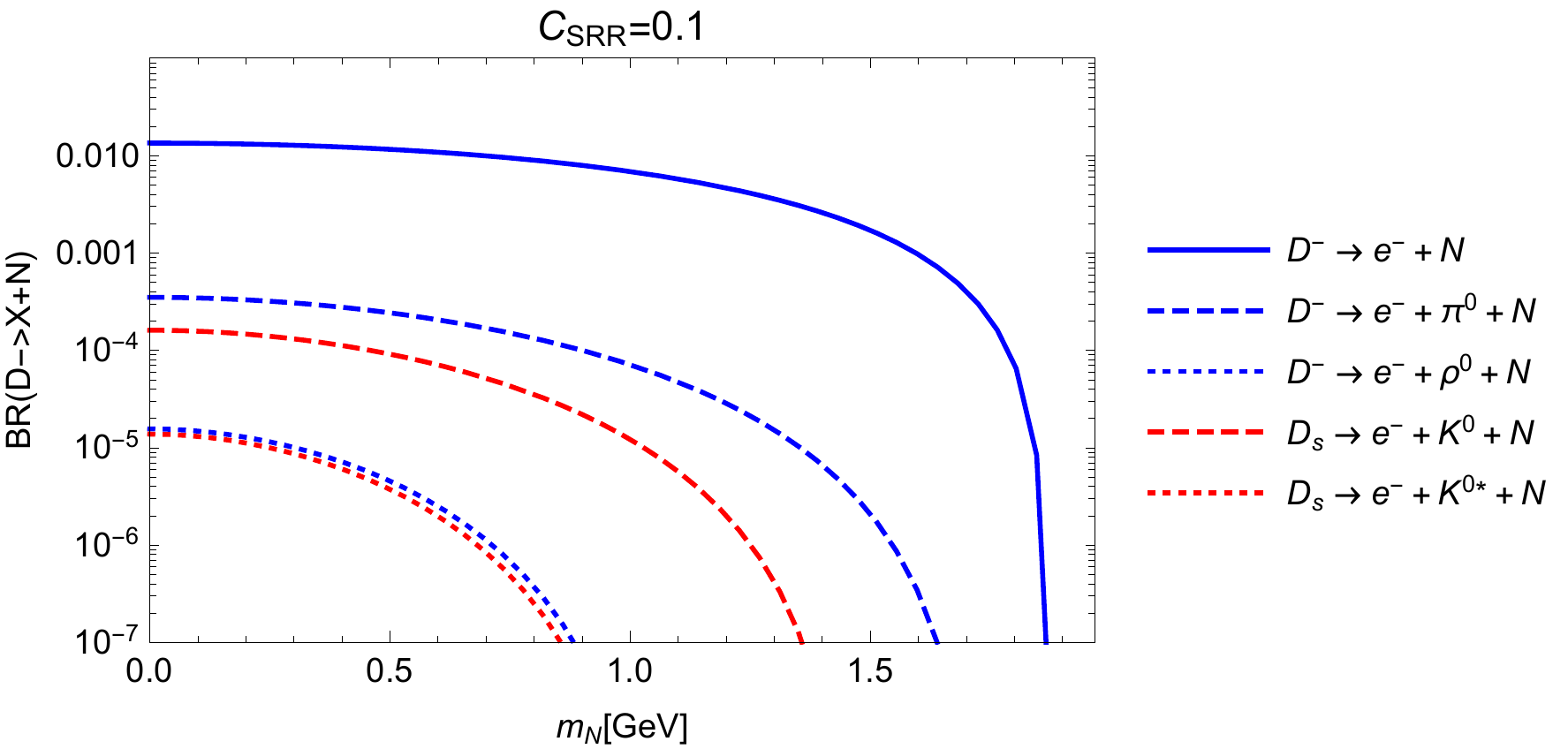}
\includegraphics[width=0.49\textwidth]{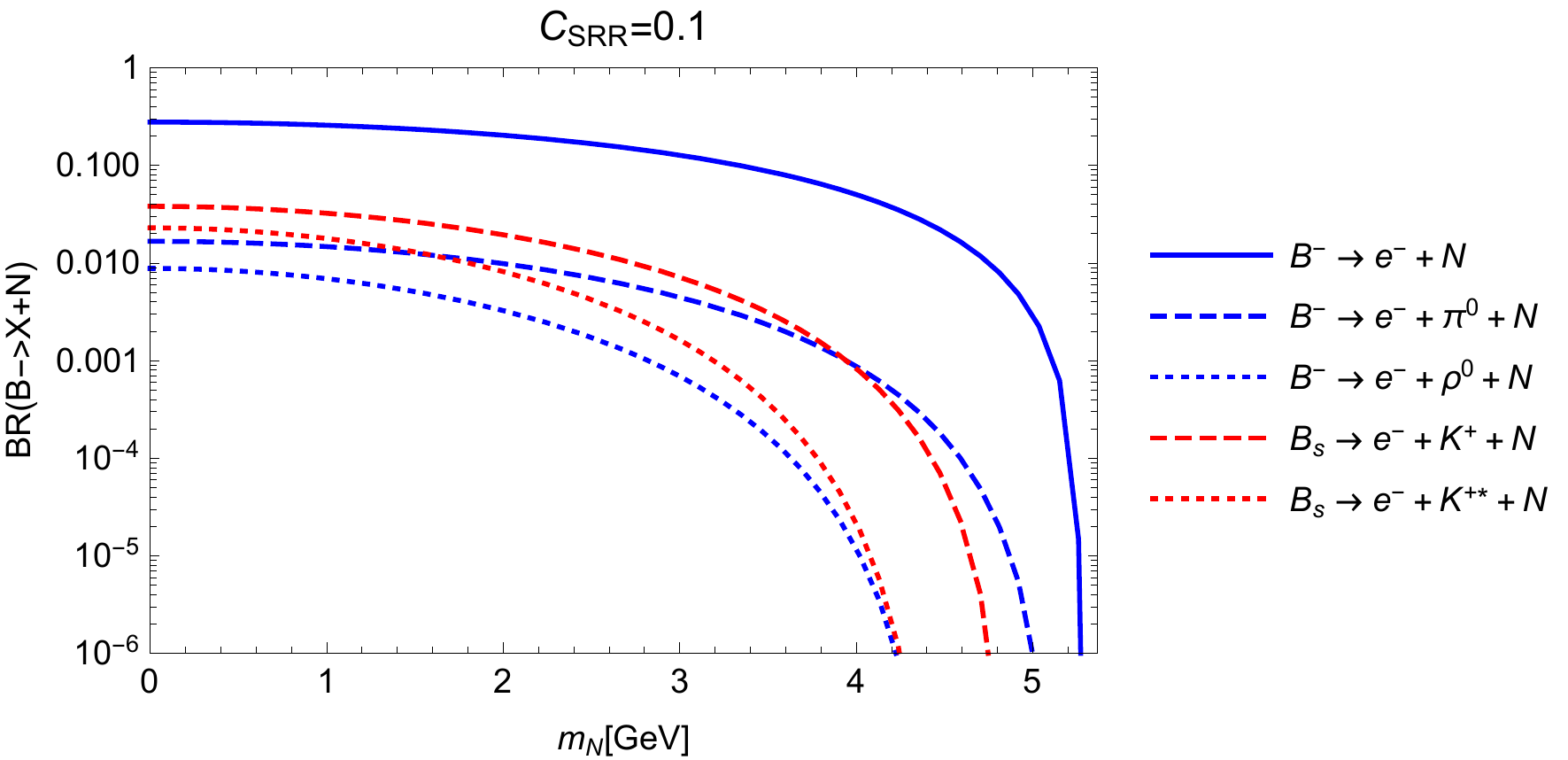}
\includegraphics[width=0.49\textwidth]{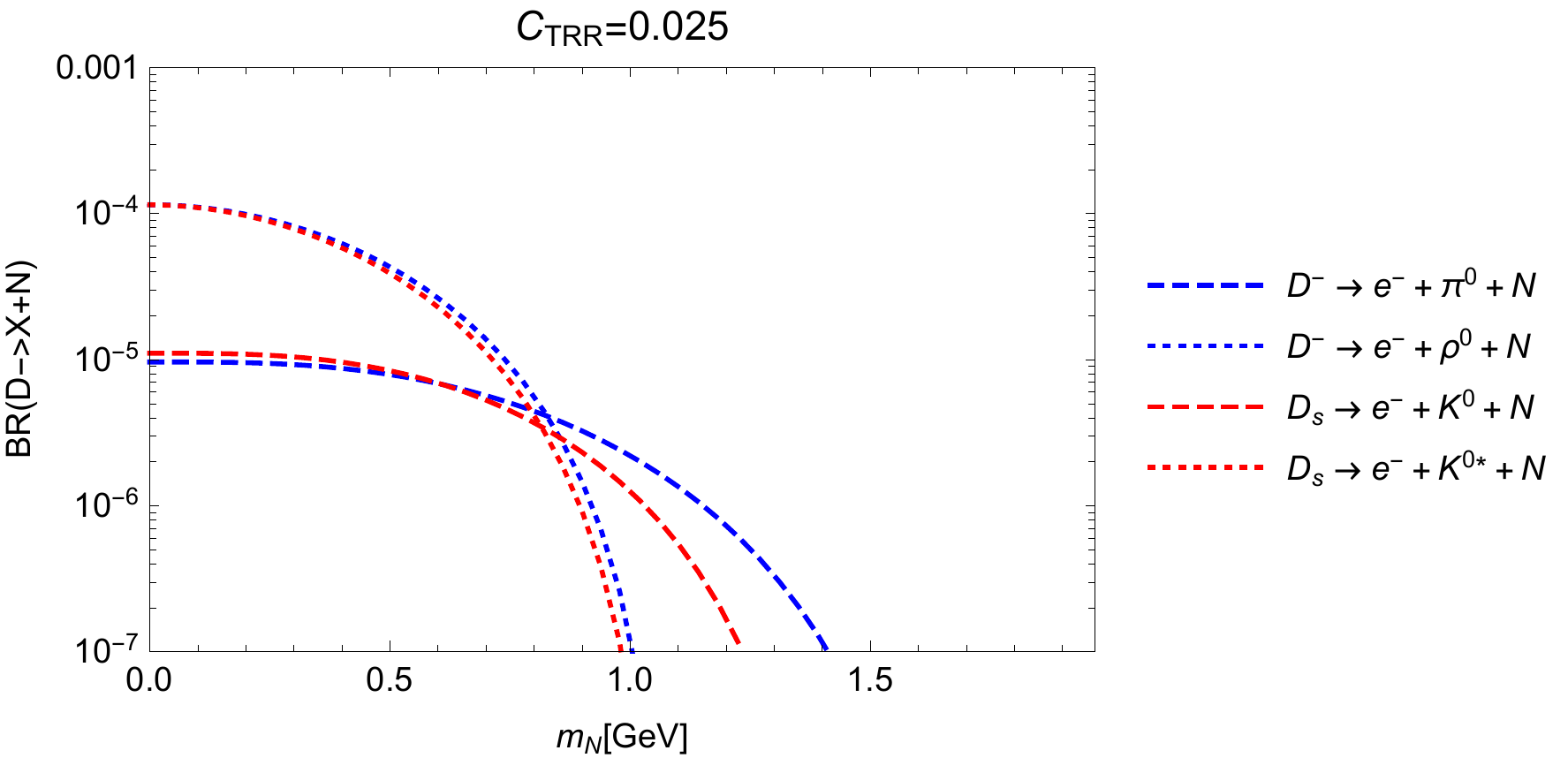}
\includegraphics[width=0.49\textwidth]{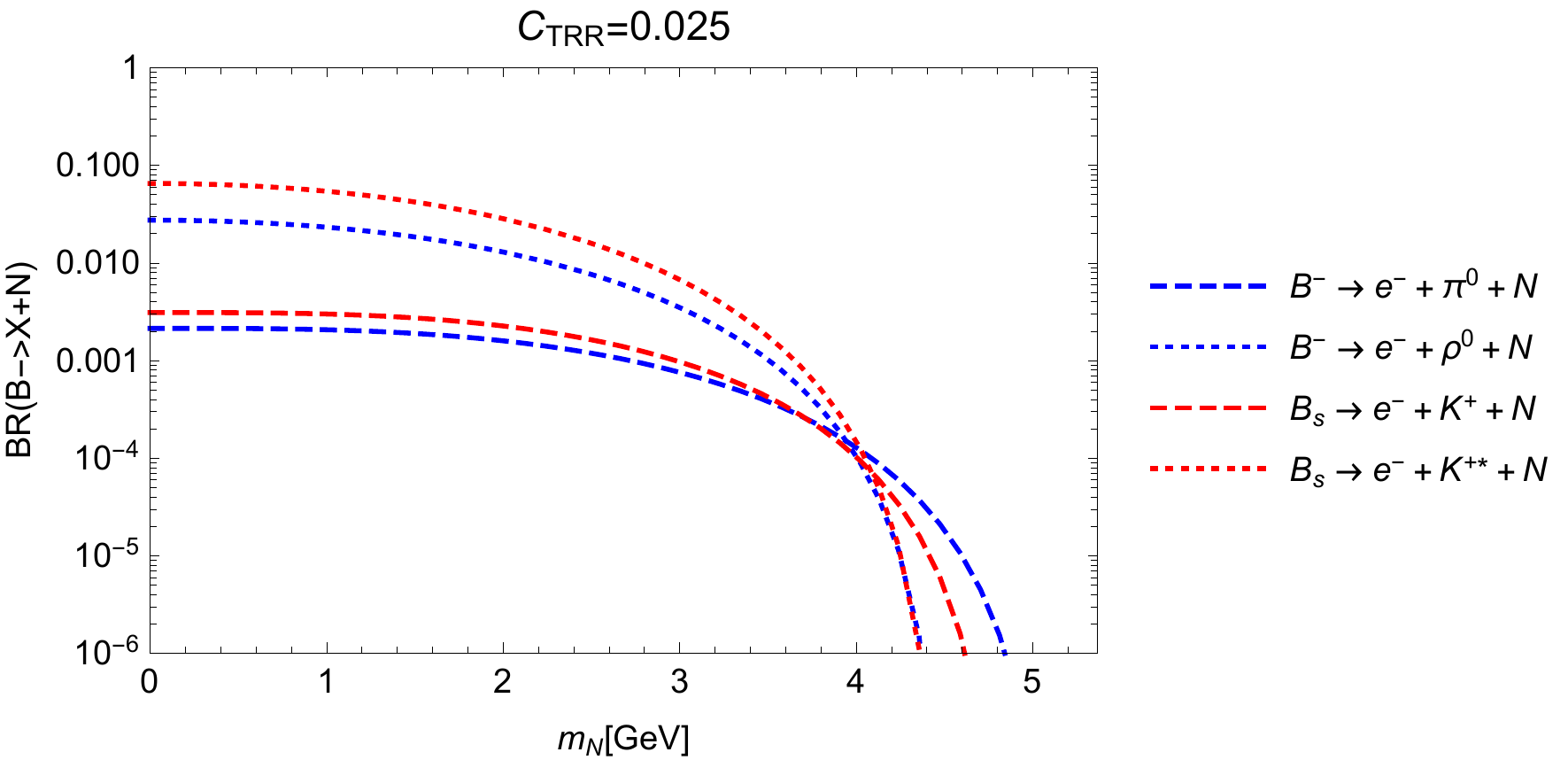}
\includegraphics[width=0.49\textwidth]{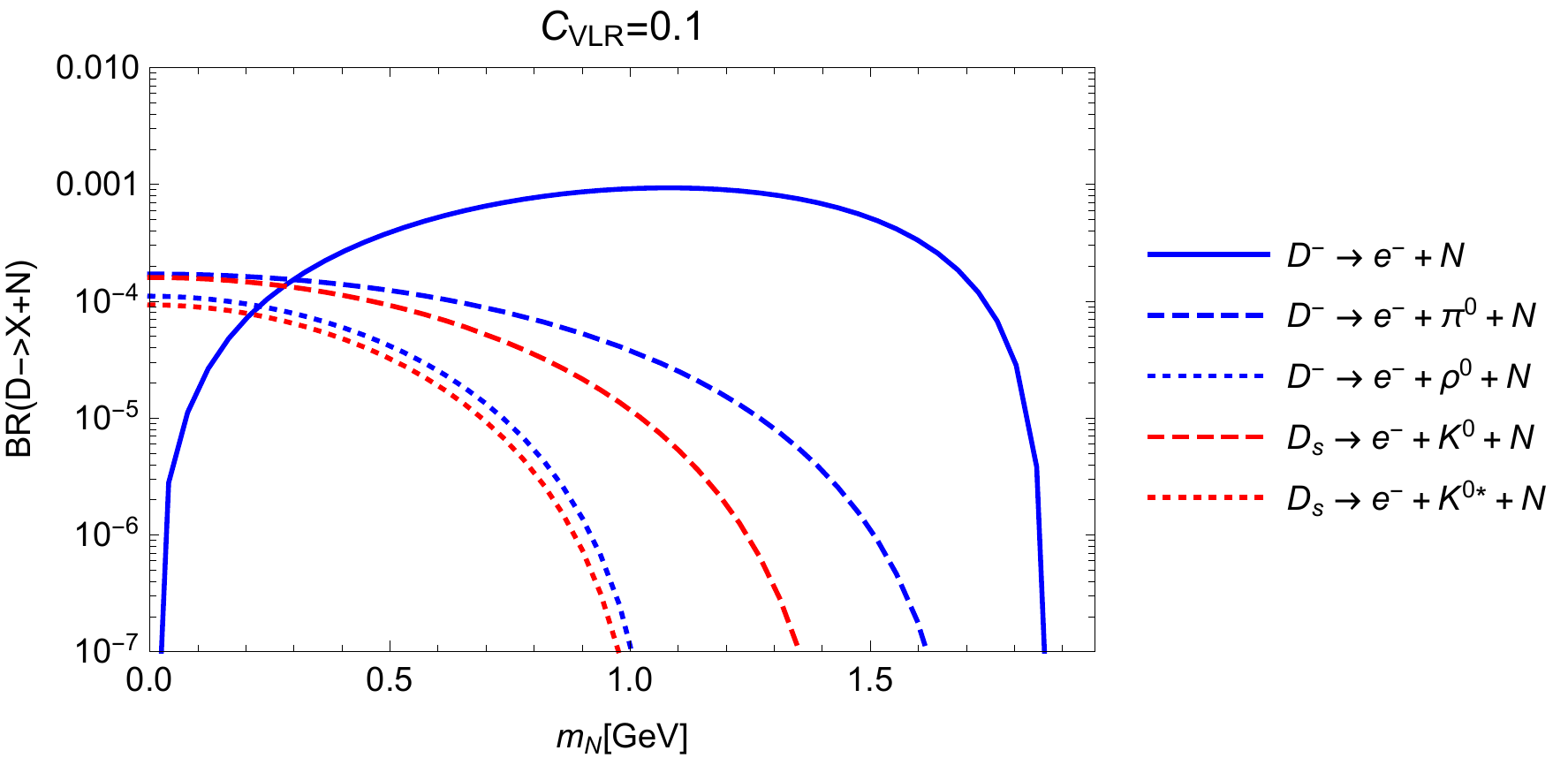}
\includegraphics[width=0.49\textwidth]{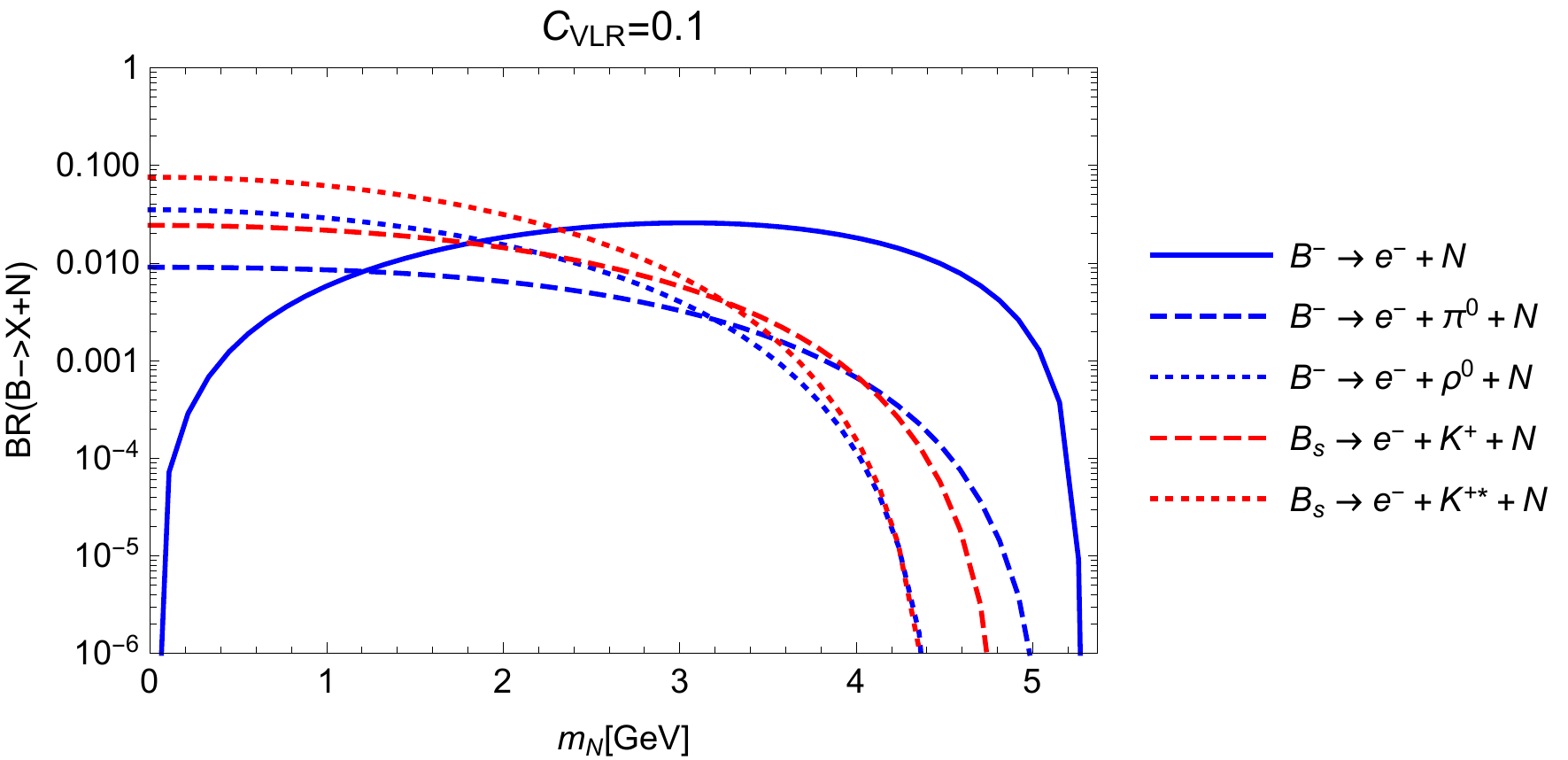}
	\caption{Branching ratios of $D$ and $B$ mesons in the left and right panels, respectively. Left: from top to bottom the figures correspond to $(C_{\rm SRR}^{(6)})_{21e4}=0.1$,  $(C_{\rm TRR}^{(6)})_{21e4}=0.025$ and $(C_{\rm VLR}^{(6)})_{21e4}=0.1$, respectively. Right:  from top to bottom the figures correspond to $(C_{\rm SRR}^{(6)})_{13e4}=0.1$, $(C_{\rm TRR}^{(6)})_{13e4}=0.025$ and $(C_{\rm VLR}^{(6)})_{13e4}=0.1$, respectively}.
	\label{fig:eftBR}
\end{figure}

For higher-dimensional operators the quark flavor structure of the Wilson coefficients is unknown in contrast to the minimal 
case where the CKM matrix provides the relation between processes involving different quarks. As such, each flavor structure 
is independent unless model assumptions are used. This leads to a large number of possible cases corresponding to several 
flavor structures for each effective operator in Eq.~\eqref{final67}. All branching ratios can be calculated from the expressions 
given in the appendices. 

Here we discuss a few cases only. We consider the operators with Wilson coefficients $C_{\rm SRR}^{(6)}$,  $C_{\rm TRR}^{(6)}$, 
and $C_{\rm VLR}^{(6)}$. We consider the flavor structures $\{ijkl\} = {13e4}$ and $\{ijkl\} = {21e4}$ that allow for leptonic decays 
$B \rightarrow N + e$ and  $ D \rightarrow N + e$, respectively, if the Lorentz structure permits this. These choices also allow for 
semi-leptonic decays of the form $B \rightarrow N + e + X$ and  $ D \rightarrow N + e + X$ where $X$ is a pseudoscalar or vector 
meson consisting of just up, down, and strange quarks (strange quarks only if the decaying meson contains a strange quark as is 
the case for $B_s$ and $D_s$ mesons). For the plots in this section, we assume the weak interaction is turned off and consider 
only one non-zero EFT operator at a time. The results are depicted in Fig.~\ref{fig:eftBR}.

The three chosen operators correspond to quark bilinears with different Lorentz structures (scalar, tensor, and vector, respectively). 
In the scalar case, leptonic decays are allowed and these dominate over semi-leptonic decay modes for all considered values of 
the sterile neutrino mass. For the tensor operator, however, the leptonic decay mode is forbidden and a final-state meson must be 
produced. In these cases, the dominant decay modes are those with a final-state vector meson. Finally, the $C_{\rm VLR}^{(6)}$ 
vector operator has a similar Lorentz structure as the SM charged weak current, but with different flavor structure. As was the case 
in Fig.~\ref{fig:BRminimal}, depending on the sterile neutrino mass, either leptonic (solid lines)  or semi-leptonic processes 
(dashed, dotted, and dot-dashed lines) can dominate the production of sterile neutrinos, and must all be included.

\section{Decay of Sterile Neutrinos}\label{sec:decays-of-HNLs}

\subsection{Sterile Neutrino Decays in Minimal Models}

We begin by considering the minimal scenario, where we assume that the only non-zero term in Eq.~\eqref{final67} is 
$(C_{\rm VLL}^{(6)})_{ijk4}=-2 V_{ij}\,U_{k4}$. For concreteness we consider decays of Majorana sterile neutrinos into final-state electrons and set 
$U_{e4}\neq 0$ and $U_{\mu4}=U_{\tau 4}=0$. In addition, we consider the SM weak neutral current (see 
Eq.~\eqref{SMNeutralCurrent}), which leads to $N \rightarrow \nu + f +\bar f$ decays where $f$ denotes any SM fermion 
that is kinematically allowed (in case of quarks, we consider a final-state neutral meson). These decay rates have all 
been calculated in the literature, see \textit{e.g.} 
Refs.~\cite{Johnson:1997cj,Gribanov:2001vv,Gorbunov:2007ak,Atre:2009rg,Bondarenko:2018ptm,Coloma:2020lgy}. 
Most results agree with each other and with our findings given in the Appendix, with the exception for decay processes 
into final-state neutral mesons where some differences appear. For these cases, our results agree with Ref.~\cite{Coloma:2020lgy}.

We consider $N \rightarrow \mathrm{leptons}$ through both charged and neutral weak currents. The latter leads to the 
invisible three-neutrino decay mode. We include decays into a single pseudoscalar ($\pi$, $K$, $\eta$, $\eta'$, $D$, $D_s$, 
$\eta_c$) and vector meson ($\rho$, $\omega$, $K^*$, $\phi$, $D^*$, $D_s^*$, $\mathrm{J}/\Psi$). This effectively also takes 
into account decay modes into two pions through the intermediate decays of $\rho$ mesons 
\cite{Bondarenko:2018ptm}, assuming $m_N > m_\rho$. For heavier sterile neutrinos other multi-meson final states become 
relevant and summing exclusive channels becomes impractical. Instead we follow Refs.~\cite{Bondarenko:2018ptm} and estimate 
the total hadronic decay width by calculating the decay width to spectator quarks times appropriate loop corrections. The loop 
corrections are taken from a comparison to hadronic $\tau$ decays
\begin{equation}
1+\Delta_{QCD}(m_\tau) \equiv \frac{\Gamma(\tau\rightarrow \nu_e +\mathrm{hadrons})}{\Gamma_{\mathrm{tree}}(\tau
\rightarrow \nu_\tau +\bar{u}+D))}\,,
\end{equation} 
where $D$ denotes a $d$ or $s$ quark and
\begin{equation}
\Delta_{QCD} = \frac{\alpha_s}{\pi} + 5.2 \frac{\alpha_s^2}{\pi^2} + \dots\,,
\end{equation}
where dots denote higher-order corrections. This gives a good description of the inclusive hadronic $\tau$ decay rate and 
we assume this to hold for sterile neutrino decays in the minimal scenario as well. That is, we use
\begin{equation}\label{Ndecayhadronic}
1+\Delta_{QCD}(m_N) \equiv \frac{\Gamma(N\rightarrow e^-/\nu_e +\mathrm{hadrons})}{\Gamma_{\mathrm{tree}}(N\rightarrow e^-/\nu_e +\bar{q} q)}\,,
\end{equation} 
to calculate the inclusive hadronic sterile neutrino decay rate through both charged and neutral weak currents.

\begin{figure}[t]
	\centering
	\includegraphics[scale=0.5]{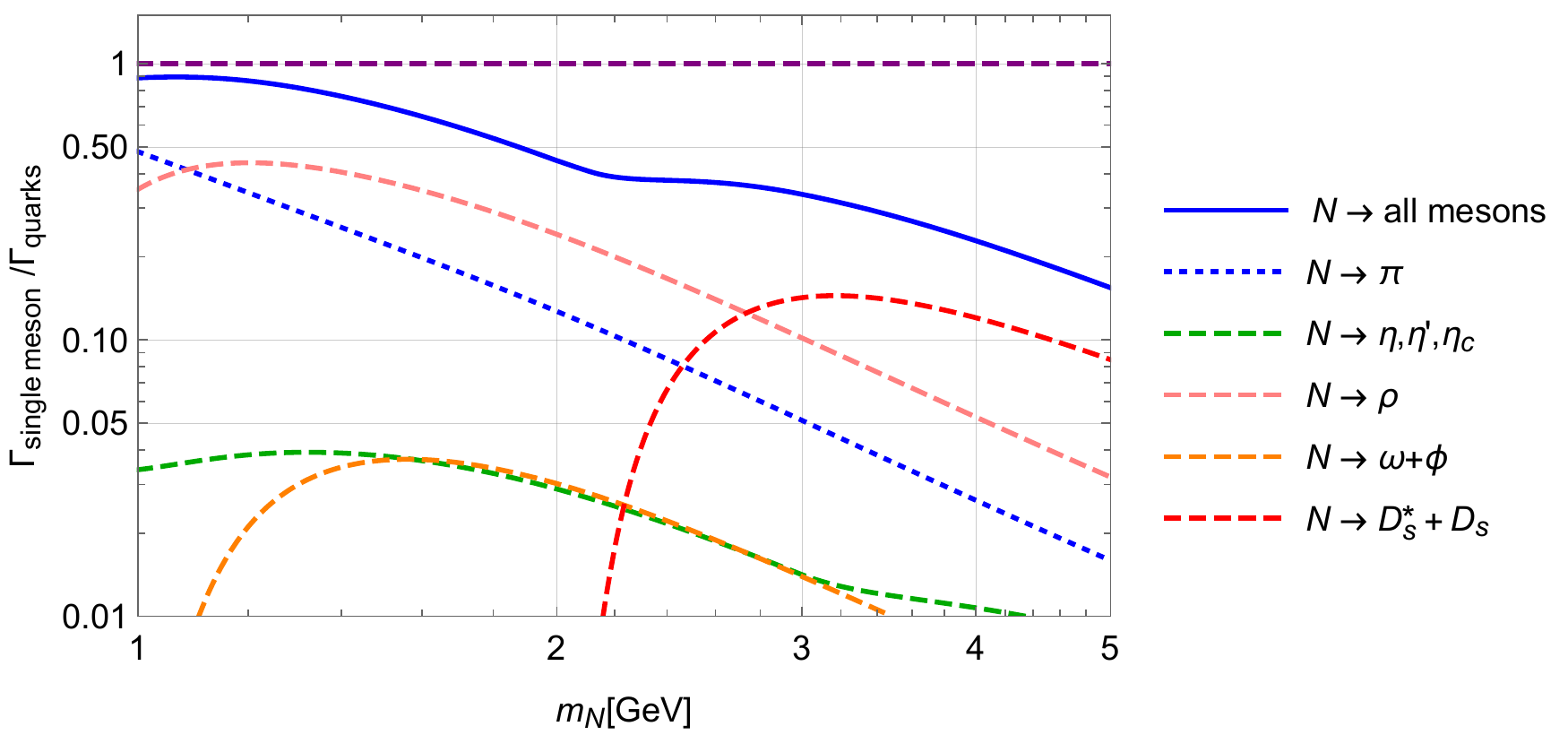}
	\includegraphics[scale=0.5]{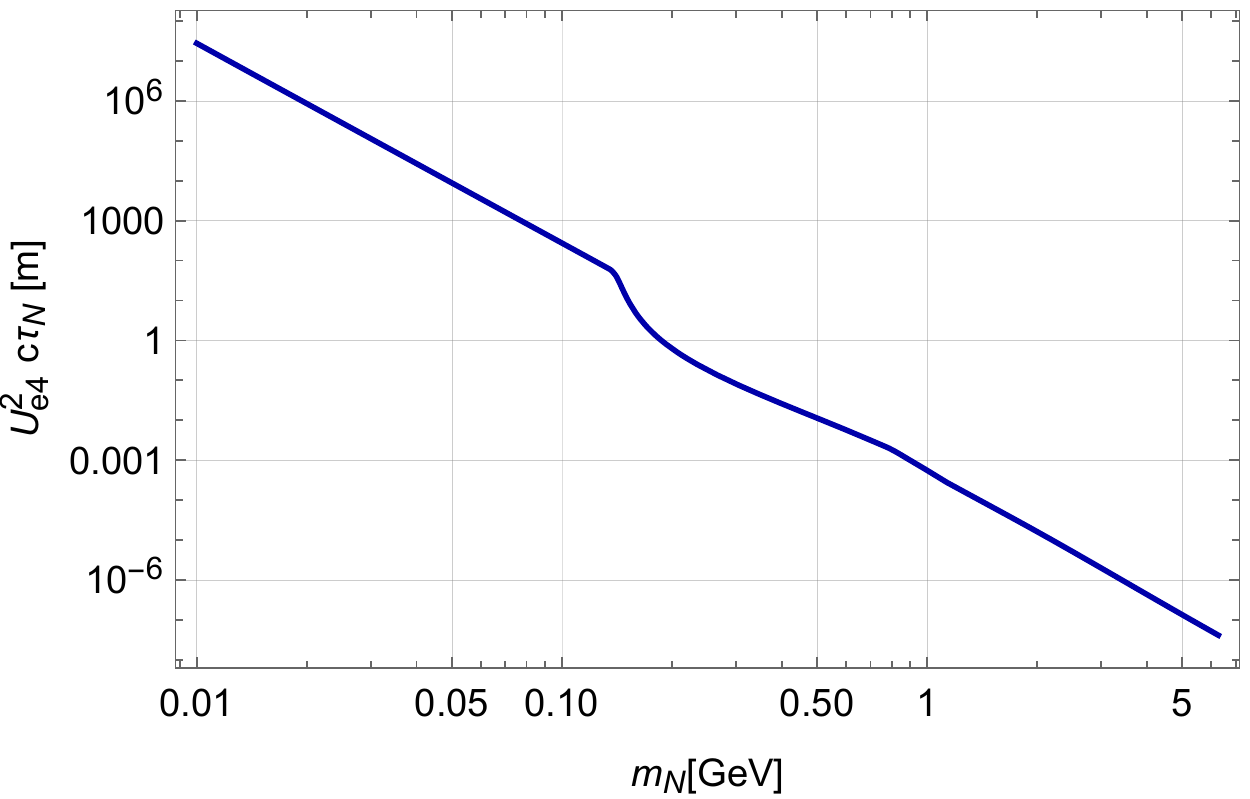}
	\caption{Left: Comparison of branching ratios to individual mesons divided by the inclusive hadronic branching ratio calculated via Eq.~\eqref{Ndecayhadronic}.
		Right: Decay length of the sterile neutrino in minimal scenarios.}.
	\label{fig:MesontoQuarks}
\end{figure}

We find that single meson channels dominate for $m_N \lesssim 1$ GeV, while the decay to quarks become relevant for larger 
masses, indicating that multi-meson final states become significant.
We demonstrate this in the left plot of Fig.~\ref{fig:MesontoQuarks}, which shows branching ratios to individual mesons, compared to the sum of all single-meson final states, and compared to quarks, for $m_N \gtrsim 1$ GeV.
At $m_N = 5$\,GeV, the single meson final states make up roughly $20\%$ of the hadronic decay rate.
Our results are in good agreement with Ref.~\cite{Bondarenko:2018ptm}, apart from decays to neutral vector mesons, which only play a small role.

\begin{figure}[t]
	\centering
	\includegraphics[width=0.49\linewidth]{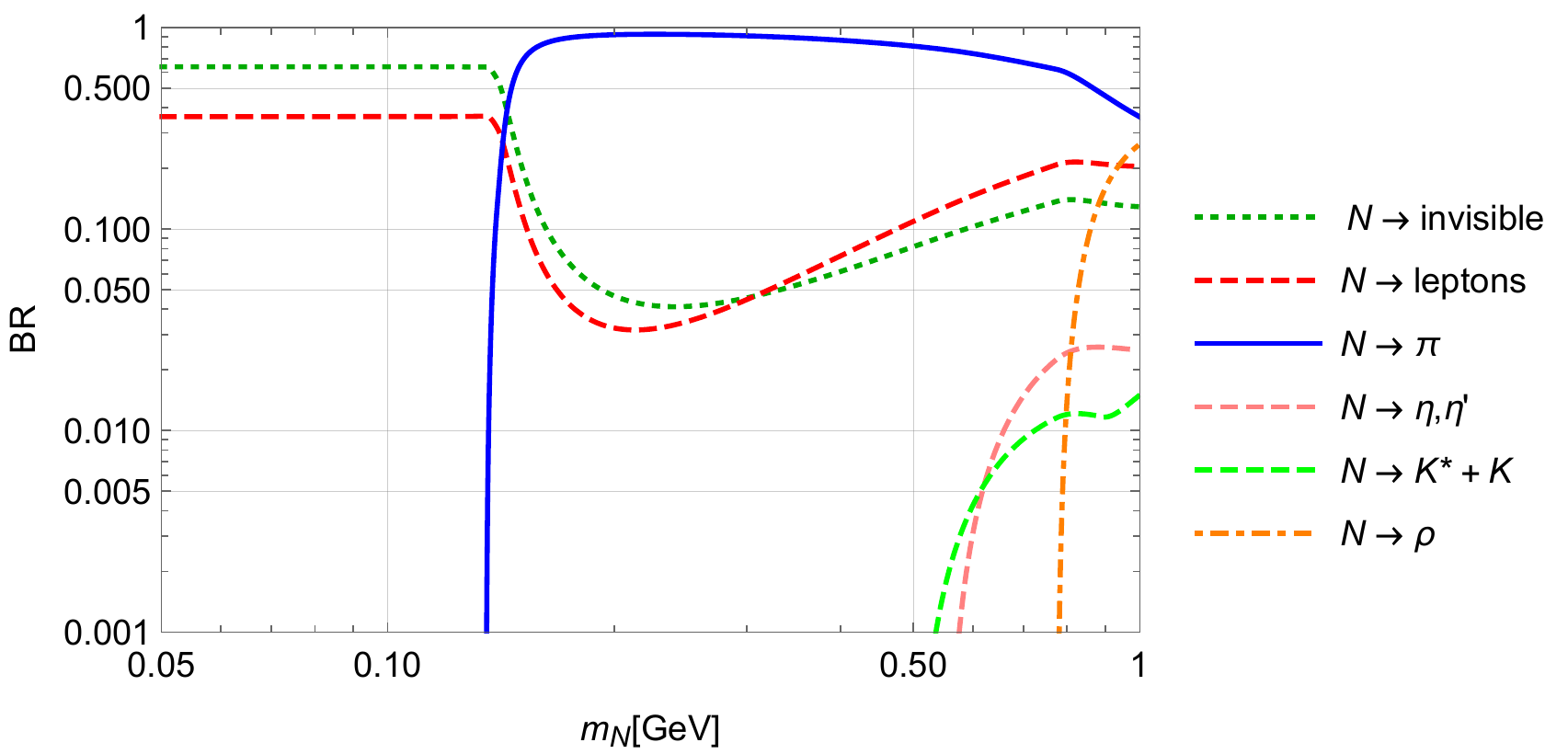}
	\includegraphics[width=0.49\linewidth]{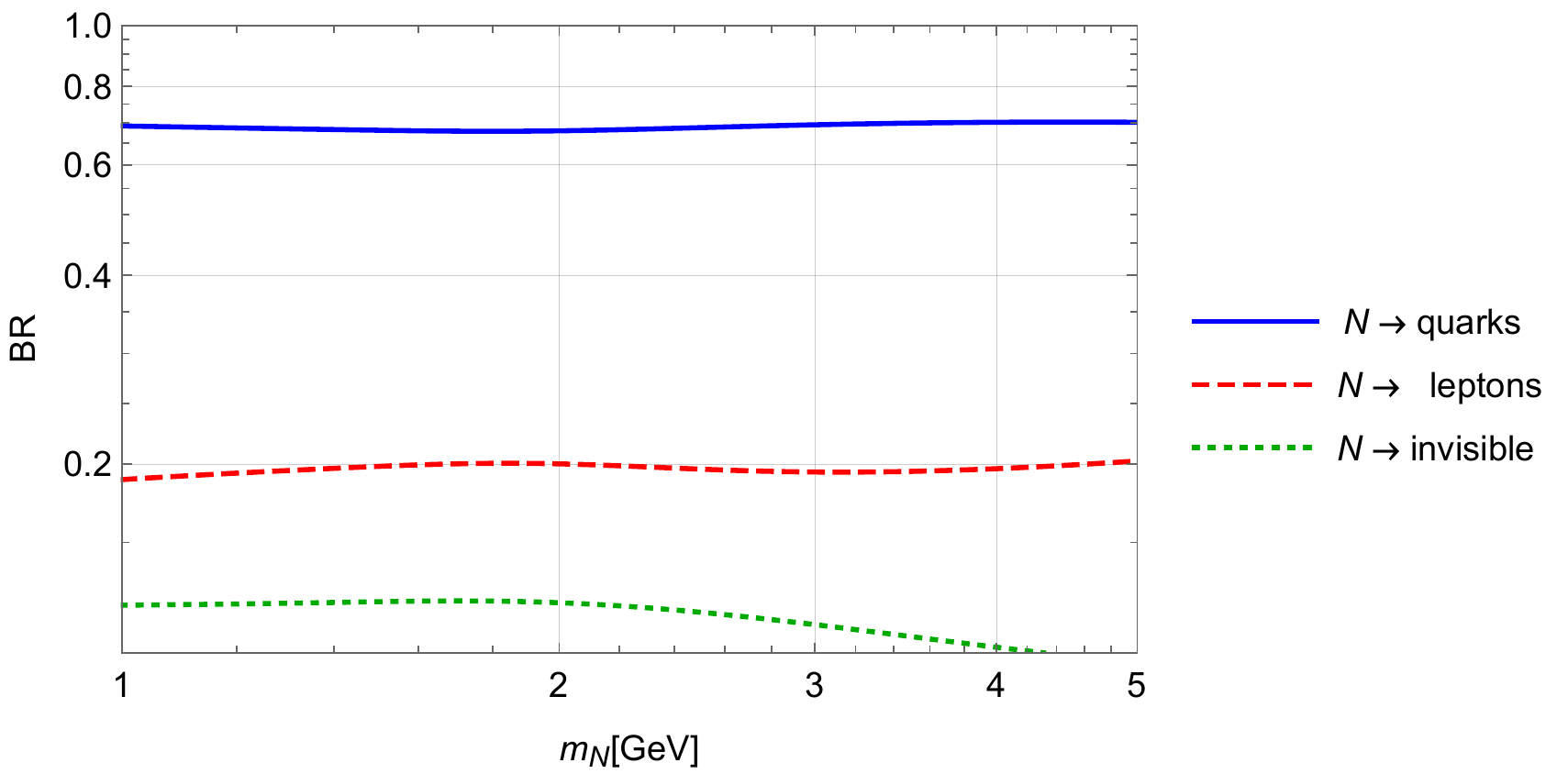}
	\caption{Branching ratios in the minimal scenario for $U_{e4} =1$ and $U_{\mu4}=U_{\tau 4}=0$. Left: $m_N <1$\,GeV with decays to individual mesons.
		Right: $m_N > 1$ GeV and the decays to quarks correspond to the total hadronic branching ratio.}
	\label{fig:NBR}
\end{figure}

We write the total decay rate as
\begin{eqnarray}
\Gamma_{N}&=&\theta(1\,\mathrm{GeV}-m_N)\Gamma_{N\rightarrow \text{single meson}}+\theta(m_N-1\,\text{GeV})\left[
1+\Delta_{QCD}(m_N)\right]\Gamma_{N\rightarrow \bar{q}q}\nonumber\\
&&+\Gamma_{N\rightarrow \text{leptons}}\,.
\end{eqnarray}
In the right panel of Fig.~\ref{fig:MesontoQuarks}, we show a plot of the scaled proper decay length, $U^2_{e4}c\tau_N$, as a function of $m_N$ in the minimal scenario, where $c$ is the speed of light and $\tau_N$ is the proper lifetime of $N$. 
The branching ratios to individual mesons, leptons, and three neutrinos (invisible) for $m_N < 1$\,GeV are shown in the left panel of Fig.~\ref{fig:NBR} while the branching ratios to quarks, leptons, and invisible for $m_N > 1$ GeV are shown in the right panel of the same figure. 

\subsection{Sterile Neutrino Decays from Higher-Dimensional Operators}

The EFT operators in Eq.~\eqref{final67} involve two quarks and a charged lepton and induce new channels for sterile neutrinos 
to decay into hadrons. Depending on the flavor structure of the operators, typically only one or two single-meson decay modes 
are relevant. For instance for an operator $(C_{\rm VLR}^{(6)})_{ije4}$ with $ij =11$ we consider decays into a single charged 
pion or $\rho$ meson, while for $(C_{\rm SRR}^{(6)})_{11e4}$ only pions are relevant. As is the case for the minimal scenario, 
one can imagine multi-meson states to become relevant for larger sterile neutrino masses (for this flavor choice, such states 
would be three- or more pions). We do not include such states here, although we find that decays to quarks become dominant 
around $m_N\gtrsim 2$\,GeV for scalar operators, as we do not have the benchmark of hadronic $\tau$ decays to verify our 
results for non-SM currents. We only consider decays into individual mesons. This leads to a potential underestimate 
of the sterile neutrino decay width and consequently renders our sensitivity limits for future experiments conservative in the large 
decay length regime. As all our results below are given on log scales, we do not expect significant deviations from our findings. 
\begin{figure}[t]
	\centering
	\includegraphics[width=0.89\linewidth]{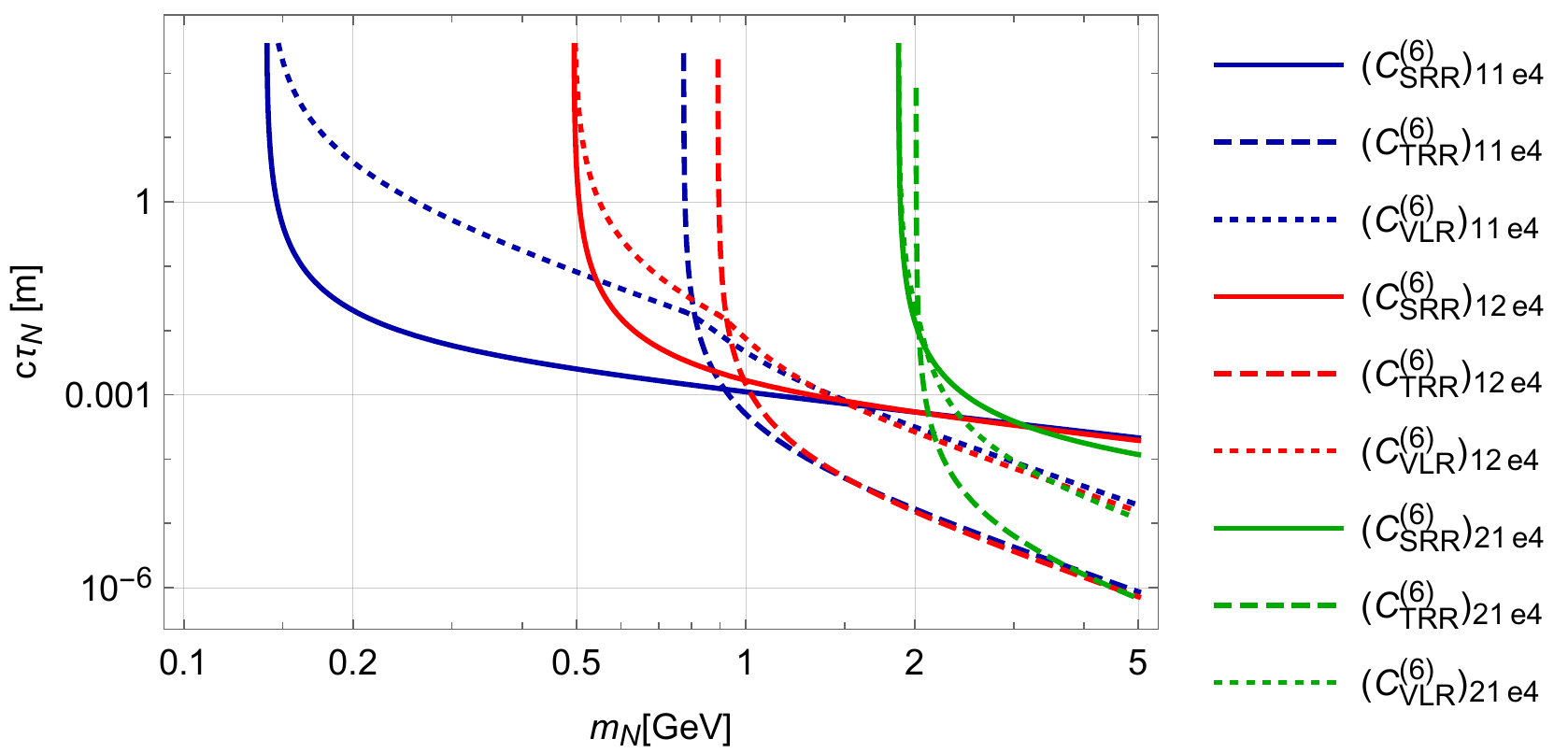}
	\caption{Proper decay length of sterile neutrinos for various choices of EFT operators and flavor assignments. The Wilson coefficients are set to unity and the minimal active-sterile mixing is turned off. 
	}
	\label{fig:Neftlength}
\end{figure}
In Fig.~\ref{fig:Neftlength} we plot the proper decay length of $N$, $c\tau_N$, against its mass for various choices of Wilson coefficients and flavor assignments. We consider one effective coupling at the time, turn off minimal mixing, and the Wilson coefficients are set to unity.

\section{Theoretical Scenarios}
\label{sec:scenarios}

In this work we focus on the production of sterile neutrinos through the decays of $B$ and $D$ mesons. We consider three 
classes of scenarios that are representative of the effective Lagrangian in Eq.~\eqref{final67}. We always consider the case 
of a single sterile neutrino which is mixed with the active electron neutrino. The three classes of scenarios are listed below:
\begin{enumerate}
\item Here we consider the minimal scenario without higher-dimensional operators and 1 sterile neutrino (a $3+1$ model). 
Because of minimal sterile-active mixing, the sterile neutrino only interacts through charged and neutral SM weak 
interactions. In the $3+1$ minimal seesaw model, the mixing angle is related to the ratio of active and sterile neutrino 
masses $|U_{e4}|^2 \sim m_{\nu}/m_N $, but we treat the mixing angle as a free parameter. We stress that the minimal 
$3+1$ model leads to two massless active neutrinos and is thus ruled out by neutrino oscillation experiments, but with its 
simplicity it provides a useful benchmark. 

\item In this scenario we extend the minimal $3+1$ model by  interactions generated by the exchange of leptoquarks. All possible 
representations of LQs are summarized in Ref.~\cite{Dorsner:2016wpm}. We focus on the representation $\tilde R\left({\bf 3},~{\bf 
2},~1/6\right)$, which can couple to (sterile) neutrinos through the Lagrangian
\begin{equation}\label{eq:LQlag}
{\cal L}_{\rm LQ}=-{y}^{RL}_{jk}\bar{d}_{Rj}\tilde R^a\epsilon^{ab}L_{Lk}^{b}+y^{\overline {LR}}_{il}\bar{Q}^{a}_{Li}\tilde R^a\nu_{Rl} 
+{\rm h.c.}\,,
\end{equation}
where $a,b$ are $SU(2)$ indices and $i,j,k,l$ are flavor indices, respectively. Note that $l=1$ in the $3+1$ model. LHC constraints 
force the leptoquark mass, $m_{\rm LQ}$, to be above a few TeV and for low-energy purposes we can integrate it out. At tree level 
this leads to the effective operator 
\begin{align}
{\cal L}^{(6)}_{\nu_R}=\left(C^{(6)}_{LdQ\nu}\right)_{ijkl}\left(\bar{L}^a_k d_j \right)\epsilon^{ab}\left(\bar{Q}^b_i \nu_{Rl} \right)+{\rm h.c.}\,,
\end{align}
where 
\begin{equation}
\left(C^{(6)}_{LdQ\nu}\right)_{ijkl}=\frac{1}{m^2_{\rm LQ}}y^{\overline{LR}}_{il}y^{RL*}_{jk}\,.
\end{equation}
We read from Eq.~(\ref{match6LNC})
\begin{equation}
\bar{c}^{(6)}_{\mathrm SR} = 4 \bar{c}^{(6)}_{\mathrm T} = \frac{v^2}{2}C^{(6)}_{LdQ\nu}\,.
\end{equation}
Finally, going to the neutrino mass basis and focusing on the couplings to electrons and sterile neutrinos, we obtain the 
matching contributions to the effective operators in Eq.~\eqref{final67}
\begin{equation}
\left(C^{(6)}_{\rm VLL}\right)_{ije4}=-2 V_{ij} U_{e4}\,, \hspace{0.5cm}
\left(C^{(6)}_{\rm SRR}\right)_{ije4}=4\left(C^{(6)}_{\rm TRR}\right)_{ije4}= \left(\bar{c}^{(6)}_{\rm SR}\right)_{ije1} U^*_{44}\,,
\end{equation}
where $i$ and $j$ denote the up- and down-quark generation, respectively, and $V_{ij}$ elements of the CKM matrix. In this scenario, 
we have contributions from minimal mixing proportional to the mixing angle $U_{e4}$ and from leptoquark interactions proportional 
to $U^*_{44}$. We will use the canonical see-saw relations
\begin{equation}
U_{e4} \simeq \sqrt{\frac{m_{\nu}}{m_N}}\,,\qquad U_{44} =1\,,
\end{equation}
and set $m_{\nu} = 0.05$ eV as representative for the active neutrino masses. For a specific quark flavor choice of $i$ and $j$, this 
reduces the effective number of free parameters to two: the sterile neutrino mass $m_N$ and the combination of the LQ couplings and 
mass $y^{\overline{LR}}_{i1}y^{RL*}_{je}/m_{LQ}^2$. 

\item The final scenario we consider is inspired by models, such as left-right symmetric models, with right-handed charged gauge bosons, 
which can mix with $W^\pm$. Instead of implementing the full left-right symmetric model, we take a simplified scenario and consider the 
effects of a nonzero $C^{(6)}_{\rm VLR}$ in Eq.~\eqref{final67} in combination with the SM left-handed weak interactions. That is, we 
consider
\begin{equation}
\left(C^{(6)}_{\rm VLL}\right)_{ije4}=-2 V_{ij} U_{e4}\,,\qquad \left(C^{(6)}_{\rm VLR}\right)_{ije4} > 0\,.
\end{equation}
In left-right symmetric scenarios nonzero $C^{(6)}_{\rm VRR}$ and $C^{(6)}_{\rm VRL}$ are generally induced as well. The resulting 
phenomenology is very similar to $C^{(6)}_{\rm VLL}$ and $C^{(6)}_{\rm VLR}$ and for simplicity we do not consider these effective 
operators here. They can be easily added to the analysis if so required. For the active-sterile neutrino mixing parameter, we follow the 
leptoquark scenario and set $U_{e4} = \sqrt{m_{\nu}/m_N}$. In minimal left-right symmetric models with exact $P$ or $C$ symmetry, 
the dependence of the effective operators on the quark flavor indices $ij$ can be calculated. We do not consider this here and consider 
one particular choice of $ij$ at a time. It is straightforward to generalize this choice.

\end{enumerate}

\section{Collider and Fixed-target Analysis}\label{sec:exp}

We proceed to explore the search possibilities for the above scenarios at the LHC, considering both existing and proposed 
experiments, as well as the proposed fixed-target experiment \texttt{SHiP} at the CERN SPS 
\cite{Calviani:2063300, Anelli:2015pba,SHiP:2018yqc}. Currently at the LHC we have the operational experiments 
\texttt{ALICE} \cite{Aamodt:2008zz,Abelev:2014ffa}, \texttt{ATLAS} \cite{Aad:2008zzm}, \texttt{CMS} \cite{Chatrchyan:2008aa}, and \texttt{LHCb} \cite{Alves:2008zz,Aaij:2014jba}. Of these we shall focus on the search sensitivity for 
long-lived sterile neutrinos at \texttt{ATLAS}, which is the largest experiment in detector volume, and can thus 
in principle explore the largest decay lengths. Beyond this, a series of new experiments has recently been proposed 
at various locations near the LHC interaction points (IPs), namely in alphabetical order: \texttt{AL3X} \cite{Gligorov:2018vkc}, 
\texttt{ANUBIS} \cite{Bauer:2019vqk},  \texttt{CODEX-b} \cite{Gligorov:2017nwh}, \texttt{FASER} and \texttt{FASER2} 
\cite{Feng:2017uoz,Ariga:2018uku}, \texttt{MATHUSLA} \cite{Chou:2016lxi,Curtin:2018mvb,Alpigiani:2020tva}, and 
\texttt{MoEDAL-MAPP1} and \texttt{MoEDAL-MAPP2} of the \texttt{MoEDAL} collaboration 
\cite{Pinfold:2019nqj,Pinfold:2019zwp}. These latter experiments, including \texttt{SHiP}, are all designed to specifically 
look for neutral long-lived particles (LLPs). We shall discuss the search potential of all the above-mentioned experiments 
specifically for sterile neutrinos.

In this section, we review briefly the setup of the beam and detector geometries for each experiment and  introduce the Monte Carlo 
(MC) simulation procedure for the event simulation. We focus on the production via the rare decay of $B$- and 
$D$-mesons. This can proceed via either purely leptonic two-body decays, or semi-leptonic three-body decays, as discussed in Sec.~\ref{sec:prod-of-HNLs}.
Similarly, for the displaced decay of sterile neutrinos, we consider both the two-body decay into a charged lepton and a charged meson, and decays into multiple hadronic states plus a lepton, as explained in Sec.~\ref{sec:decays-of-HNLs}. 
It is worth mentioning that the heavy meson $M_1$ that is to decay into a sterile neutrino, is produced in the process $p p \rightarrow M_1 + X$ and decays promptly into \textit{e.g.} a sterile neutrino ($X$ denotes the remaining decay products).
Such signal events can be observed in the near detector such as \texttt{ATLAS} with \textit{e.g.} the prompt lepton accompanying the production of $M_1$.
The long-lived sterile neutrinos decay at a macroscopic distance, where the displaced vertex (DV) is reconstructed if at least two tracks are observed stemming from the same 
DV inside the detector.

We do not study the production of the sterile neutrinos from the decay of lighter mesons such as $\pi^\pm$ and kaons as these are only relevant for very light sterile neutrinos, and their simulation in \texttt{Pythia 8} \cite{Sjostrand:2006za,Sjostrand:2007gs} is insufficiently validated in the forward direction, which is relevant for the \texttt{FASER} experiments.
In fact, even for $D^\pm$ and $B^\pm$ mesons, we will use \texttt{FONLL} \cite{Cacciari:1998it,Cacciari:2001td,Cacciari:2012ny,Cacciari:2015fta} 
to correct the behavior of \texttt{Pythia 8} in the very large pseudorapidity regime. Finally, we ignore the vector mesons decays into sterile neutrinos, as their decay width is typically many orders of magnitude larger than that of pseudoscalar mesons leading to tiny branching 
ratios for sterile neutrino production. 

Instead of performing a detailed study considering different components of the detectors separately, for simplicity we will take the whole detector as the 
fiducial volume, and make a comparison between the various experiments. Since the \texttt{ATLAS} detector was not designed to look for neutral LLPs, 
we shall add an estimate for the efficiency factor based on existing neutral LLP searches, which, however, search for heavier candidates than considered here.

One potential issue relates to possible background events which, depending on the placement of the detector, may consist of 
long-lived SM hadron decays, cosmic rays, hadronic interaction with the detector material, etc. All the experiments considered 
in this study, except \texttt{ATLAS}, employ a far detector with a distance $5-500$ meters away from the IP. The space between 
the IP and the far detector is usually sufficiently large to allow for the installation of veto and shielding segments, as argued in 
Refs.~\cite{Chou:2016lxi,Curtin:2018mvb,Feng:2017uoz,Gligorov:2017nwh,Aielli:2019ivi,Gligorov:2018vkc,Alekhin:2015byh,Bauer:2019vqk,Pinfold:2019zwp}.
For \texttt{MATHUSLA} the rock and shielding below ground should remove the SM background. As for cosmic rays, directional 
cuts will be applied. To assess and compare the sensitivities of different experiments, we show 3-event isocurves which 
correspond to 95\% confidence level (C.L.) with zero background. This is not appropriate for the \texttt{ATLAS} detector which 
has an almost $4\pi$ coverage immediately around the IP, and a large irreducible SM background is expected. Depending on the 
signal type, the number of such background events may vary. Instead of performing an estimate of background events for each 
scenario, we shall implement an estimate for the efficiency, which we discuss below.

In addition, since the detector efficiency of the future experiments is unknown, in order to make a fair comparison, we assume a 100\% reconstruction 
efficiency for all the experiments except \texttt{ATLAS}.

The search potential of these experiments, but also of other fixed-target experiments, has been investigated for example for neutralinos 
\cite{Dreiner:2002xg,Dedes:2001zia,deVries:2015mfw,Dercks:2018eua,Dercks:2018wum}. For various other theoretical scenarios, see Ref.~\cite{Alimena:2019zri} 
for a recent review.

We start with a brief introduction of the fixed-target experiment \texttt{SHiP}, as its beam setup is different from the other experiments, and then discuss 
the other experiments, which are all associated to either the \texttt{ATLAS}, \texttt{CMS}, or \texttt{LHCb} IP and the LHC accelerator\footnote{In this work, we consider the LHC center-of-mass energy at 14 TeV for all experiments. We assume an LHC upgrade before 
the experiments are online. Changing it to 13 TeV would only have a small effect on our results.}. Since these experiments differ in the projected 
integrated luminosity, we summarize them in Table~\ref{table:lumi}. 

\begin{table}[t]
\begin{tabular}{ c | c  c  c  c  c}
	Experiment &\texttt{SHiP}  & \texttt{ATLAS}  &\texttt{AL3X} &\texttt{ANUBIS}& \texttt{CODEX-b} \\[2mm]
	Int. Lumi. & $2\times 10^{20}$ POT& 3000 fb$^{-1}$ &100 or 250 fb$^{-1}$& 3000 fb$^{-1}$ & 300 fb$^{-1}$ \\[2mm]
	Angular Cov. & 0.89\% & 100\% & 13.73\%  &1.79\%& 1\%\\
	\hline \hline \\ &&&&& \\[-6mm]
	Experiment &\texttt{FASER}&\texttt{FASER2}&\texttt{MAPP1}&\texttt{MAPP2} &\texttt{MATHUSLA}\\[2mm]
	Int. Lumi.  &150 fb$^{-1}$&3000 fb$^{-1}$&30 fb$^{-1}$&300 fb$^{-1}$&3000 fb$^{-1}$ \\[2mm]
	Angular Cov. & $1.1\times 10^{-8}$ &  $1.1\times 10^{-6}$ &0.17\%&0.68\%& 3.8\%
\end{tabular}
\caption{Summary of integrated luminosities for the various experiments. ``POT'' for \texttt{SHiP} stands for ``Protons 
on Target''. We also list the simple geometric coverage for each experiment.}\label{table:lumi}
\end{table}

\subsection{SHiP}

The \texttt{SHiP} facility was proposed to make use of the high-intensity CERN SPS beam of 400 GeV protons incident on a fixed target made of 
\textit{e.g.} a hybrid material composed of (solid) molybdenum alloy and pure tungsten \cite{Calviani:2063300, Anelli:2015pba,SHiP:2018yqc}. 
It has not been approved yet. With a center-of-mass energy of approximately 27\,GeV, large production rates of $D$- and $B$-mesons are expected.
The \texttt{SHiP} experiment is proposed to have a cylindrical detector downstream  at roughly 70\,m away 
from the IP. The experiment is specifically designed for detecting long-lived neutral particles, which are produced from \textit{e.g.} charm or bottom 
meson rare decays, fly an extended length, and then decay inside the detector chamber downstream, especially if the lab-frame decay length of the 
LLP lies within the \texttt{SHiP} sensitivity range. For the lifetime of the \texttt{SHiP} project, a total of $2\times 10^{20}$ protons on target (POT) are planned.

At \texttt{SHiP}, the initial meson $M_{1}$ of sterile neutrinos is produced in a hadronic collision between the beam protons and the target material. The 
differential production cross section is strongly forward peaked, and the $M_{1}$ will have a significant forward boost. This will be passed onto the 
decay products, including $N$, the sterile neutrino. The active decay chamber is 68.8\,m downstream of the target. It has a cyclindrical shape with a length of 60\,m, 
where the first 5\,m are to be used for placing background suppression vetoes. The front surface has an elliptical shape with semi-axes of 5\,m and 
2.5\,m. The optimal sensitivity is for particles with
\begin{equation}
68.8\,\mathrm{m} < \beta_N^z \gamma_N c \tau_N < 123.8\,\mathrm{m}\,,
\end{equation}
where $\beta_N^z,\,\gamma_N$ are the relativistic speed along the beam axis and the Lorentz boost factor of $N$, respectively.

In order to study sterile neutrinos produced from the decays of $D$- and $B$-mesons, we need to know the total number of these mesons expected at 
\texttt{SHiP} in its 5 year lifetime: $N_{D^\pm}$, $N_{D^0}$, $N_{D_s}$, $N_{B^\pm}$, $N_{B^0}$, and $N_{B_s}$, respectively. These 
numbers can be estimated by following Ref.~\cite{Bondarenko:2018ptm}. See also the earlier work in Ref.~\cite{deVries:2015mfw}.
The $c\bar{c}$ ($b\bar{b}$) production rate is $1.7\times 10^{-3}$   ($1.6\times 10^{-7}$) per collision. After the fragmentation 
factors are taken into account, the numbers of the heavy-flavor mesons can be estimated and 
reproduced below from Ref.~\cite{Bondarenko:2018ptm}:
\begin{eqnarray}
	N_{D^\pm}^{\texttt{SHiP}} = 1.4\times 10^{17},& \ \ N_{D^0}^{\texttt{SHiP}}= 4.3\times 10^{17}, & \ \ N_{D_s}^{\texttt{SHiP}}= 6.0\times 10^{16},\\[1mm]
	N_{B^\pm}^{\texttt{SHiP}}= 2.7\times 10^{13},& \ \ N_{B^0}^{\texttt{SHiP}} = 2.7\times 10^{13}, & \ \ N_{B_s}^{\texttt{SHiP}} = 7.2\times 10^{12}.
\end{eqnarray}
We note that $B_c$ mesons are in principle also produced and may extend the upper reach in $m_N$. However, given the much smaller production 
cross section, we do not take it into account. Similarly for the other LHC experiments, as discussed below.

\subsection{Experiments at the LHC}

For  \texttt{ATLAS} and  extended programs at the \texttt{ATLAS}/\texttt{CMS}/\texttt{LHCb} sites, we consider $pp$ collisions at 
$\sqrt{s}=14\,$TeV with equal beam energies. These experiments benefit from a beam energy that is orders 
of magnitude higher compared to \texttt{SHiP}. As a result, the mesons and the therefrom produced sterile neutrinos are much more boosted, leading to good sensitivities even for detectors
that are to be installed hundreds of meters away from the IP.

To perform the sensitivity estimate for experiments at \texttt{ATLAS}/\texttt{CMS}/\texttt{LHCb}, it is necessary to know the inclusive production 
rate of the heavy-flavor mesons at the HL-LHC with up to 3\,ab$^{-1}$ integrated luminosity. To achieve this, we follow the procedure  in 
Refs.~\cite{deVries:2015mfw,Dercks:2018eua, Dreiner:2020qbi}. The \texttt{LHCb} collaborations reported $D$-meson and $B$-meson 
production cross sections for a 13\,TeV $pp$-collider, for certain kinematic range. Using the simulation tools \texttt{FONLL} 
\cite{Cacciari:1998it,Cacciari:2001td,Cacciari:2012ny,Cacciari:2015fta} and \texttt{Pythia 8} we can extrapolate these cross sections to the whole 
kinematic range. We find that for 3\,ab$^{-1}$ integrated luminosity over the full $4\pi$ solid angle the following list of numbers of the produced 
charm and bottom mesons: 
\begin{eqnarray}
	N_{D^\pm}^{\text{HL-LHC}} = 2.04\times 10^{16}, & \ \ N_{D^0}^{\text{HL-LHC}} = 3.89\times 10^{16}, & \ \ N_{D_s}^{\text{HL-LHC}} = 6.62\times 10^{15},\\[1mm]
	N_{B^\pm}^{\text{HL-LHC}} = 1.46\times 10^{15}, & \ \ N_{B^0}^{\text{HL-LHC}} = 1.46\times 10^{15}, & \ \ N_{B_s}^{\text{HL-LHC}} = 2.53\times 10^{14}.
\end{eqnarray}
For the estimate of $N_{D^\pm}^{\text{HL-LHC}}$, the decay branching ratio of $D^{*\pm}$ into $D^\pm$ is included, 
and for the number of $B$-mesons the corresponding fragmentation factors determined by \texttt{Pythia 8} are used. 
These numbers will be used not only for the evaluation at the \texttt{ATLAS} experiment, but also for all the extended 
programs discussed below with a possible overall re-scaling by the integrated luminosity.

\subsubsection{FASER}

At the \texttt{ATLAS} IP, the differential cross section for the production of GeV-scale mesons is strongly peaked at large pseudorapidities, \textit{i.e.} close to the beam pipe in both directions.
The production rate of light neutral LLPs, resulting from the decay of the mesons, should then also be peaked in the large pseudorapidity regime.
A far detector known as \texttt{FASER} (Forward Search ExpeRiment) \cite{Feng:2017uoz,Ariga:2018uku} has been proposed, which is designed to specifically make use of this feature.
It has been officially approved by CERN and should be collecting data during Run~3 of the LHC.\footnote{See the official website of the experiment \texttt{FASER}: \url{https://faser.web.cern.ch/}.}
It is placed in the existing TI12 tunnel at a distance of 480 m from the \texttt{ATLAS} IP and has a cylindrical shape exactly aligned with the beam collision axis, but slightly off of the beam due to the curvature of the accelerator.
The \texttt{FASER} detector has a cylindrical radius of $10\,$cm and a length of $1.5\,$m, and the expected integrated luminosity is 150\,fb$^{-1}$.
The corresponding angular coverage is $\eta\in[9.17,+\infty]$ in pseudorapidity and full $2\pi$ in the azimuthal angle. 

After Run 3 is finished, \texttt{FASER} is currently planned to be upgraded to a larger version known as 
\texttt{FASER2}, to be under operation during the HL-LHC era, collecting up to 3\,ab$^{-1}$ of data \cite{Ariga:2018uku}, 
and located in the same place. Its geometry is specified as a cylinder with 1\,m radius and 5\,m length, which is 300x 
larger by volume, than \texttt{FASER}. The pseudorapidity coverage is correspondingly enlarged to $\eta\in[6.86,+\infty]$ while 
the azimuthal angle remains fully covered. In this work, we will study the search potential of both \texttt{FASER} and 
\texttt{FASER2} for sterile neutrinos as neutral LLPs.

As mentioned earlier and discussed in Ref.~\cite{Dercks:2018eua}, \texttt{Pythia 8} is not well validated in the large 
pseudorapidity regime for the differential production cross section of charm and bottom mesons. To solve this issue, 
we re-scale the meson production cross section in \texttt{Pythia 8} at different ranges of the transverse momentum and 
pseudorapidity by using the more reliable results given by \texttt{FONLL}.

\subsubsection{MATHUSLA}

A much larger experiment called ``\texttt{MATHUSLA}'' (MAssive Timing Hodoscope for Ultra Stable neutraL pArticles) 
has been suggested to be constructed at the surface above the \texttt{CMS} experiment 
\cite{Chou:2016lxi,Curtin:2018mvb,Alpigiani:2020tva}.\footnote{See also the webpage of the experiment: 
\url{https://mathusla-experiment.web.cern.ch/}.} It is proposed to be built for the HL-LHC phase with 3\,ab$^{-1}$ integrated 
luminosity. With a box shape, it has a base area of 100 m$\times$ 100 m and a height of 25 m. Relative to the \texttt{CMS} 
IP, the front edge of the detector should be horizontally shifted along the beam axis by 68 m, and vertically upwards by 60 m. 
Despite its huge size, the large distance of \texttt{MATHUSLA} from the IP still leads to a small geometric coverage.
It corresponds to a solid angle or geometric coverage of about 3.8\%.\footnote{The center of the \texttt{MATHUSLA} detector is at about 132 m from the IP. The area of the enclosing sphere is about $2.2 \cdot 10^{5}$ m$
^2$. The \texttt{MATHUSLA} detector has an base area of about $10^4$\,m$^2$ and is tilted roughly at 63$^\circ$ relative 
to the radial direction. $10^4\,$m$^2\,\cos(63^\circ)/(2.2\cdot 10^{5}\,\mathrm{m}^2)\approx2.1$\%. In a more precise Monte 
Carlo integration we found a coverage of 3.8\%.} Nevertheless, it has been shown to be one of the far detectors at the LHC 
that may have the strongest reach in the very small active-sterile neutrino mixing at $|U_{e 4}|^2 \sim 10^{-9}$, when 
only the weak interaction is considered \cite{Alpigiani:2020tva}. 

\subsubsection{ANUBIS}

It has recently been proposed \cite{Bauer:2019vqk} to construct a detector, named \texttt{ANUBIS} (AN Underground 
Belayed In-Shaft search experiment), in one of the service shafts just above the \texttt{ATLAS} or \texttt{CMS} IP. 
Consequently, a total of 3 ab$^{-1}$ integrated luminosity from the LHC is projected. Similarly to the detectors discussed 
above, it also has a cylindrical shape however with the axis oriented in the vertical direction. The cyclinder diameter is 
18\,m and the length is 56\,m. The axis of the cylinder runs from the middle point of the bottom: $(x_b, y_b,z_b)=(0,24
\,\mathrm{m}, 14\,\mathrm{m})$ to the top $(x_t, y_t,z_t)=(0,80\,\mathrm{m}, 14\,\mathrm{m})$, where $z$ is along the 
beam axis, $x$ is horizontally transverse, and $y$ is vertically transverse, and the IP is at $(x,y,z)=(0,0,0)$. 
Please refer to Refs.~\cite{Bauer:2019vqk,Hirsch:2020klk} for a sketch. MC integration finds the angular coverage of 
\texttt{ANUBIS} to be $1.79\%$ \cite{Dreiner:2020qbi}.

\subsubsection{CODEX-b}

An external detector extension has also been proposed at the IP8 of the \texttt{LHCb} experiment. \texttt{CODEX-b} 
(COmpact Detector for EXotics at LHCb) \cite{Gligorov:2017nwh} is designed as a cubic box of dimensions 
10\,m$\,\times\,$10\,m$\,\times\,$10\,m. Occupying an empty space with a distance $\sim$25\,m from the \texttt{LHCb} 
IP, it covers the pseudorapidity range of $\eta\in[0.2,0.6]$ with the azimuthal angle coverage of $\sim 6.4\%$. The total geometric coverage of the solid-angle is about 1\%.

\subsubsection{MoEDAL-MAPP}

\texttt{MAPP} (MoEDAL's Apparatus for Penetrating Particles) is proposed as one sub-detector of the \texttt{MoEDAL} 
experiment \cite{Pinfold:2019nqj,Pinfold:2019zwp} at the IP8 of \texttt{LHCb}, and designed for searching for neutral LLPs. 
Similar to \texttt{FASER}, the \texttt{MAPP} experiment will have a significant upgrade in a second version. \texttt{MAPP1} 
is a rather small detector of volume $\sim130$\,m$^3$, currently under deployment and expected to collect data during 
LHC Run 3 with up to 30~fb$^{-1}$ integrated luminosity. It is roughly 55\,m from the IP8 at a polar angle $5^\circ$ from the 
beam. During the HL-LHC era, the \texttt{MAPP2} program is planned to be under operation at IP8 until the end of Run~5, 
accumulating up to 300\,fb$^{-1}$ of data. It is designed to occupy almost the whole of the UGC8 gallery in the LHC tunnel 
complex, taking up a volume of about 430\,m$^{3}$. With a larger integrated luminosity and a bigger volume, \texttt{MAPP2} 
is predicted to have higher sensitivity. The two detectors cover about 0.17\% and 0.68\% of the total solid angle 
\cite{Dreiner:2020qbi}, respectively.

\subsubsection{AL3X}

\texttt{AL3X} (A Laboratory for Long-Lived eXotics) was proposed in Ref.~\cite{Gligorov:2018vkc} to be built at the LHC IP2 where the \texttt{ALICE} experiment sits. Placed at a horizontal distance along the beam line of 5.25 m from the IP, the 
detector has a cylindrical shape aligned with and surrounding the beam line, corresponding to a full azimuthal coverage. The 
proposed inner (outer) radius is 0.85\,m (5\,m) with the length 12\,m. This gives a pseudo-rapidity coverage of $\eta\in[0.9,3.7]$. 
The authors of Ref.~\cite{Gligorov:2018vkc} proposed two values of the integrated luminosity, 100~fb$^{-1}$ and 250~fb$^{-1}$, in 
order to accommodate practical concerns including the move of the IP, beam quality, and investigation of background events. 
Here we focus on the benchmark value of 250~fb$^{-1}$ integrated luminosity.

\subsubsection{ATLAS}

In the previous sections we have discussed proposed new experiments which are specifically designed to look for long-lived 
particles. They are typically placed some distance away from the various IPs at the LHC or, in the case of \texttt{SHiP},
designed as a fixed-target experiment with a long decay path. In \textit{e.g.} Ref.~\cite{deVries:2015mfw} some of us 
considered the search for light long-lived particles at \texttt{ATLAS}.\!\footnote{We focus here on the \texttt{ATLAS} experiment, 
which is by volume the larger experiment. In principle this study could be performed for \texttt{CMS}, as well. See 
Ref.~\cite{Ilten:2015hya} for a related study on dark photons at LHCb.}
\texttt{ATLAS} can study decays upto a length of about $\sqrt{11^2+(43/2)^2}=24$\,m, where 11\,m is the cylindrical radius and 24\,m is half the detector length.
Over this length, as we discuss below in more detail, the fraction $1-\exp[-24\,\mathrm{m}/(\beta\gamma c\tau)]$ of the LLP's decay where $\tau$ 
and $\gamma$ denote the LLP lifetime and boost factor, respectively, and $\beta$ is the relativistic speed of the particle. However, 
\texttt{ATLAS} offers almost 4$\pi$ in angular coverage, which is significantly larger than the other detectors at the LHC. In 
Ref.~\cite{deVries:2015mfw}, we found that for 100\% signal efficiency, an integrated luminosity of 250\,fb$^{-1}$, and assuming 
zero background, \texttt{ATLAS} is competitive with or slightly better than \texttt{SHiP} for the LLPs produced from $B$-mesons 
decays, and was somewhat worse in the case of $D$-mesons.

Here we describe in more detail the geometrical parameters we use for the fiducial volume of the \texttt{ATLAS} detector.
It has a cylindrical shape with inner (outer) radius of 0.0505 m (11 m) corresponding to the beginning of the inner detector (the 
end of the muon spectrometer), and a length of 43 m. In principle, even if a DV is inside the beam pipe, 
as long as its distance from the IP is larger than the detector spatial resolution and consists of displaced tracks, it can be 
reconstructed. However, to be more conservative, we choose to include only DVs that are located inside the detector volume. We 
take 3\,ab$^{-1}$ as the benchmark value for the integrated luminosity over the lifetime fo the HL-LHC.

As we have mentioned, all the above proposed new experiments are specifically designed to look for light long-lived particles.
They are thus further away from the LHC IPs and shielded from many SM backgrounds generated at the IP. All the same they 
will all have separate background issues, depending on where they are located. \texttt{MATHUSLA} is to be installed on the 
ground and thus highly susceptible to cosmic rays. \texttt{ANUBIS} is in a shaft which has minimal overhead shielding. 
\texttt{FASER} is far down along the tunnel, but still close to the beam pipe. In order to compare these experiments we have for 
now assumed that they can tackle the issues concerning background, and assumed zero background for all. The precise design 
of these detectors is also not yet well-established, except for the first phase of \texttt{FASER}. We thus furthermore assume 
100\% signal efficiency.

All this does not hold for \texttt{ATLAS} (or \texttt{CMS}, of course).  \texttt{ATLAS} is a well-established experiment, operating 
in the immediate vicinity of the IP. There is purposely no shielding, as \textit{a priori} all events are of interest. With respect to 
light long-lived particles there are thus large backgrounds which must be dealt with.  Any cuts which are imposed to reduce the background, will also affect the signal efficiency, most likely in a significant manner. In order to compare a search
at \texttt{ATLAS} with the other experiments we must impose a realistic signal efficiency, to take this into account. To our 
knowledge at the moment, there is no dedicated study at \texttt{ATLAS} or \texttt{CMS} for the scenarios we are considering here. 
We discuss several related analyses and estimate a signal efficiency based on this.

In Ref.~\cite{Aad:2019tua}, \texttt{ATLAS} considered the following scenario proposed in Ref.~\cite{Falkowski:2010cm}. A 
Higgs boson decays to two dark fermions, which in turn decay to one or two dark photons plus an invisible hidden lightest stable 
particle. The dark photons are the long-lived particles and decay either to a pair of muons, or a pair of electrons/pions (light hadrons).
Dark photon masses between 0.4 and 3.5\,GeV are considered, \textit{i.e.} similar to our sterile neutrino mass range. However the Higgs 
boson masses are 125 or 800\,GeV and thus the dark photons would have a much higher boost than our sterile neutrino's. For the lighter 
Higgs mass and the purely hadronic decay of the dark photon \texttt{ATLAS} finds a signal efficiency of at best 5$\cdot10^{-5}$ 
for $c\tau\approx35\,$mm, and dropping off rapidly for smaller or larger $c\tau$. 

In Ref.~\cite{Aad:2019xav}, \texttt{ATLAS} considered the following scenario. A Higgs boson decays to a pair of neutral long-lived 
scalars, which in turn each decay to 2 jets, one in the inner detector and one in the muon spectrometer, thus utilizing different 
parts of the \texttt{ATLAS} detector. For the lightest mass scenario the Higgs boson is 125\,GeV and the scalar is 8\,GeV.
A detector efficiency of 3$\cdot10^{-5}$ is found, similar to the other analysis. In Ref.~\cite{Aaboud:2019opc} \texttt{ATLAS} 
considered a related scenario with scalar masses down to 5\,GeV. For a Higgs boson of 125\,GeV, and a scalar of 5\,GeV with a 
decay length of $75\,$cm they find a signal efficiency of 7$\cdot10^{-4}$, somewhat better than in the other studies.

In all these analyses, \texttt{ATLAS} searches for pairs of produced particles.
This reduces background but also efficiency.
Overall we have applied a from the \texttt{ATLAS} point of view somewhat optimistic flat signal efficiency factor of $10^{-3}$ to all the \texttt{ATLAS} searches.
In addition we assume that the corresponding cuts reduces the background to zero.

\subsection{Monte-Carlo Simulation}

We perform the MC simulation with the tool \texttt{Pythia~8.243} \cite{Sjostrand:2006za,Sjostrand:2007gs}, in order to extract the kinematics and to estimate the number
of signal  events. We express the number of sterile neutrinos, $N$, produced as
\begin{eqnarray}
	N_{N}^{\text{prod}} =\sum_i N_{M_{i}} \cdot \text{Br}(M_{i}\rightarrow N + X), 
\end{eqnarray}
where $M_{i}$ is summed over all mesons that can decay to $N$ in a given scenario. ``Br'' stands for decay branching ratio. $N_{M_i}$ is the number of initial mesons, $M_i$, produced in the initial collisions at the LHC or for \texttt{SHiP}.

The number of sterile neutrino decays in a detector volume $N^{\text{dec}}_N$ can then be estimated by taking into account the 
boost factor and traveling direction of $N$, geometries of the detector, and the decay branching ratio of sterile neutrinos into 
the signal final-state particles. For the latter, we consider all the decay channels except for the fully invisible state, 
which contains solely three neutrinos, and is mediated by the SM weak interaction.
We use the expressions
\begin{eqnarray}
N^{\text{dec}}_N &=& N_{N}^{\text{prod}}\cdot \langle P[N \text{ in f.v.}] \rangle\cdot \text{Br}(N\rightarrow \text{signal}),\label{eqn:decaynumber}\\
\langle P[N \text{ in f.v.}] \rangle &\equiv& \frac{1}{N^{\text{MC}}_N}\sum_{i=1}^{N^{\text{MC}}_N}P[N_i \text{ in f.v.}],
\end{eqnarray}
where $\langle P[N \text{ in f.v.}] \rangle$ denotes the average probability of all the simulated sterile neutrinos to decay inside 
the fiducial volume (``f.v.'') and $P[N_i \text{ in f.v.}]$ is the individual probability of the $i-$th simulated sterile neutrino to decay 
in the f.v., discussed in more detail below. $N^{\text{MC}}_N$ is the total number of sterile neutrinos $N$ generated in the simulation. For all 
the experiments except \texttt{MAPP1} and \texttt{MAPP2}, we use formulas for calculating the individual decay probability 
with the exponential decay distribution, extracted from existing references 
\cite{deVries:2015mfw, Dercks:2018eua, Dercks:2018wum, Hirsch:2020klk}. As input we use the boosted decay length and 
the traveling direction of each simulated sterile neutrino, denoted by $\lambda_i=\beta_i \, \gamma_i  \, c \, \tau_N$, where $\beta_i$ is the speed and $\gamma_i$ the boost factor.

For the \texttt{MoEDAL-MAPP} detectors, following Ref.~\cite{Dreiner:2020qbi} we implement a code which determines 
whether each simulated sterile neutrino travels in the direction pointing towards the detector and if so returns $L_{1i}$ and $L_{2i}$, 
where $L_{1i}$ denotes the distance from the IP to the position where the $i-$th sterile neutrino would enter the detector, and $L_{i2}$ 
the distance the $i-$th sterile neutrino would travel across the detector, if it leaves the detector without having decayed. If the travel 
direction of the long-lived sterile neutrino points towards the detector, we compute $P[N_i \text{ in f.v.}]$ for the \texttt{MoEDAL-MAPP} 
detectors through
\begin{eqnarray}
	P[N_i \text{ in f.v.}] &=& e^{-L_{1i}/\lambda_i}\cdot (1-e^{-L_{2i}/\lambda_i}).
\end{eqnarray}

We use the modules ``\texttt{HardQCD:hardccbar}'' and ``\texttt{HardQCD:hardbbbar}'' of \texttt{Pythia~8} to simulate the 
production of the charm and bottom mesons, respectively, including the processes $q\bar{q}, gg\rightarrow c\bar{c}/b\bar{b}$. For each benchmark scenario, we simulate 10$^6$ events of the corresponding process, and fix the initial-state mesons to decay to the various channels, mediated by both the SM weak interaction and the EFT operators, with 
the corresponding decay branching ratios. From the MC simulation with \texttt{Pythia 8}, we obtain the average decay 
probability from which we calculate $N^{\text{dec}}_N$ in Eq.~\eqref{eqn:decaynumber}.

We emphasize that for the partial decay widths of the heavy mesons into $N$ and of the $N$ into light mesons, we take into account the weak interaction via active-sterile neutrino mixing, the EFT operator(s), and also the interference between them. The computation for the production and decay processes of sterile neutrinos are presented in Secs.~\ref{sec:prod-of-HNLs} and \ref{sec:decays-of-HNLs} and in the appendices.

\section{Numerical Results}\label{sec:flavorbenchmarks} 

To present the results of this collider study in a compact and representative manner we consider a subset of possible EFT operators and focus on the three scenarios described in Sec.~\ref{sec:scenarios}.
For scenario 1, the minimal scenario, there is no EFT operator involved, and we consider the full standard model charged- and neutral-currents mediated via the active-sterile neutrino mixing.
For scenarios $2$ and $3$, which involve EFT operators, we must specify their flavor.
In these scenarios we consider 5 different flavor assignments.
Each assignment involves two EFT operators of different quark flavors. All EFT operators are dimension-6, and we drop the $(6)$ superscript in the following.
First, we fix the production mode via the ``production operator" $\left(C_{\mathrm {P}}\right)_{ij}$, with associated Wilson coefficients.
We shall consider two cases for the indices $ij$, which indicate the up- and down-type quark generations considered in a flavor benchmark, respectively.
The choices of the flavors should lead to charm and bottom meson decays:
\begin{eqnarray}
	\mathrm{Sterile \;Neutrino \; Production \;Modes:}\quad \left\{ \begin{array}{cc} (C_{\mathrm{P}})_{21}: & D\to N + e \,(+X),  \\[1mm]
		(C_{\mathrm{P}})_{13}: & B\to N + e \,(+X).
\end{array}
\right. 
\label{eq:CP}
\end{eqnarray}
Here $X$ indicates a potential final state meson. See examples below.
Second, we simultaneously turn on another EFT operator\footnote{It is possible to have only one non-vanishing operator \textit{e.g.} $C_{\text{SRR}}^{11} = 4 C_{\text{TRR}}^{11}$ or $C_{\text{VRR}}^{11}$, leading to $\rho^\pm$ decaying to $N+e^\pm$ and $N$ decaying to $\pi^\pm+e^\mp$.
However, such scenarios probe only a small sterile neutrino mass range and in addition require the decaying meson to be a vector particle. This results in a too small sensitivity reach because of the small production rate and large 
decay width of the vector mesons.} the ``decay operator" $\left(C_{\mathrm {D}}\right)_{ij}$, which leads to the decay of sterile neutrinos via semi-leptonic processes.
Here, the $ij$ indicate the flavor content of the final state meson.
We shall consider several decay modes
\begin{eqnarray}
\mathrm{Sterile \;Neutrino \; Decay \; Modes:}\quad \left\{ \begin{array}{cl} (C_{\mathrm{D}})_{11}: & N\to \pi^\pm+ e^\mp,\,\rho^\pm+e^\mp\,, \\[1mm]
(C_{\mathrm{D}})_{12}: & N\to K^\pm+ e^\mp,\,K^{*\pm}+e^\mp\,,  \\[1mm]
(C_{\mathrm{D}})_{21}: & N\to D^\pm+ e^\mp,\,D^{*\pm}+e^\mp \,. 
\end{array}
\right. 
\label{eq:CD}
\end{eqnarray}
Both the production and decay will be induced by various operators with associated Wilson coefficients, which we discuss in detail below.
For all scenarios we include the production and decay of sterile neutrinos via the SM weak interaction (see Sect.~\ref{sec:scenarios} for details) and the corresponding interference terms with the EFT operators. For the theoretical scenarios 2 and 3 we impose the type-I Seesaw relation to include weak interaction contributions through minimal mixing, see Sect.~\ref{sec:scenarios}. 

As indicated in Eq.~(\ref{eq:CP}), we only include sterile neutrino production via $B$ and $D$ decays,  also for lighter sterile neutrinos with masses $m_N < m_K,\,m_\pi$, the kaon and pion masses, respectively.
We do not include possible production via kaon or pion decays for the following reasons.
Despite the larger production rates of these mesons, the simulation of soft pions is not well validated in quick MC simulation tools.
Furthermore, the kaons are long-lived, leading to further complications.
Finally, sterile neutrinos produced from pions and kaons are necessarily light, resulting in limited sensitivity reach in $m_N$.
To summarize, while we show results for light sterile neutrinos in the following, these must be taken as conservative as we underestimate the number of the produced sterile neutrinos from pion and kaon decays both via the SM weak interaction as well as via the ``decay operator" $C_{\mathrm{D}}$.

\subsection{The Minimal Scenario}

\begin{figure}[t!]
	\centering
	\includegraphics[width=0.6\linewidth]{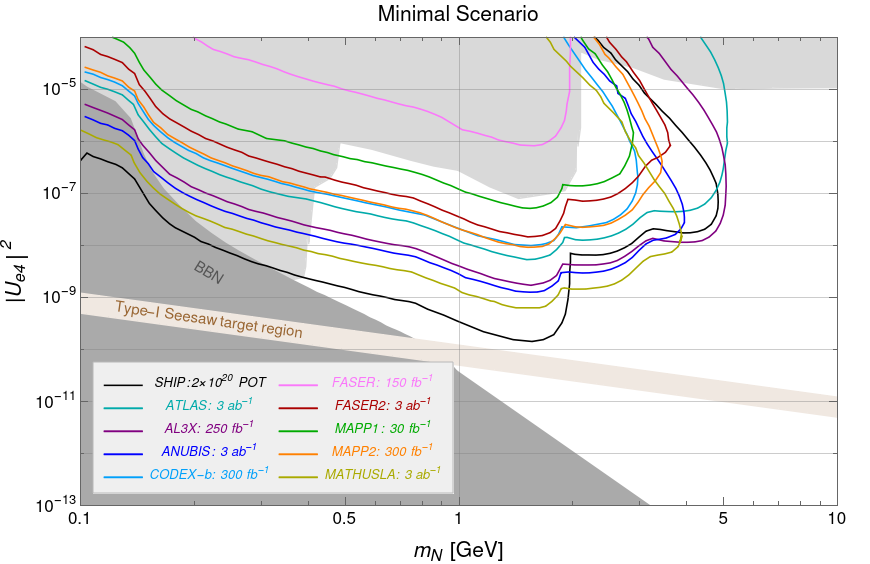}
	\caption{Results for the minimal scenario with the sterile neutrino mixed solely with the electron neutrino.
	}
	\label{fig:min_scen}
\end{figure}

In the minimal scenario, the interactions are purely mediated by the $W$- and $Z$-bosons via the active-sterile neutrino mixing.
Using the analytical expressions given in the previous sections for the production and decay of the sterile neutrino, we estimate the sensitivity reach of the experiments discussed in Sec.~\ref{sec:exp}.
We present the results in Fig.~\ref{fig:min_scen}, shown in the plane $|U_{e4}|^2$ vs. 
$m_N$. Here, we lift the requirement of the type-I seesaw relation $|U_{e 4}|^2 \simeq  m_{\nu_e}/m_N$, and treat the mixing angle and the sterile neutrino mass as two independent free parameters.
The light gray area shows the present bounds obtained by various experiments including the searches from \texttt{CHARM} \cite{Bergsma:1985is}, \texttt{PS191} \cite{Bernardi:1987ek}, \texttt{JINR} \cite{Baranov:1992vq}, and \texttt{DELPHI} \cite{Abreu:1996pa}.
The dark gray area corresponds to the part excluded by big bang nucleosynthesis (BBN) \cite{Sabti:2020yrt,Boyarsky:2020dzc}.
We also show a brown band of ``Type-I Seesaw target region'' for $m_{\nu_e}$ between 0.05 eV and 0.12 eV with the relation $|U_{e4}|^2 \simeq  m_{\nu_e}/m_N$.
These two limits are derived from neutrino oscillation and cosmological observations, respectively.
The former finds that there is at least one active neutrino mass eigenstate of mass at least 0.05 eV \cite{Canetti:2010aw} while the latter imposed an upper limit of 0.12 eV for the sum of the active neutrino masses \cite{Aghanim:2018eyx}.


The sensitivity reaches of the various experiments  have been determined in the literature 
\cite{Hirsch:2020klk,Dercks:2018wum, Helo:2018qej, Ariga:2018uku, Curtin:2018mvb, SHiP:2018xqw}, except for the \texttt{MAPP1} and 
\texttt{MAPP2} experiments, for which we present the estimate for the sensitivity reach for the first time. Our estimate for the \texttt{ATLAS} 
experiment considers an integrated luminosity of 3\,ab$^{-1}$ and takes 3000 signal events \textit{before} taking into account an universal 
efficiency factor $10^{-3}$ as the $95\%$ C.L. exclusion limit. To the best of our knowledge, a similar estimate for sterile neutrinos in the minimal 
scenario produced from heavy-flavor mesons rare decays has not been conducted for \texttt{ATLAS}. See Ref.~\cite{Aad:2020fzq} for a 
related study within the minimal model, with the sterile neutrino produced from $W$-decays, however with promptly decaying sterile neutrinos. Furthermore, in 
Ref.~\cite{Aad:2019kiz} \texttt{ATLAS} investigated a sterile neutrino mixing with $\nu_\mu$ and with a delayed decay, \textit{i.e.} a displaced vertex, 
as well as a promptly decaying sterile neutrino, which mixes with $\nu_e$ however only for $m_N\gsim6\,$GeV. For the other experiments, we assume 
zero background and 100\% detector efficiency. Hence we take 3 signal events as the 95\% C.L. exclusion limit. Comparing the sensitivity reach 
of the experiments shown in Fig.~\ref{fig:min_scen} with those from the literature, we find a good agreement in most cases. For the 
\texttt{ANUBIS} exclusion limits, our results shown in Fig.~\ref{fig:min_scen} are inferior by a factor $\sim 3.5$, to those given in 
Ref.~\cite{Hirsch:2020klk}. This difference is due to a corrected meson-production-rate and sterile-neutrino-decay-width calculation.

Given the general agreement with the existing results in the literature, we proceed to evaluate the sensitivities of these experiments to a set of benchmark scenarios, where the sterile neutrino interactions with the SM particles are enhanced by heavy new physics, encoded by EFT operators. 
There are a large number of possibilities for the flavor structure of the EFT operators. Here we consider a few representative flavor choices, to get an understanding of the general features and the sensitivity reach of various experiments. The calculations can easily be repeated for other flavor choices, for instance those inspired from specific UV-complete scenarios.
	
We note that in the following EFT flavor benchmarks, with the choice of the canonical type-I Seesaw relation, $m_N\lesssim 1$ GeV appears to be disfavored by BBN considerations.
However, the inclusion of the EFT operator $C_\text{D}$ can reduce the lifetime of the sterile neutrinos, possibly circumventing BBN constraint and leading to a potential lower bound on the EFT Wilson coefficients.
Different flavor assignments result in different final-state particles, affecting the primordial helium and deuterium abundances to different extents. A detailed study is necessary to investigate the limits that can be set from BBN consideration on EFT operators and we do not present BBN exclusion bands below.

\subsection{Flavor Benchmark 1}

\begin{table}[t]
	\begin{center}
		\begin{tabular}{r||c|c}
			& Flavor benchmark 1.2 & Flavor benchmark 1.3 \\
			\hline
			production operator: $C_\text{P}$ & $C_{\text{SRR}}^{21}=4C_{\text{TRR}}^{21}$ & $C_{\text{VLR}}^{21}$ \\[1.7mm]
			decay operator: $C_\text{D}$ & $C_{\text{SRR}}^{11}=4C_{\text{TRR}}^{11}$ & $C_{\text{VLR}}^{11}$ \\[1.7mm]
			production process via $C_\text{P}$	& \multicolumn{2}{c}{$D^\pm/D^0/D_s \rightarrow N + e^\pm (+ X) $} \\[-3.2mm] &\multicolumn{2}{c}{{\color{white}.}} \\
			decay process via $C_\text{D}$   	& \multicolumn{2}{c}{$N \rightarrow \pi^\pm + e^\mp, \; \rho^\pm + e^\mp$}  
		\end{tabular}
	\end{center}
	\caption{Summary of flavor benchmark 1 for the theoretical scenarios 2 and 3 of Sec.~\ref{sec:scenarios}, respectively. $X$ denotes 
		any additional final state particles.}
	\label{tab:scenario1}
\end{table}

We consider the theoretical scenarios 2 and 3 of Sec.~\ref{sec:scenarios} for a specific flavor choice.
We focus on sterile neutrino production through decay of $D$-mesons, and thus consider the production operator $\left(C_{\mathrm {P}}\right)_{21}$. 
For scenario 2, the leptoquark scenario, for $C_{\text{P}}$ and $C_{\text{D}}$  we consider scalar and tensor operators that are related through 
$C_{\rm SRR}=4 C_{\rm TRR}$. For scenario 3 corresponding to anomalous vector interactions, the production and decay operators are 
both $C_{\rm VLR}$.  The following decays then produce sterile neutrinos
\begin{eqnarray}
	D^\pm \rightarrow e^\pm + N, & \ \ D^\pm \rightarrow \pi^0 + e^\pm + N, & \ \ D^\pm \rightarrow \rho^0 + e^\pm + N,\\[1mm]
	D^0 \rightarrow \pi^\pm + e^\mp + N, & \ \ D^0 \rightarrow \rho^\pm + e^\mp + N, & \ \ D_s \rightarrow K^{(*)0} + e^\pm + N\,.
\end{eqnarray}
We are sensitive to sterile neutrino masses $m_N < m_D - m_e$. For the decay operator we choose 
$\left(C_{\mathrm {D}}\right)_{11}$ resulting in the decays
\begin{eqnarray}
	N \rightarrow e^\pm+\pi^\mp\,,\quad  \text{and} \quad N \rightarrow e^\pm+\rho^\mp\,.
\end{eqnarray}
The essential features of this benchmark are summarized in Table~\ref{tab:scenario1}.
In addition, we include production and decay modes via minimal mixing which extends both the upper and lower reach in $m_N$.

\begin{figure}[th]
	\centering
	\includegraphics[width=0.49\linewidth]{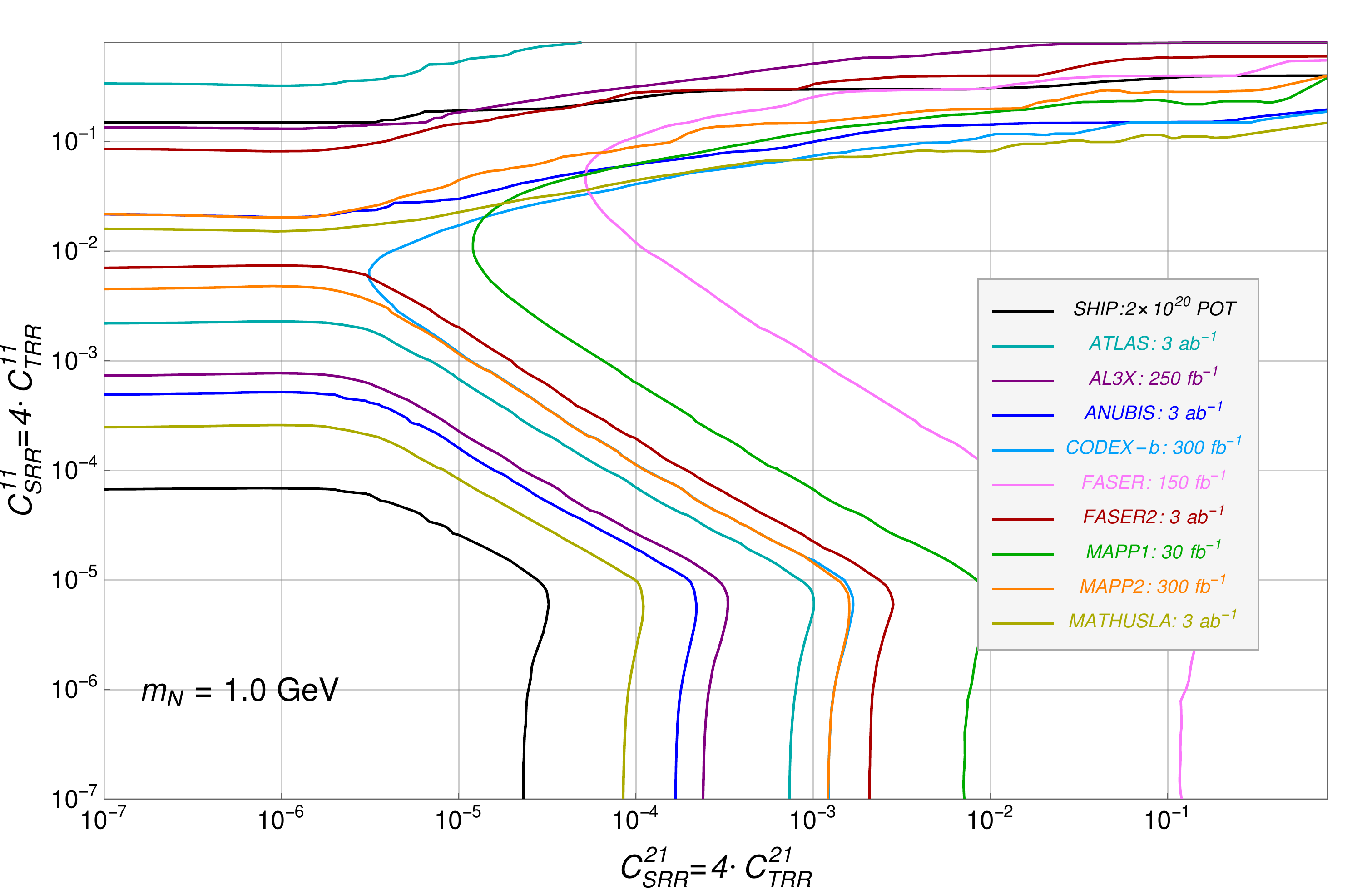}
	\includegraphics[width=0.49\linewidth]{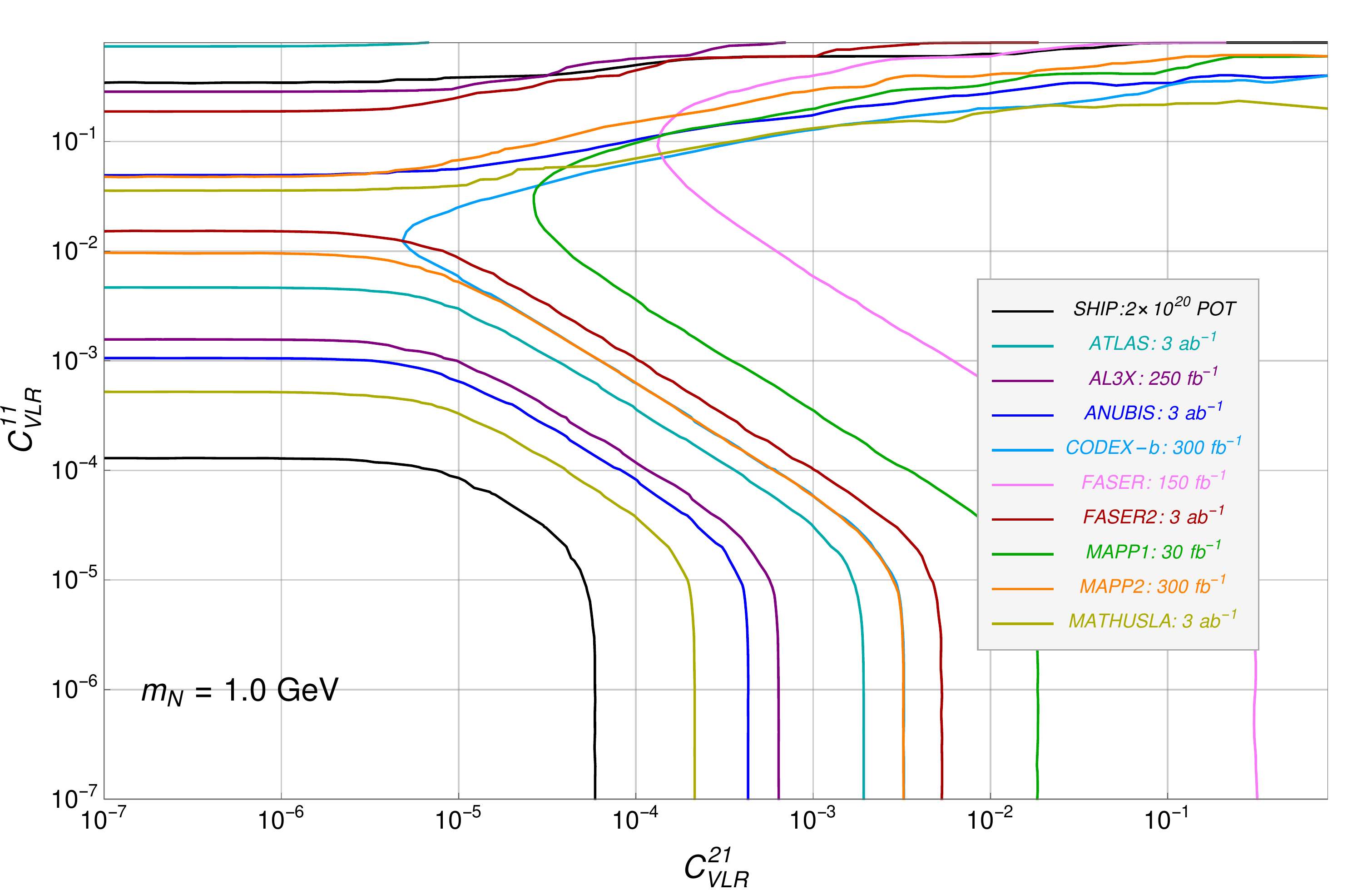}\\
	\includegraphics[width=0.49\linewidth]{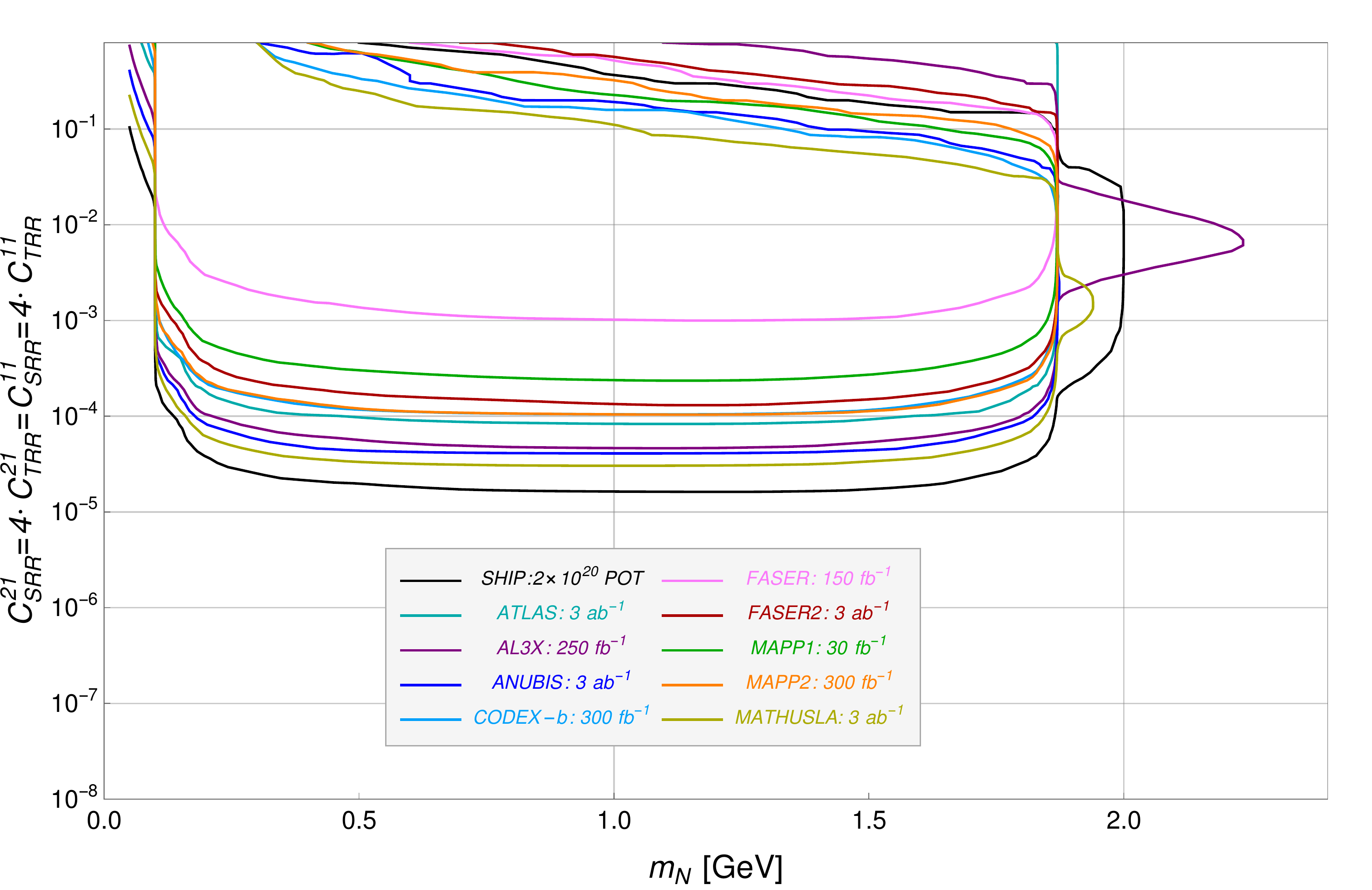}
	\includegraphics[width=0.49\linewidth]{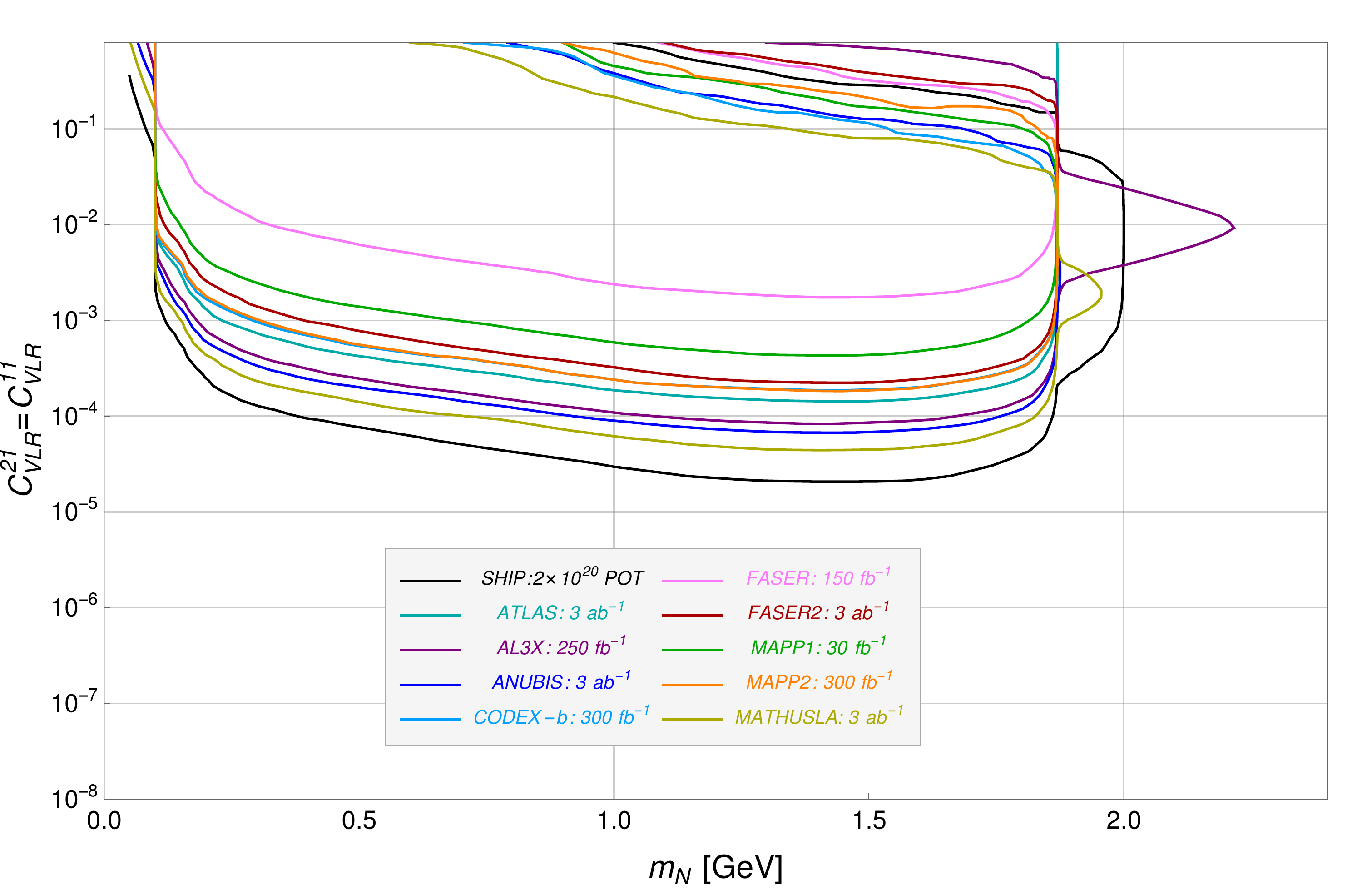}
	\caption{Results for the sensitivity reach for flavor benchmark 1. On the left we have the leptoquark-like case, and on the right the VLR case.
		For each case the upper figure shows $C_{\mathrm{D}}$ vs. $C_{\mathrm{P}}$, where as the lower case is for $C_{\mathrm{P}}=C_
		{\mathrm{D}}$ and shows $C_{\mathrm{P}}$ vs. $m_N$. The color code for the various experiments is shown in each figure.
	}
	\label{fig:scen1}
\end{figure}

Fig.~\ref{fig:scen1} presents the sensitivity reach of all considered experiments for the flavor benchmark~1.
The left panels correspond to theory scenario 2 (leptoquark) and the right panels to scenario 3 (anomalous vector interactions).
In the upper row we plot the value of the decay operator $C_\text{D}$ vs. the production operator $C_{\text{P}}$ for a fixed sterile neutrino mass $m_N =1$ GeV.
In the bottom row we have fixed $C_{\text{D}}=C_{\text{P}}$ and show the dependence of the sensitivity of the experiments on $C_D=C_P$ and the sterile neutrino mass. 
Both the top and bottom panels show that the sensitivity reach in scenarios 2 and 3 are rather similar, indicating that the specific Lorentz structure of the EFT interactions does not greatly affect the overall sensitivity.
In the upper-left and lower-right part of the top plots the curves become horizontal and vertical, respectively.
In this part of the parameter space either the production (horizontal, upper left) or decay (vertical, lower right) of sterile neutrinos through EFT operators becomes sub-leading with respect to the contributions from minimal mixing.
This roughly happens for couplings $C_{\text{P,D}} \leq  10^{-5}$, indicating that EFT operators can dominate over minimal interactions for a new physics scale of $\Lambda\sim\sqrt{v^2/C_{\mathrm{P,D}}} = \mathcal O(80)\,$TeV.
This scale does not include possible small dimensionless couplings or loop suppressions of the EFT operators and is thus only indicative of the sensitivity range.

For some experiments (\texttt{CODEX-b}, \texttt{MAPP1}, and \texttt{FASER}), there is no sensitivity in the upper left corner (small $C_\text{P}$ and large $C_\text{D}$).
This is caused by the rather weak detector acceptance of these experiments for the light sterile neutrinos produced from $D$-mesons decays.

In the lower set of plots, we assume equal $C_\text{P}=C_\text{D}$, and vary the sterile neutrino mass $m_N$ and jointly the Wilson coefficients. 
The plots for scenario 2 and 3 look rather similar although in scenario 2, the sensitivity to smaller sterile neutrino masses is a bit better.
The most sensitive experiment would clearly be \texttt{SHiP}, which reaches roughly $C_\text{P}=C_\text{D}\sim2 \cdot 10^{-5}$ in the range $0.5\,\mathrm{GeV} 
< m_N < 1.8\,$GeV.
For couplings at this level, both the production and decay of sterile neutrinos are still dominated by the EFT operators and minimal mixing plays a sub-leading role.
The sensitivity then depends a lot on the experimental setup under consideration. \texttt{FASER} reaches couplings at the $10^{-3}$ level (corresponding to scales of roughly $8\,$TeV in $\Lambda$).
Next in sensitivity is \texttt{MAPP1} at $3\cdot 10^{-4}$, and then \texttt{FASER2}, \texttt{MAPP2}, \texttt{CODEX-b}, and \texttt{ATLAS} at roughly the $10^{-4}$ level.
\texttt{MATHUSLA}, \texttt{ANUBIS}, and \texttt{AL3X}, should be sensitive to couplings down to around $5\cdot 10^{-5}$, and finally \texttt{SHiP} at the aforementioned $C_{\text{P,D}} \leq 2\cdot 10^{-5}$ level.
The hierarchy in sensitivity reach shown by the various experiments is essentially the same in scenarios 2 and 3, and is very similar to the hierarchy in the minimal scenario (see Fig.~\ref{fig:min_scen}) for masses $m_N < m_D$. 

Again, the sensitivity reach in $m_N$ goes beyond the kinematical thresholds set by the pion and $D$-meson masses.
For $m_N < m_\pi$, sterile neutrinos can still decay leptonically via the weak interaction.
Thus larger $C_\text{P}$ values can still lead to detectable rates of sterile neutrino production.
We stress again that for $m_N < m_\pi$, we underestimate the production of sterile neutrinos by omitting production via pions and kaons. 
For $m _N > m_D$, sterile neutrinos  for this benchmark can still be produced from the $B$-meson decays via the weak current.
If $C_\text{P}$ and $C_\text{D}$ are large enough, sufficiently many sterile neutrinos are produced.
Furthermore, specifically for \texttt{AL3X}, and \texttt{SHiP}, and to a lesser extent \texttt{MATHUSLA}, the boosted decay lengths of these sterile neutrinos can fall into the respective geometric sensitivity ranges.
This corresponds to the extended sensitive parameter regions, as shown on the right-hand side of the two lower plots of Fig.~\ref{fig:scen1}.

\begin{table}[t]
	\begin{center}
		\begin{tabular}{r||c|c}
			& Flavor benchmark 2.2 & Flavor benchmark 2.3 \\
			\hline
			production operator: $C_\text{P}$ & $C_{\text{SRR}}^{21}=4C_{\text{TRR}}^{21}$ & $C_{\text{VLR}}^{21}$\\[1.7mm]
			decay operator: $C_\text{D}$ & $C_{\text{SRR}}^{12}=4C_{\text{TRR}}^{12}$ & $C_{\text{VLR}}^{12}$\\[1.7mm]
			production process via $C_\text{P}$	& \multicolumn{2}{c}{$D^\pm/D^0/D_s \rightarrow N + e^\pm(+X)$} \\[-3.2mm] &\multicolumn{2}{c}{{\color{white}.}} \\
			decay process via $C_\text{D}$	& \multicolumn{2}{c}{$N\rightarrow K^\pm+ e^\mp,\;  K^{*\pm} +e^\mp$ } 
		\end{tabular}
	\end{center}
	\caption{Summary of flavor benchmark 2.}
	\label{tab:scenario2}
\end{table}

\subsection{Flavor Benchmark 2}

In flavor benchmark 2 we choose a different flavor-structure for the decay Wilson coefficient.
For the production operator we take again $\left(C_{\mathrm {P}}\right)_{21}$ but for the decay now set $\left(C_{\mathrm {D}}\right)_{12}$.
This leads to sterile neutrino decay processes
\begin{eqnarray}
N \rightarrow e^\pm+K^\mp\,,\quad \mathrm{and}\quad N \rightarrow e^\pm+K^{*\,\mp}\,.
\end{eqnarray}
Table~\ref{tab:scenario2} summarizes the details of this scenario.

\begin{figure}[H]
	\centering
	\includegraphics[width=0.49\linewidth]{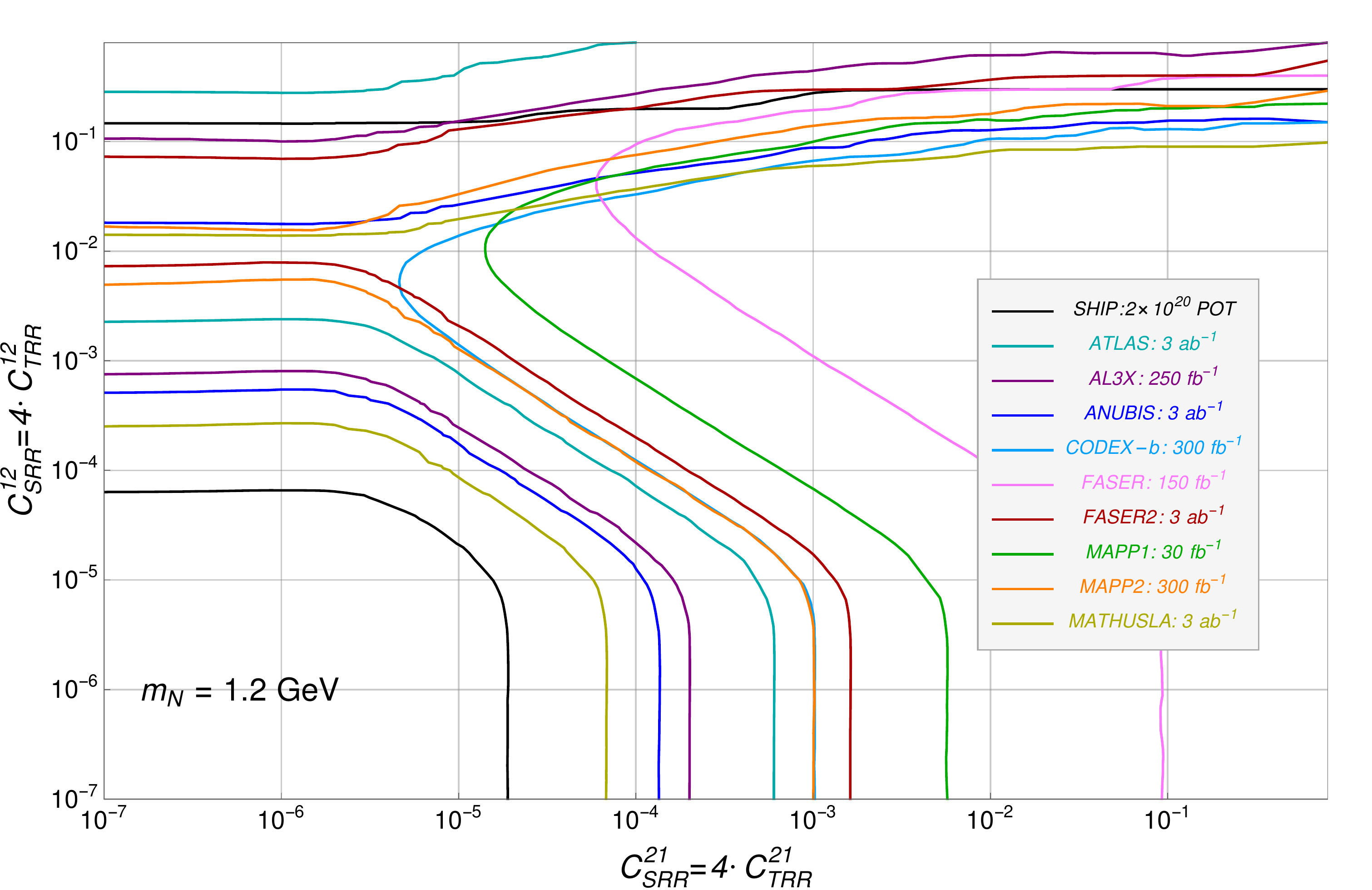}
	\includegraphics[width=0.49\linewidth]{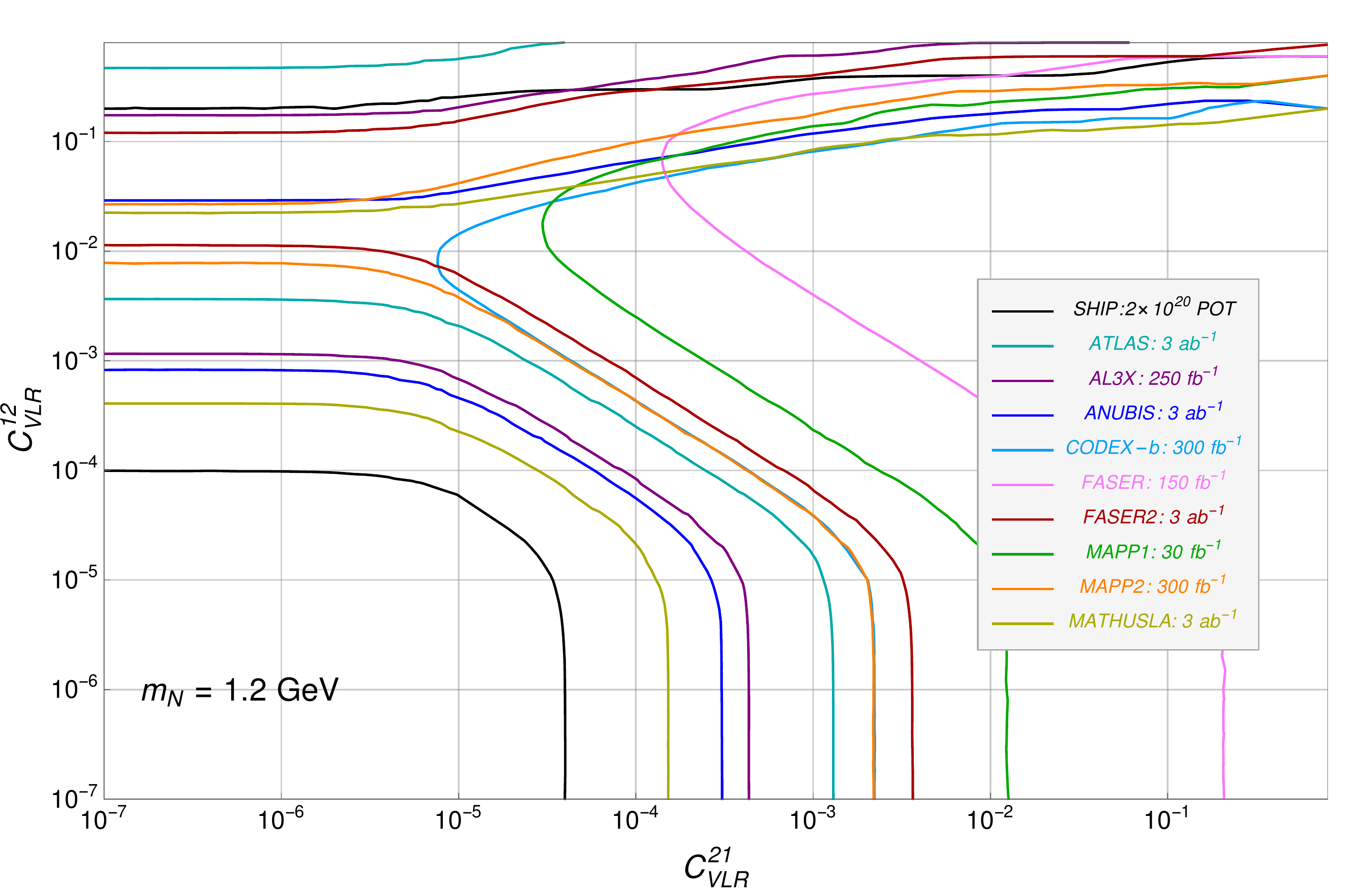}\\
	\includegraphics[width=0.49\linewidth]{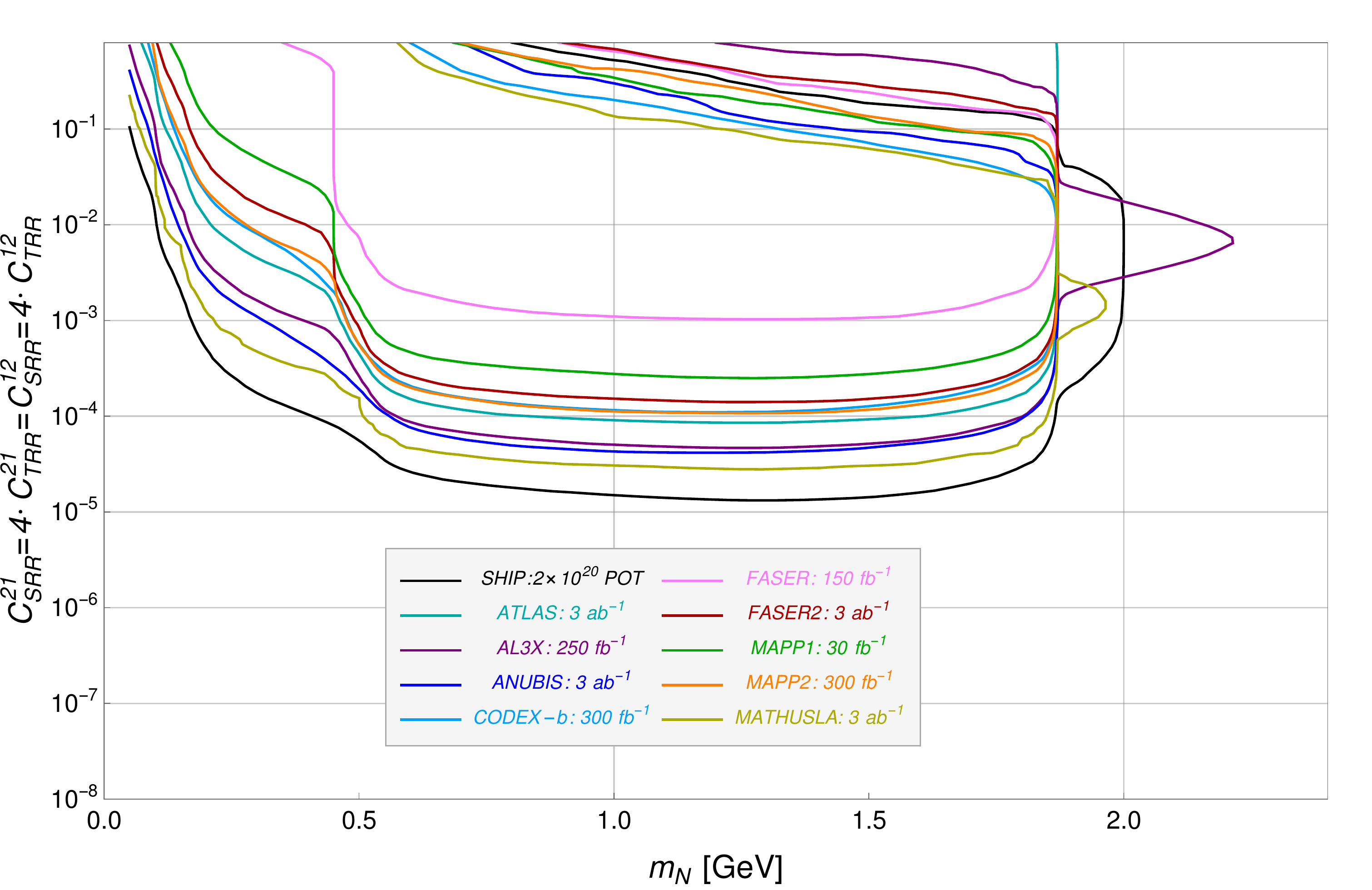}
	\includegraphics[width=0.49\linewidth]{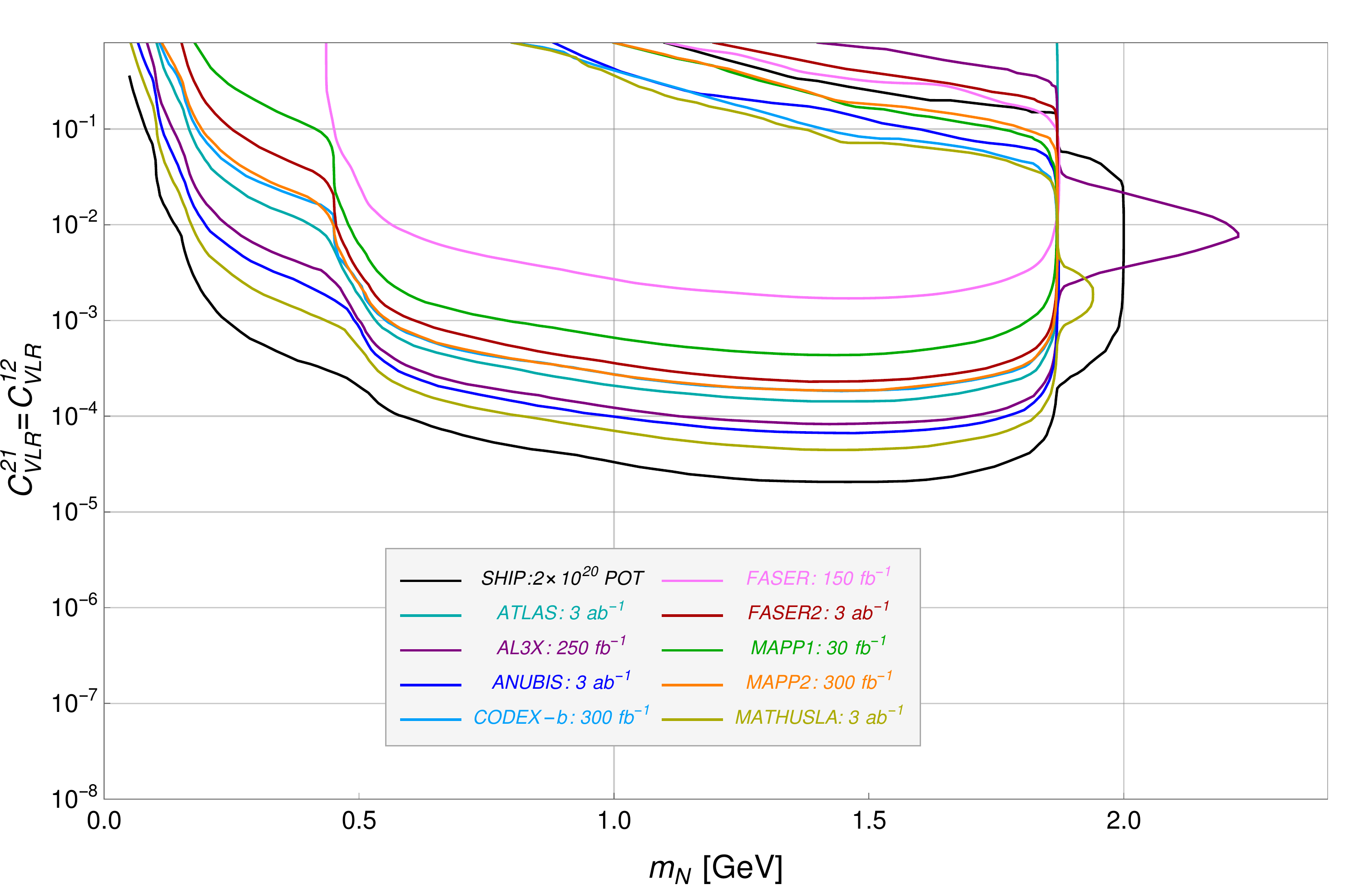}
	\caption{Results for flavor benchmark 2.
		The format is the same as in Fig.~\ref{fig:scen1}.
	}
	\label{fig:scen2}
\end{figure}

The sensitivity limits for this scenario are shown in Fig.~\ref{fig:scen2} with the same format as in Fig.~\ref{fig:scen1}. On the left we consider $C_\text{P}=C_{\text{SRR}}^{21}=4C_{\text{TRR}}^{21}$, and on the right $C_\text{P}=C_{\text{VLR}}^{21}$.
Similarly for the decay we have on the left $C_\text{D}=C_{\text{SRR}}^{12}=4C_{\text{TRR}}^{12}$ and on the right $C_\text{D}=C_{\text{VLR}}^{12}$.
In the upper row, we show plots in the plane $C_\text{D}$ vs. $C_\text{P}$ with $m_N$ at 1.2 GeV.
Similar features as in the previous scenario are observed and there seems hence to be little sensitivity in the event rates to the specific final-state meson.

The hierarchy in sensitivity of the different experiments is also very 
similar.
In the lower panels we see some differences compared to flavor benchmark 1.
The sensitivity to lighter $m_N$ is reduced due to the need to produce a heavier kaon in the final state.
For $m_N < m_K$ the EFT operators no longer contribute to the decay rate and the SM weak interaction becomes the only mechanism for sterile neutrinos to decay and be detected.
This leads to a further reduction in sensitivity.
We stress again that our results in this regime are conservative as we have not considered sterile neutrino production via kaon decays.

\subsection{Flavor Benchmark 3}

\begin{table}[t]
	\begin{center}
		\begin{tabular}{r||c|c}
			& Flavor benchmark 3.2 & Flavor benchmark 3.3  \\
			\hline
			production operator: $C_\text{P}$ & $C_{\text{SRR}}^{13}=4C_{\text{TRR}}^{13}$ & $C_{\text{VLR}}^{13}$ \\[1.7mm]
			decay operator: $C_\text{D}$ & $C_{\text{SRR}}^{11}=4C_{\text{TRR}}^{11}$ & $C_{\text{VLR}}^{11}$ \\[1.7mm]
			production process via $C_\text{P}$	& \multicolumn{2}{c}{$B^\pm/B^0/B_s \rightarrow N + e^\pm(+X)$} \\[-3.2mm] &\multicolumn{2}{c}{{\color{white}.}} \\
			decay process via $C_\text{D}$	& \multicolumn{2}{c}{$N \rightarrow \pi^\pm + e^\mp, \; \rho^\pm + e^\mp$  }
		\end{tabular}
	\end{center}
	\caption{Summary of flavor benchmark 3.}
	\label{tab:scenario3}
\end{table}

We proceed to study a scenario where sterile neutrinos are mainly produced through decays of $B$-mesons. Compared to flavor benchmark 1, 
we keep the same flavor structure for the decay Wilson coefficient, but turn on $C_\text{P}^{13}$.
This leads to the sterile neutrino production via the decay processes
\begin{eqnarray}
	B^\pm \rightarrow e^\pm + N, & \ \ B^\pm \rightarrow \pi^0 + e^\pm + N, & \ \ B^\pm \rightarrow \rho^0 + e^\pm + N, \\
	B^0 \rightarrow \pi^\pm + e^\mp + N, & \ \ B^0 \rightarrow \rho^\pm + e^\mp + N, & \ \ B_s \rightarrow K^{(*)\pm} + e^\mp + N.
\end{eqnarray}
The relevant information is summarized in Table~\ref{tab:scenario3}. 

\begin{figure}[th]
	\centering
	\includegraphics[width=0.49\linewidth]{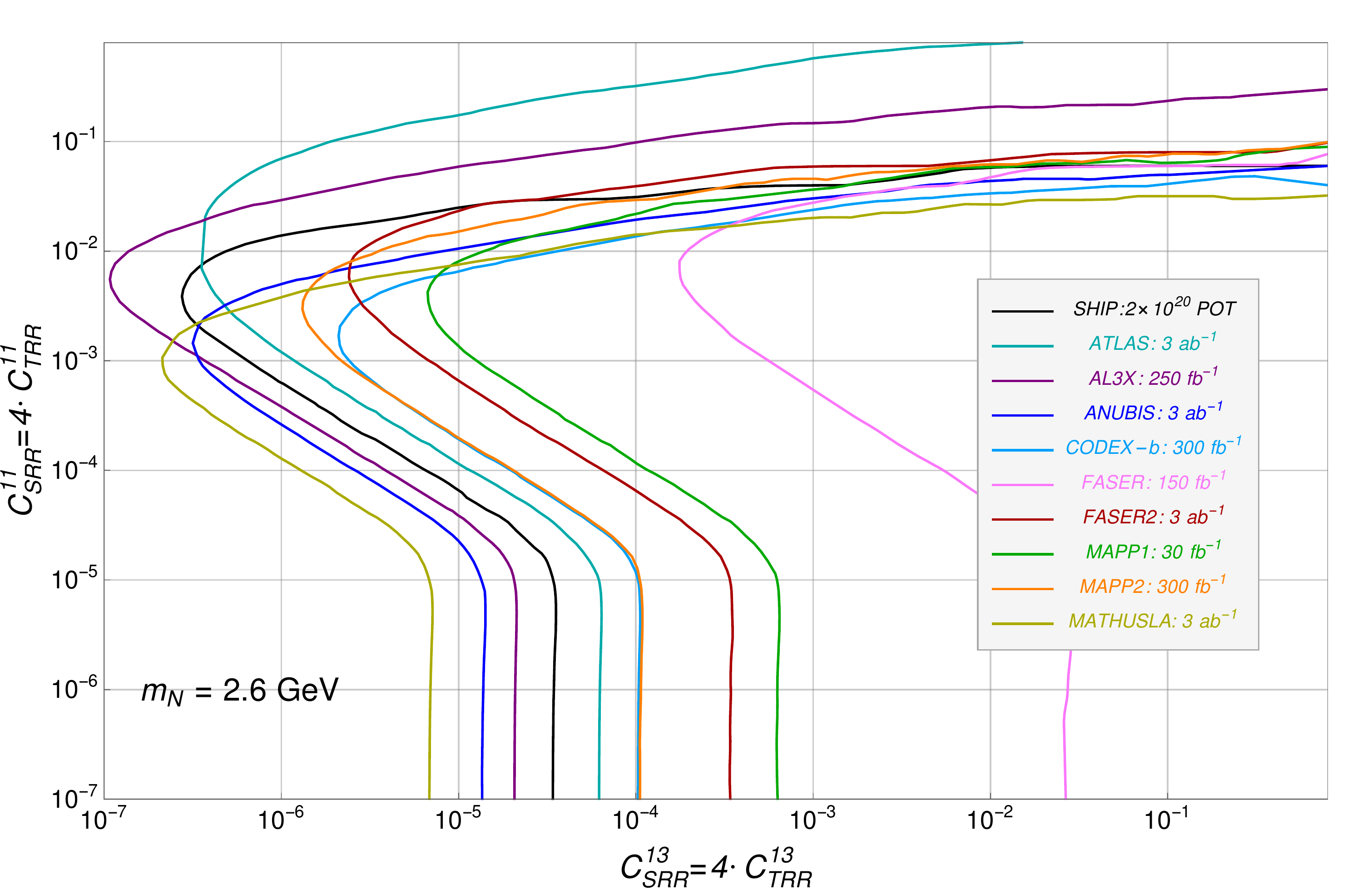}
	\includegraphics[width=0.49\linewidth]{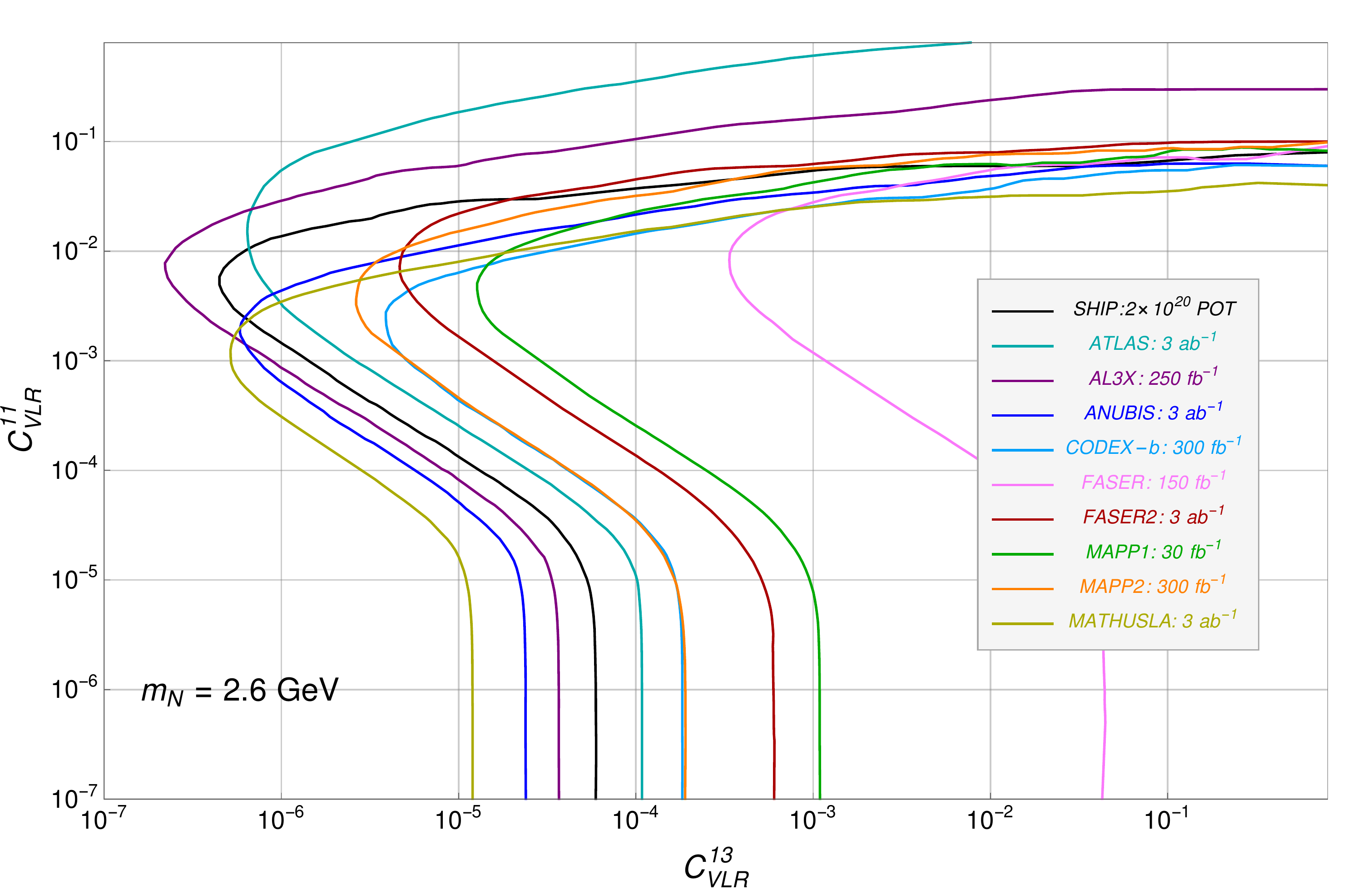}\\
	\includegraphics[width=0.49\linewidth]{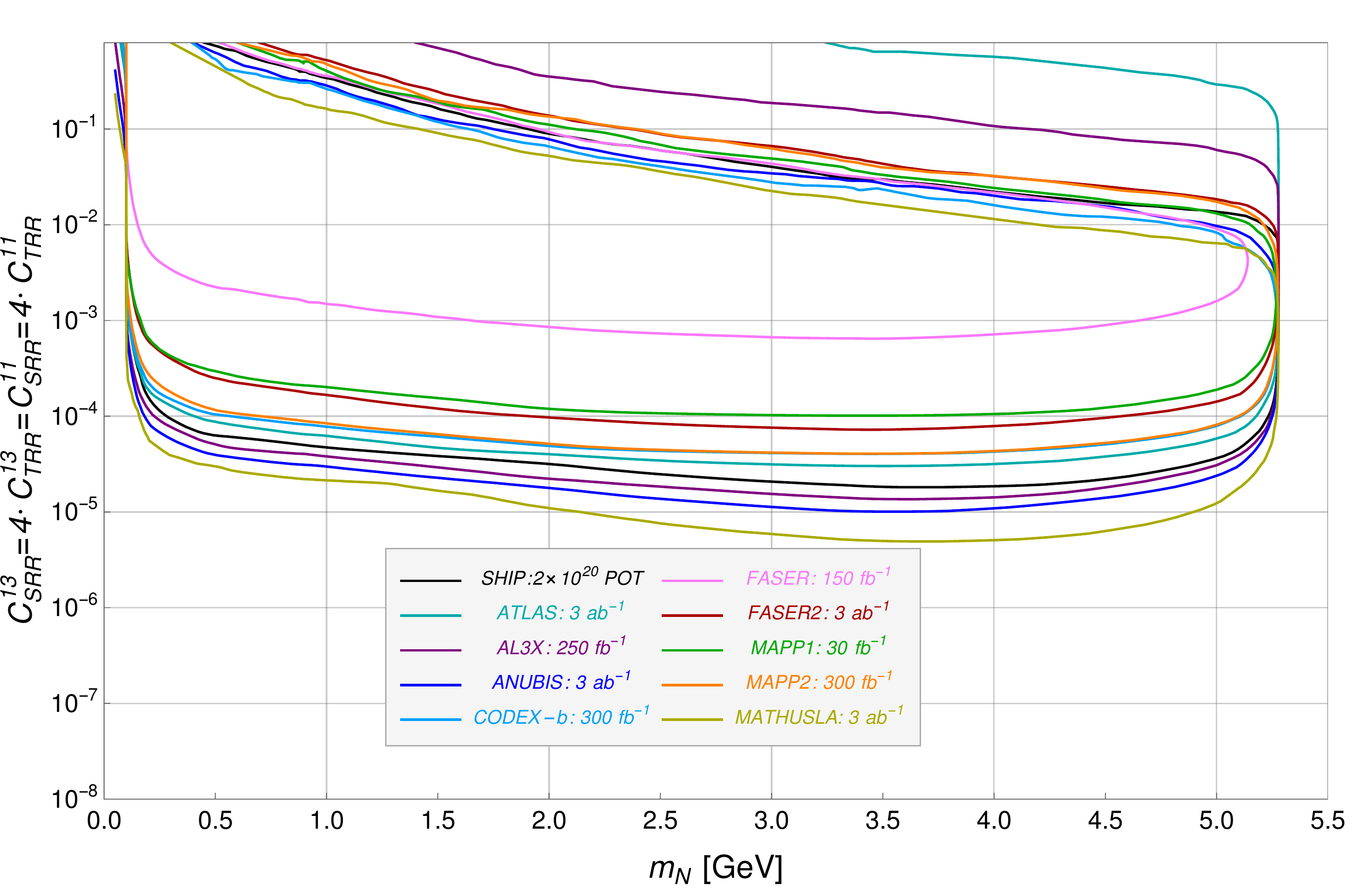}
	\includegraphics[width=0.49\linewidth]{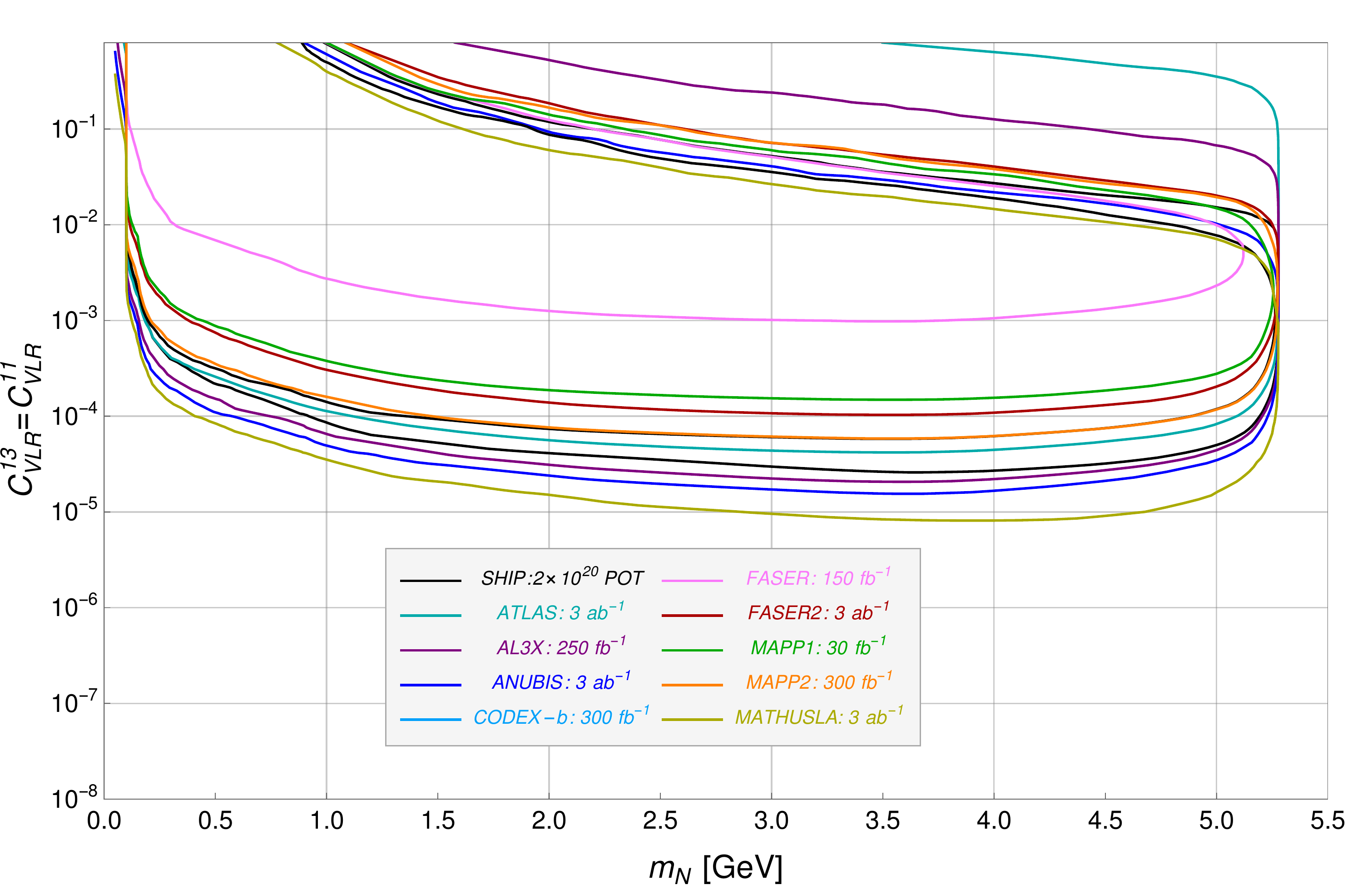}
	\caption{Results for flavor benchmark 3.
		The the scenarios and the labeling are as in Fig.~\ref{fig:scen1}.
	}
	\label{fig:scen3}
\end{figure}

Our results for the sensitivity reach for this benchmark are shown in Fig.~\ref{fig:scen3}.
The two top panels show results for the leptoquark (left) and $C_{\rm VLR}$ (right) scenarios in the $C_\text{P}$-$C_{\text{D}}$ plane for fixed $m_N = 2.6$ GeV.
The resulting curves are rather different from the scenarios, where sterile neutrinos are produced via $D$-meson decays.
In the earlier flavor benchmarks, $C_{\text{P}}$ can be turned off and sufficient sterile neutrinos will be produced via minimal mixing to still detect sterile neutrinos, as long as $C_\text{D}$ is sufficiently large to ensure sterile neutrinos decay in the respective detector volumes.
This feature has disappeared in this benchmark scenario and for $C_{\text{P}} < 10^{-7}$ no detection is possible in any of the experiments, even for large $C_\text{D}$.
This lack of sensitivity is explained by the fact that the production rates of $B$-mesons are smaller than that of $D$-meson by roughly a factor $\sim 20$ at 14 TeV $pp$ collisions and by a factor $\sim 3000$ at \texttt{SHiP}.

The two lower panels in Fig.~\ref{fig:scen3} assume $C_\text{P}=C_\text{D}$ and show sensitivity curves as a function of the sterile neutrino mass.
In the previous flavor benchmarks \texttt{SHiP} showed the strongest sensitivity, but here it performs worse than \texttt{MATHUSLA}, \texttt{ANUBIS}, and \texttt{AL3X}.
The reason is twofold.
First, the ratio between the number of $B$-mesons produced and that of $D$-mesons is much smaller at \texttt{SHiP} than at the 14 TeV $pp-$collision experiments.
Second, given their larger masses, the $B$-mesons, are less boosted in the very forward direction, leading to weakened acceptance of the \texttt{SHiP} detector for the long-lived sterile neutrinos.
The hierarchies in sensitivity of the other experiments is the same as in the minimal scenario and the other flavor benchmarks.
However, the overall reach is increased over the previous flavor benchmarks with the \texttt{MATHUSLA} sensitivity to couplings at the impressive $5\cdot 10^{-6}$ level, corresponding to scales of $\mathcal O(100)\,$TeV.

\subsection{Flavor Benchmark 4}

\begin{table}[H]
	\begin{center}
		\begin{tabular}{r||c|c}
			& Flavor benchmark 4.2 & Flavor benchmark 4.3  \\
			\hline
			production operator: $C_\text{P}$ & $C_{\text{SRR}}^{13}=4C_{\text{TRR}}^{13}$ & $C_{\text{VLR}}^{13}$ \\[1.7mm]
			decay operator: $C_\text{D}$ & $C_{\text{SRR}}^{12}=4C_{\text{TRR}}^{12}$ & $C_{\text{VLR}}^{12}$ \\[1.7mm]
			production process via $C_\text{P}$	& \multicolumn{2}{c}{$B^\pm/B^0/B_s \rightarrow N + e^\pm(+X)$} \\[-3.2mm] &\multicolumn{2}{c}{{\color{white}.}} \\
			decay process via $C_\text{D}$	& \multicolumn{2}{c}{$N \rightarrow K^{(*)\pm} + e^\mp $ }
		\end{tabular}
	\end{center}
	\caption{Summary of flavor benchmark 4.}
	\label{tab:scenario4}
\end{table}

For flavor benchmark 4 the production primarily proceeds via $B$-meson decay, as for the previous benchmark, but here sterile neutrinos decay to a kaon through $C_\text{D}^{12}$.
The relevant information is summarized in Table~\ref{tab:scenario4}. 

\begin{figure}[th]
	\centering
	\includegraphics[width=0.49\linewidth]{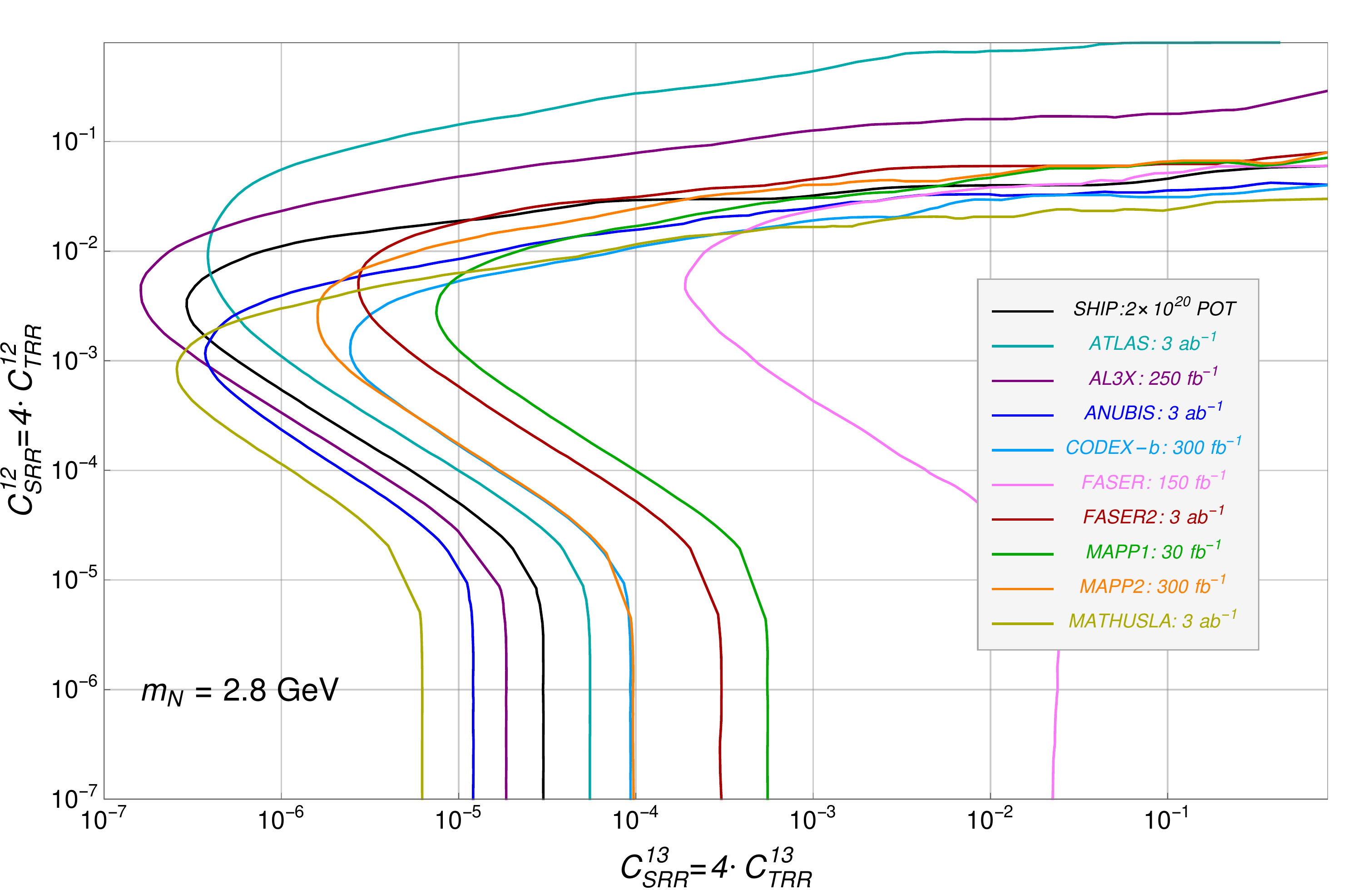}
	\includegraphics[width=0.49\linewidth]{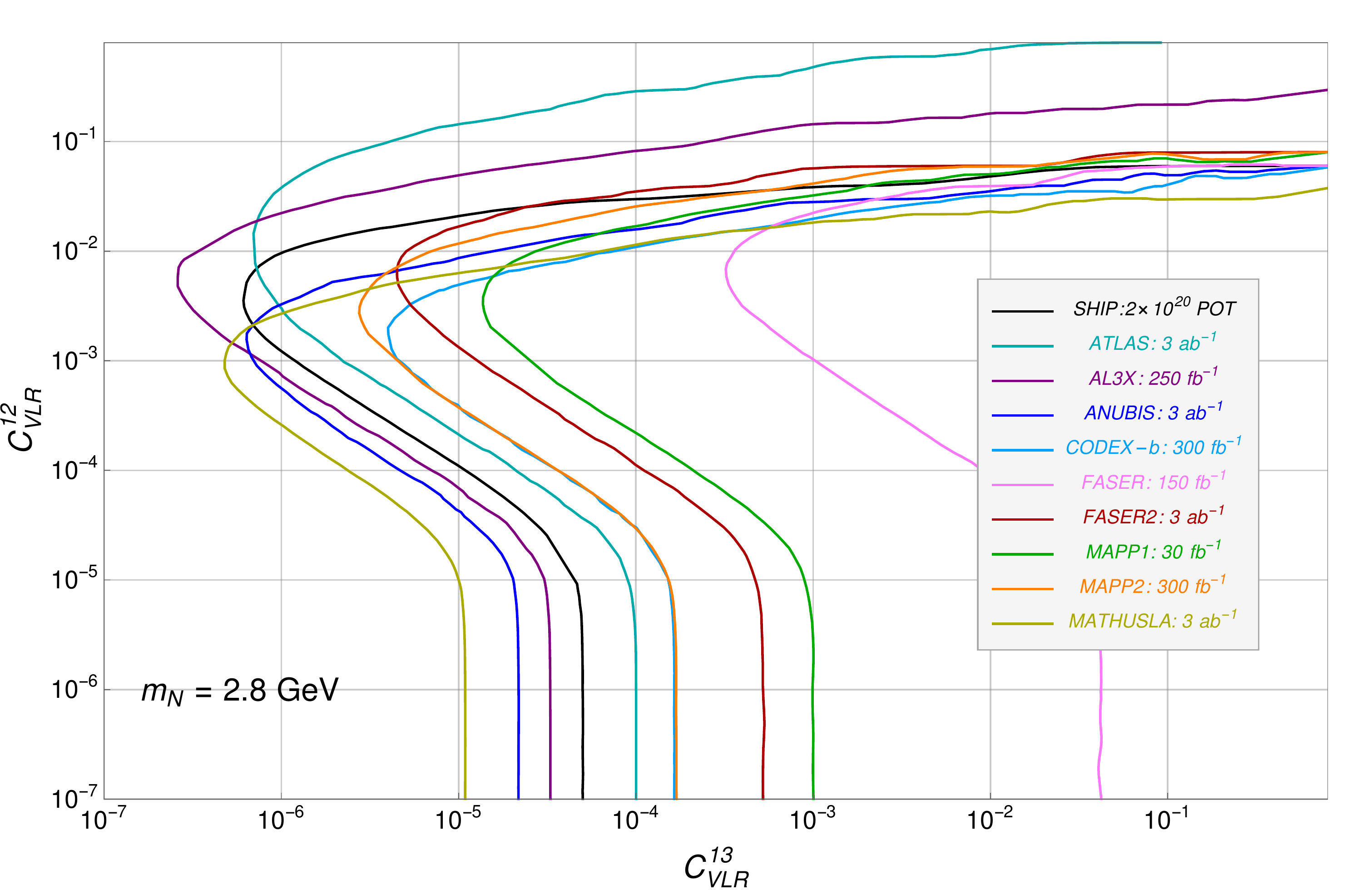}\\
	\includegraphics[width=0.49\linewidth]{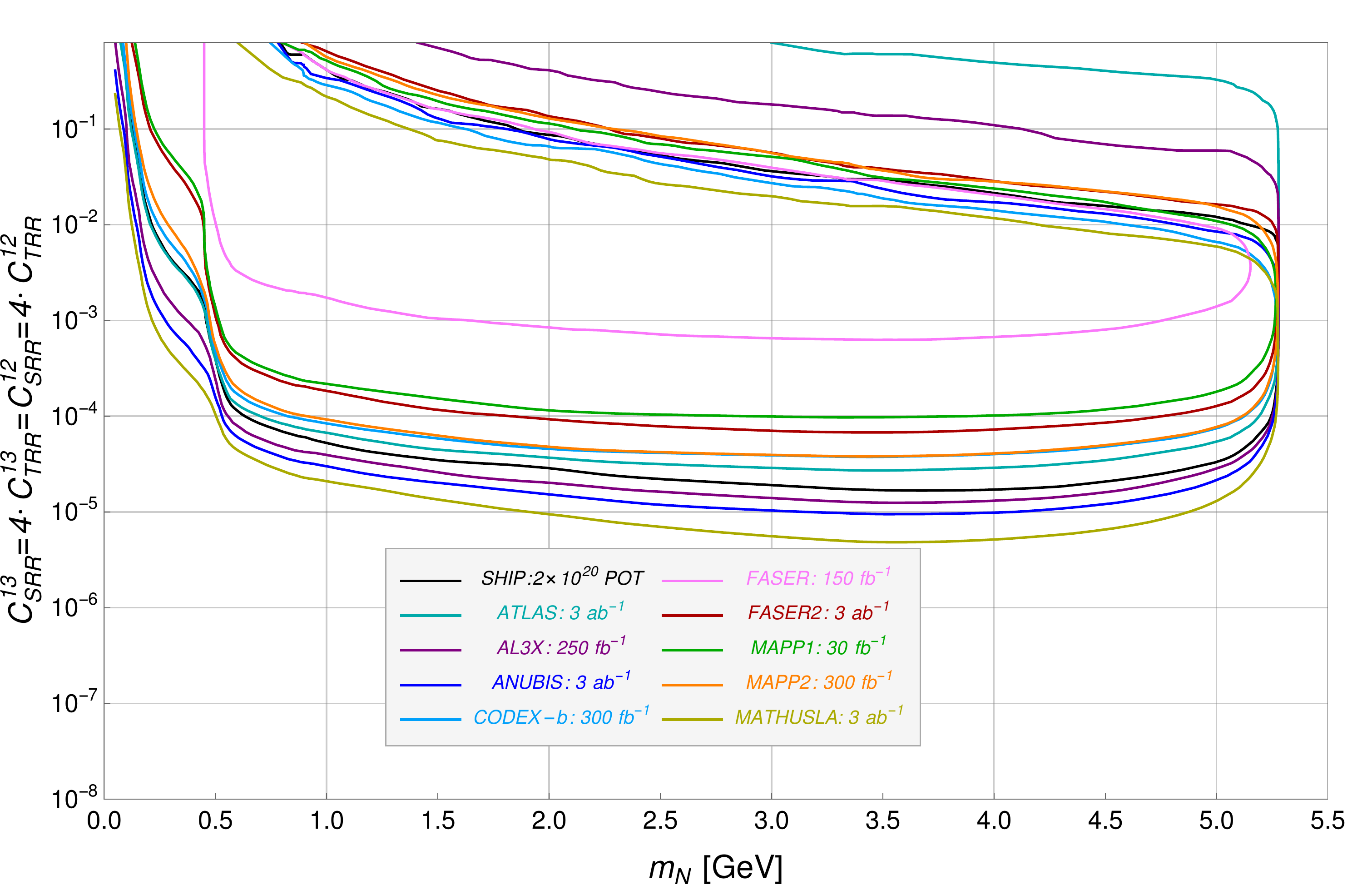}
	\includegraphics[width=0.49\linewidth]{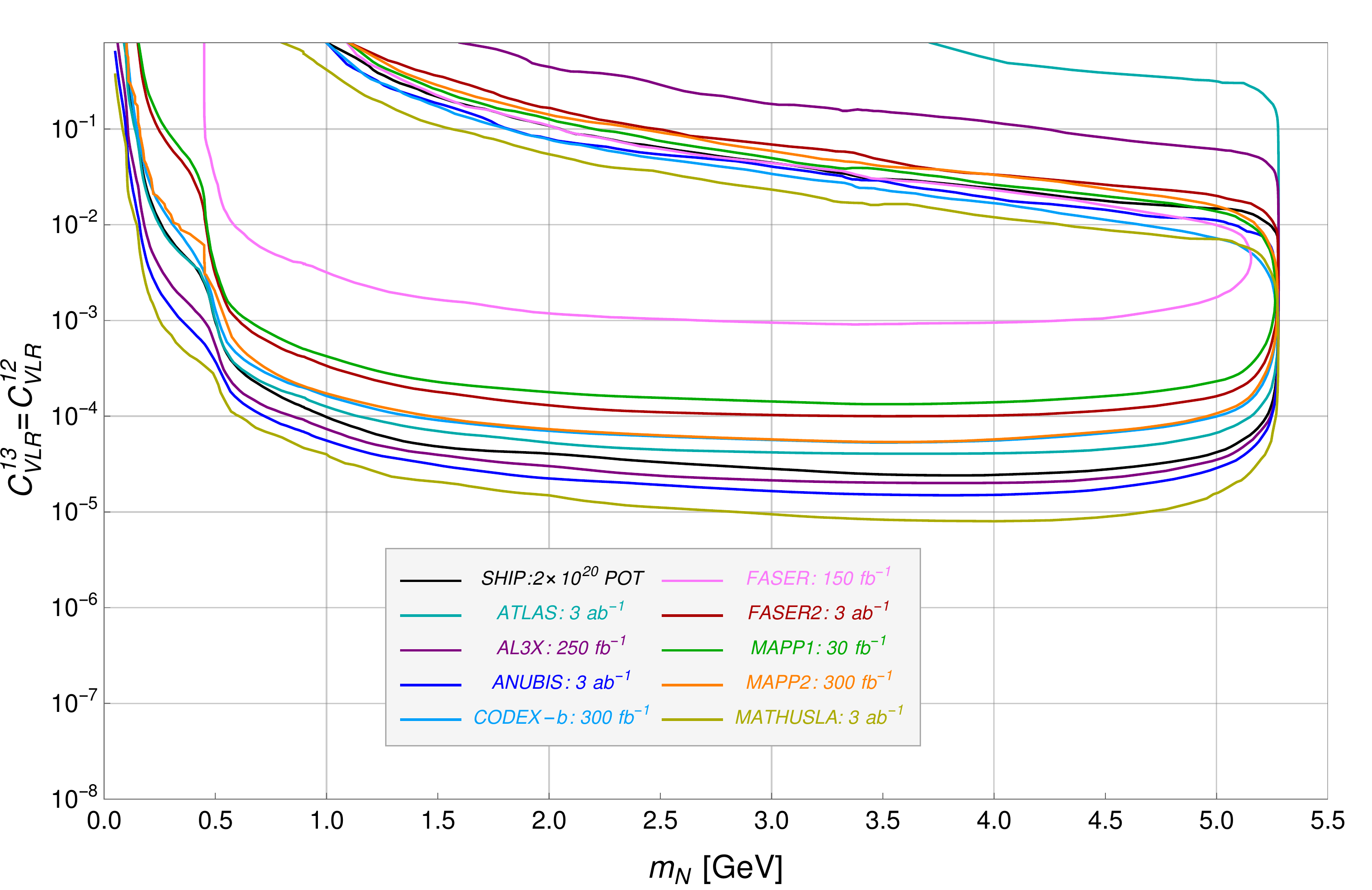}
	\caption{Results for flavor benchmark 4.
		The plot format is the same as in Fig.~\ref{fig:scen1}.
	}
	\label{fig:scen4}
\end{figure}

Fig.~\ref{fig:scen4} shows the numerical results for this benchmark. In the two top panels we show plots in the $C_{\text{P}}$-$C_{\text{D}}$ plane for fixed $m_N=2.8$~GeV.
In general these plots show very similar features as their counterparts in Fig.~\ref{fig:scen3}, except for an overall small reduction in the reach of $C_\text{P}$ because of the choice for a slightly larger mass of the sterile neutrino. The two lower plots show sensitivity curves in the plane 
$C_\text{P}=C_\text{D}$ vs. $m_N$.
Compared to the lower panels of Fig.~\ref{fig:scen3}, they show similar exclusion limits for $m_N \gtrsim m_K$.
However, for lighter sterile neutrinos, since only the decay modes via the weak current and active-sterile neutrino mixing are open, the sensitivity is significantly reduced.
The hierarchies of the various experiments is the same as in the previous benchmark for both EFT scenarios.

\subsection{Flavor Benchmark 5}

\begin{table}[h]
	\begin{center}
		\begin{tabular}{r||c|c}
			& Flavor benchmark 5.2 & Flavor benchmark 5.3 \\
			\hline
		production operator: 	$C_\text{P}$ & $C_{\text{SRR}}^{13}=4C_{\text{TRR}}^{13}$ & $C_{\text{VLR}}^{13}$\\[1.7mm]
			decay operator: $C_\text{D}$ & $C_{\text{SRR}}^{21}=4C_{\text{TRR}}^{21}$ & $C_{\text{VLR}}^{21}$\\[-3.2mm] &\multicolumn{2}{c}{{\color{white}.}} \\
			production process via $C_\text{P}$	& \multicolumn{2}{c}{$B^\pm/B^0/B_s \rightarrow N + e^\pm(+X)$} \\ [-3.3mm] &\multicolumn{2}{c}{{\color{white}.}} \\
			production process via $C_\text{D}$     & \multicolumn{2}{c}{$D^\pm/D^0/D_s \rightarrow N + e^\pm(+X)$} \\[-3.3mm] &\multicolumn{2}{c}{{\color{white}.}} \\
			decay process via $C_\text{D}$	& \multicolumn{2}{c}{$N\rightarrow D^{(*)\pm}+ e^\mp$ } 
		\end{tabular}
	\end{center}
	\caption{Summary of flavor benchmark 5.}
	\label{tab:scenario5}
\end{table}

\begin{figure}[th]
	\centering
	\includegraphics[width=0.49\linewidth]{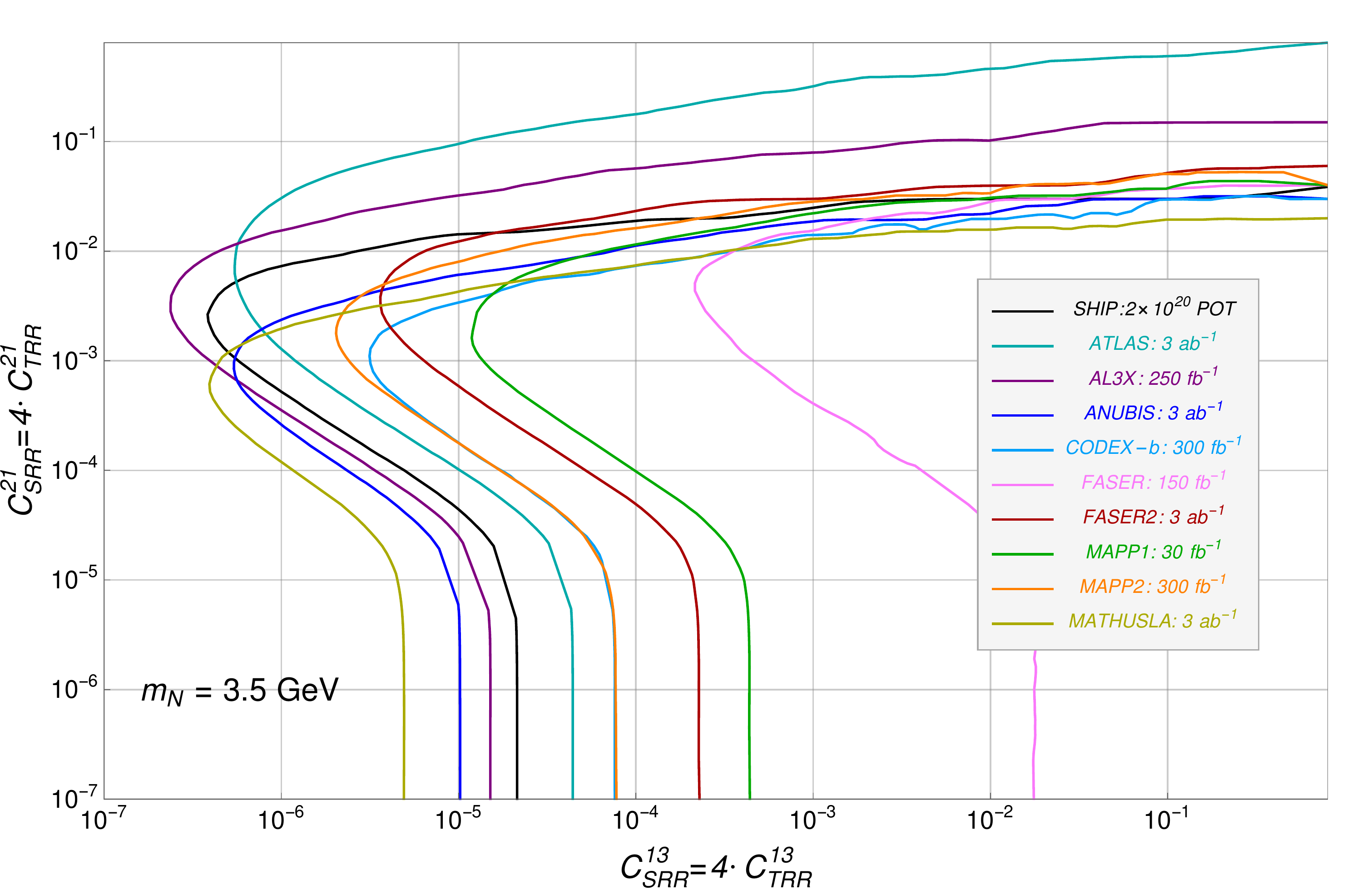}
	\includegraphics[width=0.49\linewidth]{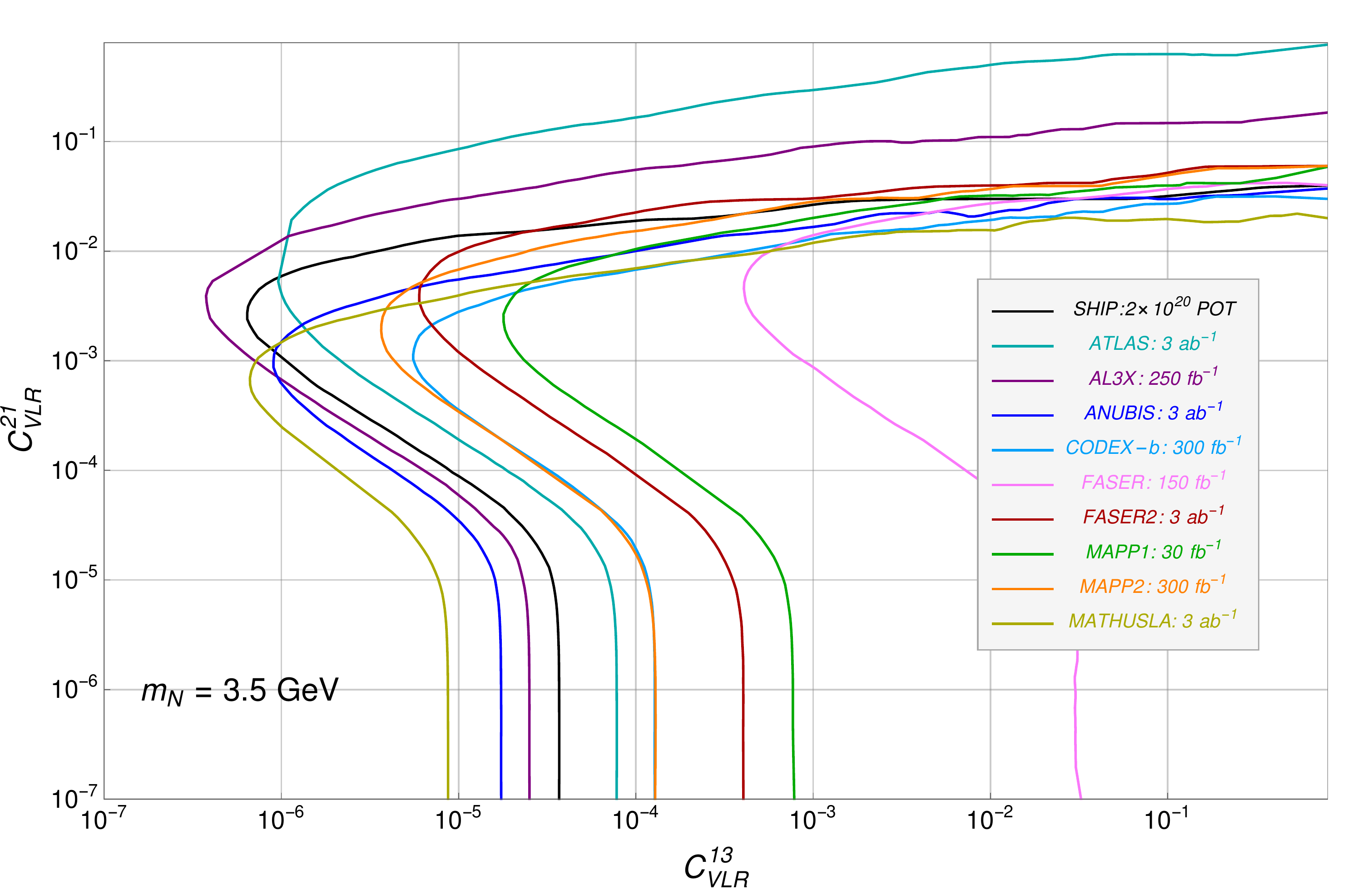}\\
	\includegraphics[width=0.49\linewidth]{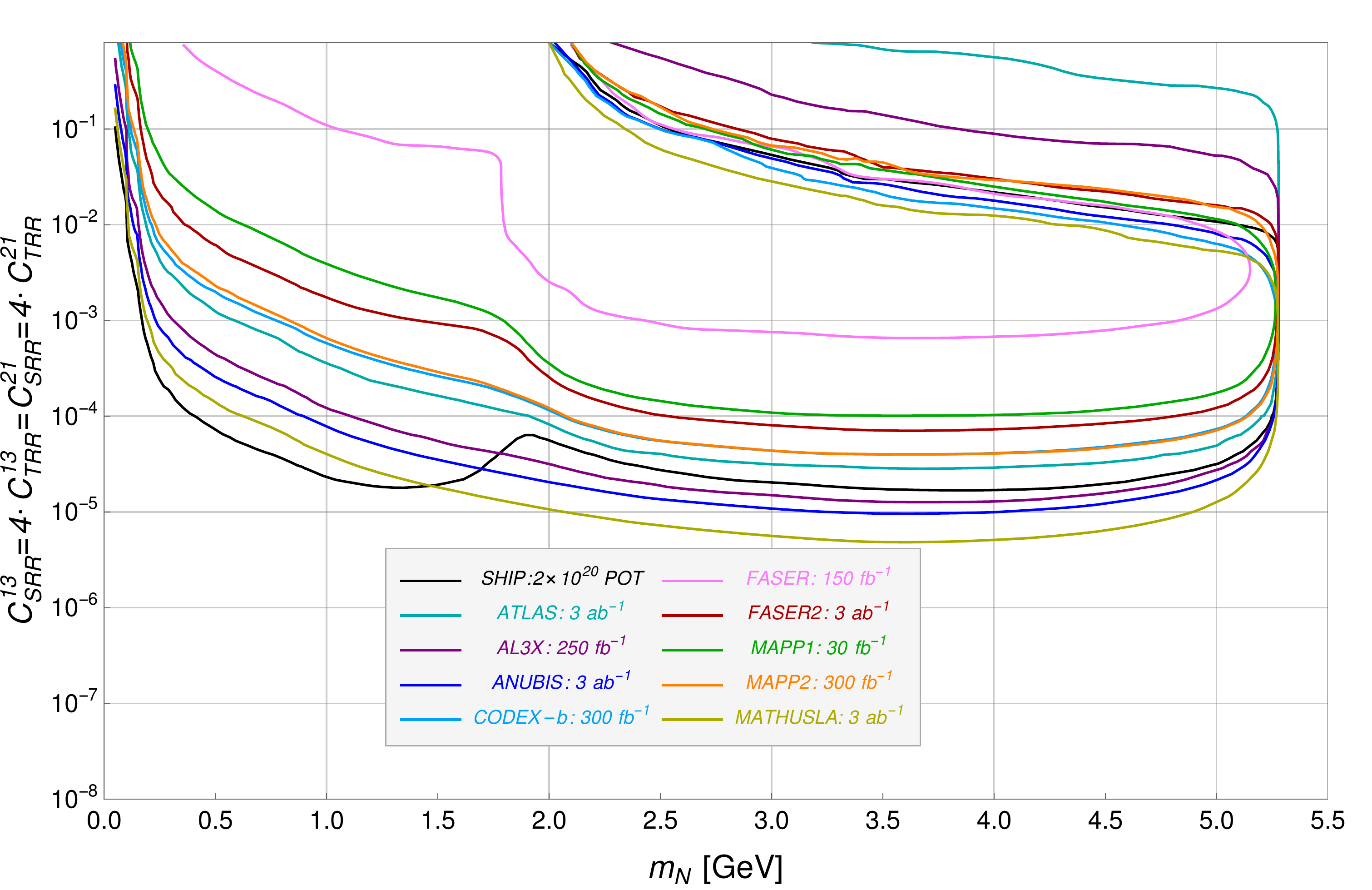}
	\includegraphics[width=0.49\linewidth]{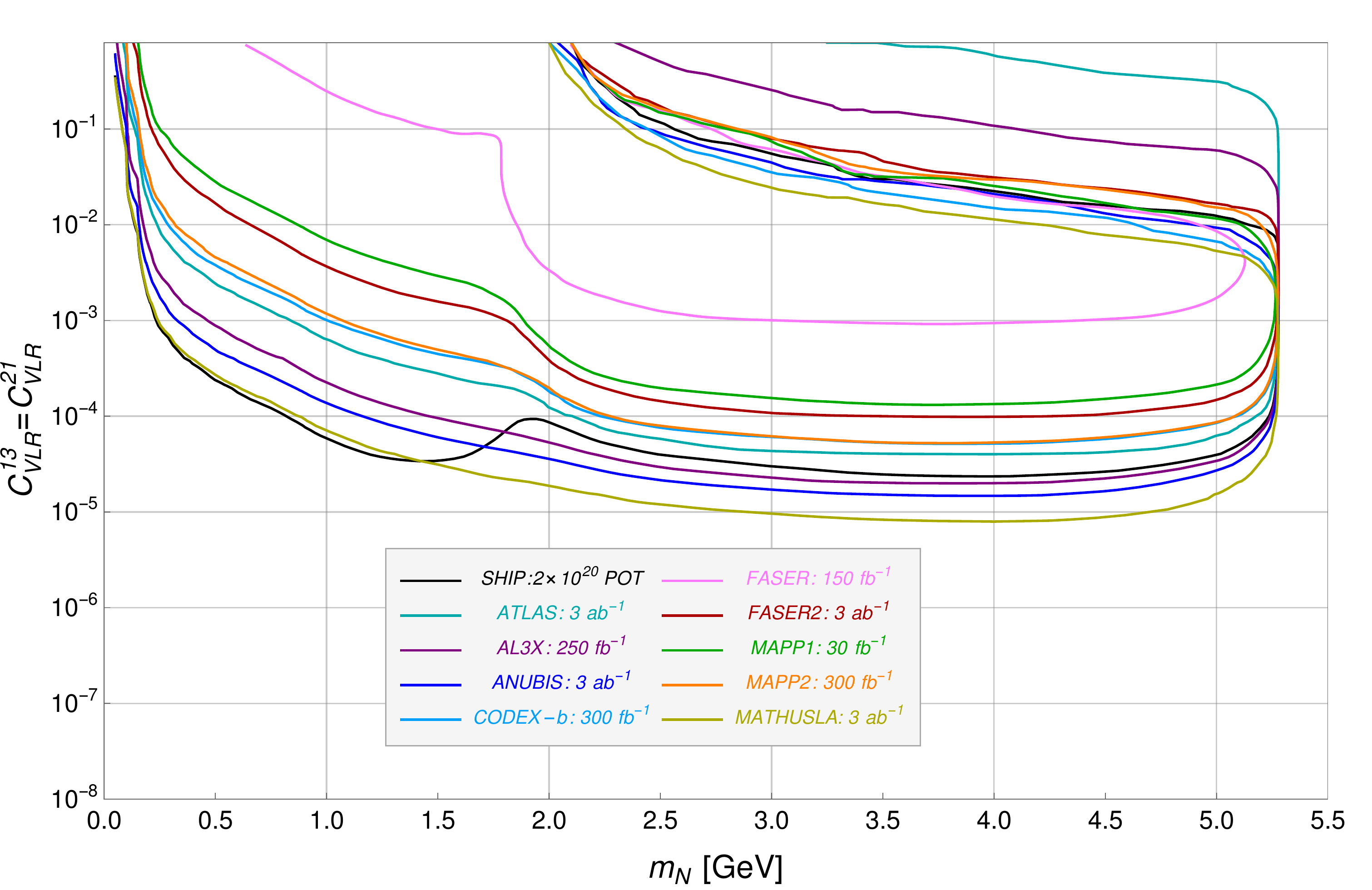}
	\caption{Results for flavor benchmark 5.
		The format is the same as in Fig.~\ref{fig:scen1}.
	}
	\label{fig:scen5}
\end{figure}

In flavor benchmark 5, we turn on the operators $C_\text{P}^{13}$ and $C_\text{D}^{21}$. In this case, the decay operator also 
leads to production of sterile neutrinos, but the resulting sterile neutrinos are restricted to a mass range where they can only decay via minimal mixing. 
We summarize the benchmark features in Table~\ref{tab:scenario5}. Fig.~\ref{fig:scen5} shows the resulting sensitivity reach. 
In the upper row we fix the sterile neutrino mass at 3.5 GeV. In general the absolute and relative sensitivities to $C_\text{P}$ and $C_
\text{D}$ are comparable to the previous flavor benchmark, but the sensitivity in the bottom panel drops a bit for $m_N < m_D$. In 
this case sterile neutrinos only decay via minimal mixing. The exception is \texttt{SHiP} for which the sensitivity grows for $m_N <m_D$, 
where \texttt{SHiP} becomes the most sensitive experiment in fact, because the production cross section difference between $D$- 
and $B$-mesons is much larger at \texttt{SHiP} than at the other experiments.

\section{Discussion and Comparison with other Probes}\label{sec:comparison}

Our results indicate that proposed experiments to detect long-lived particles are very sensitive to higher-dimensional operators in the 
$\nu$SMEFT Lagrangian. In the mass range $m_K < m_N < m_B$ the sensitivity curves are rather stable with respect to different EFT 
operators in Eq.~\eqref{final67} and the particular flavor configuration, as long as the sterile neutrino can be produced via the decay of $D$- or 
$B$-mesons, and the sterile neutrino can, in turn, decay semi-leptonically. While each particular choice of EFT operators and flavor assignment 
requires a detailed study, we find that \texttt{FASER} is sensitive to Wilson coefficient couplings of about $\sim 10^{-3}$ (this is extended 
to $\sim 10^{-4}$ for \texttt{FASER2}), while experiments such as \texttt{MATHUSLA}, \texttt{ANUBIS}, \texttt{AL3X}, and \texttt{SHiP} 
can reach down to coupling strengths of $\sim 5\cdot  10^{-6}$. From the matching relations in Eq.~\eqref{match6LNC}, we see that such 
limits can be used to constrain the $\nu$SMEFT operators $C_{H\nu e}^{(6)}$, $C_{du\nu e}^{(6)}$, $C_{L\nu Qd}^{(6)}$, $C_{LdQ\nu }
^{(6)}$, and $C_{Qu\nu L}^{(6)}$. Assuming a scaling of these Wilson coefficients as $\sim v^2/\Lambda^2$, the sensitivities range 
from $\Lambda \sim 8\,$TeV for \texttt{FASER} up to $\Lambda \sim 100\,$TeV for the larger experiments. 

The $\nu$SMEFT operators we consider here can also be probed in other experiments. These include meson and tau 
decays, elastic coherent neutrino-nucleon scattering (EC$\nu$NS), missing transverse energy searches, etc. Depending on the probe, the 
relevant  sterile neutrino mass range and the flavor assignment can differ from the cases considered in this work. For instance, 
limits from pion decay or from neutron or nuclear beta decay require sterile neutrino masses below the pion mass or the respective Q value of $\beta$ decay, considerably lower than the GeV-scale sterile neutrinos considered in this work. Here we briefly give an overview of the literature.

Refs.~\cite{Alcaide:2019pnf,Biekotter:2020tbd} investigated limits from pion decays, tau decays, and singular leptons with missing 
transverse energy. The most restrictive bounds on the new physics scale are obtained from pion decays with $\Lambda \gtrsim 36
$\,TeV. However, tau decays allow for a  neutrino mass range $m_N$ more comparable to our studies, while searches for $l+
\cancel{\it{E}}_{T}$ are largely independent of sterile neutrino masses. The latter investigations set the new physics scale to $\Lambda\gtrsim
2-5\,$TeV. We did not explicitly consider processes involving $\tau$ leptons, but there should be good sensitivity in the appropriate mass range $m_\tau + m_\pi 
< m_N  < m_B - m_\tau$. More quantitative statements require a detailed study that includes sterile neutrino production via $\tau$ decays and an efficiency factor
for reconstructing decays of $\tau$ mesons in the final states. 

Refs. \cite{Li:2020lba,Li:2020wxi} consider a larger set of pseudoscalar meson decays corresponding to several 
flavor configurations. Additionally, the effects of $\nu$SMEFT operators on lepton flavor universality (LFU), CKM unitarity, and $\beta$-decays 
are examined. The most stringent bounds are on the operators $C_{L\nu Qd}^{(6)}$, $C_{LdQ\nu }^{(6)}$, and $C_{Qu\nu L}^{(6)}$ involving 
an up quark, a down (strange) quark and an electron using LFU constraints. The new physics scale is limited by $\Lambda 
\gtrsim 74~(110)\,$TeV in the limit of massless sterile neutrinos and thus cannot be directly compared to results obtained here. Bounds on 
other operators and different flavor combinations are in the range of $\Lambda \gtrsim 0.5-8\,$TeV. Similar sensitivities are found examining anomalies in the transition $b\rightarrow c \tau\overline{\nu}$ including light sterile neutrinos ($m_N \lesssim 100$\,MeV) \cite{Mandal:2020htr}.  Further, limits from EC$\nu$NS based on the 
\texttt{COHERENT} experiment \cite{Akimov:2017ade} are considered in Refs.~\cite{Li:2020lba,Han:2020pff,Bischer:2019ttk}. Sterile neutrinos 
considered in these works are again much lighter than the GeV-scale ($m_N \lesssim 0.5$\,MeV) and the resulting bounds are at the level 
of $\Lambda\gtrsim 1\,$TeV. We conclude that the sensitivities of the experiments considered here are competitive with and complementary 
to existing constraints. Constraints on dimension-five couplings are discussed e.g. in Refs.~\cite{Caputo:2017pit,Jones-Perez:2019plk}.

In this work we have focused on Majorana neutrinos (although our sensitivity curves are not affected dramatically if we had 
considered Dirac neutrinos instead) for which strong constraints can be set from $0\nu\beta\beta$ experiments. In Ref.~\cite{Dekens:2020ttz} 
some of us developed a framework to calculate $0\nu\beta\beta$ decay rates in the presence of light sterile neutrinos and the 
$\nu$SMEFT Lagrangian. In particular, we investigated the reach of current and future experiments to probe scenario 2: the $3+1$ leptoquark 
model. As sterile neutrinos appear as virtual states, $0\nu\beta\beta$ experiments are sensitive to a broad range of neutrino masses with a peak 
sensitivity at $m_N \simeq 100$\,MeV, which drops off for larger or smaller masses. To make a comparison, we consider the case $y^{\overline{LR}}
_{11}y^{RL*}_{11} = y^{\overline{LR}}_{21}y^{RL*}_{11} = 1$ and $y^{\overline{LR}}_{11}y^{RL*}_{11} = y^{\overline{LR}}_{11}y^{RL*}_{31}=1$ and vanishing couplings for other flavors.  We can then compare flavor benchmarks $1$ and $3$ to the 
sensitivity of $0\nu\beta\beta$ experiments, which only depend on sterile neutrino couplings to first-generation quarks and leptons. 

\begin{figure}[t]
\centering
\includegraphics[width=0.49\textwidth]{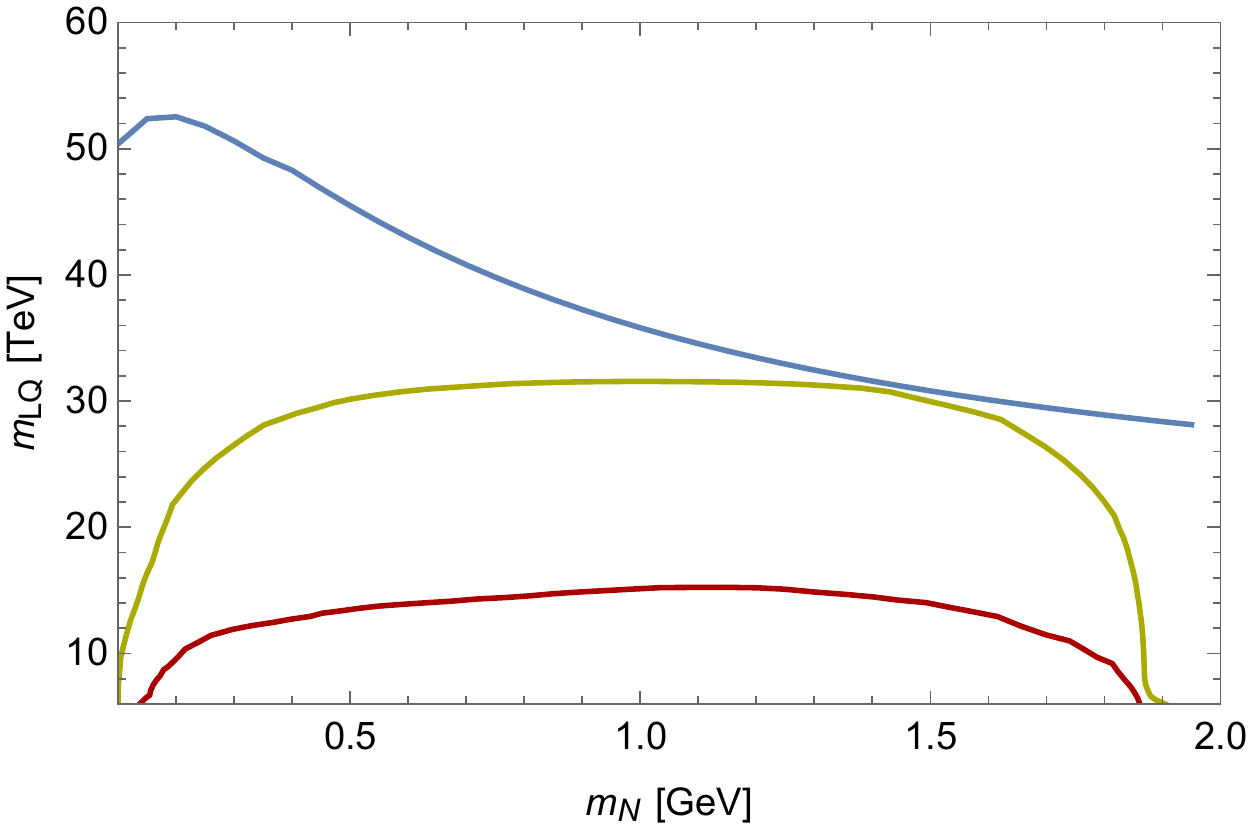}
\includegraphics[width=0.49\textwidth]{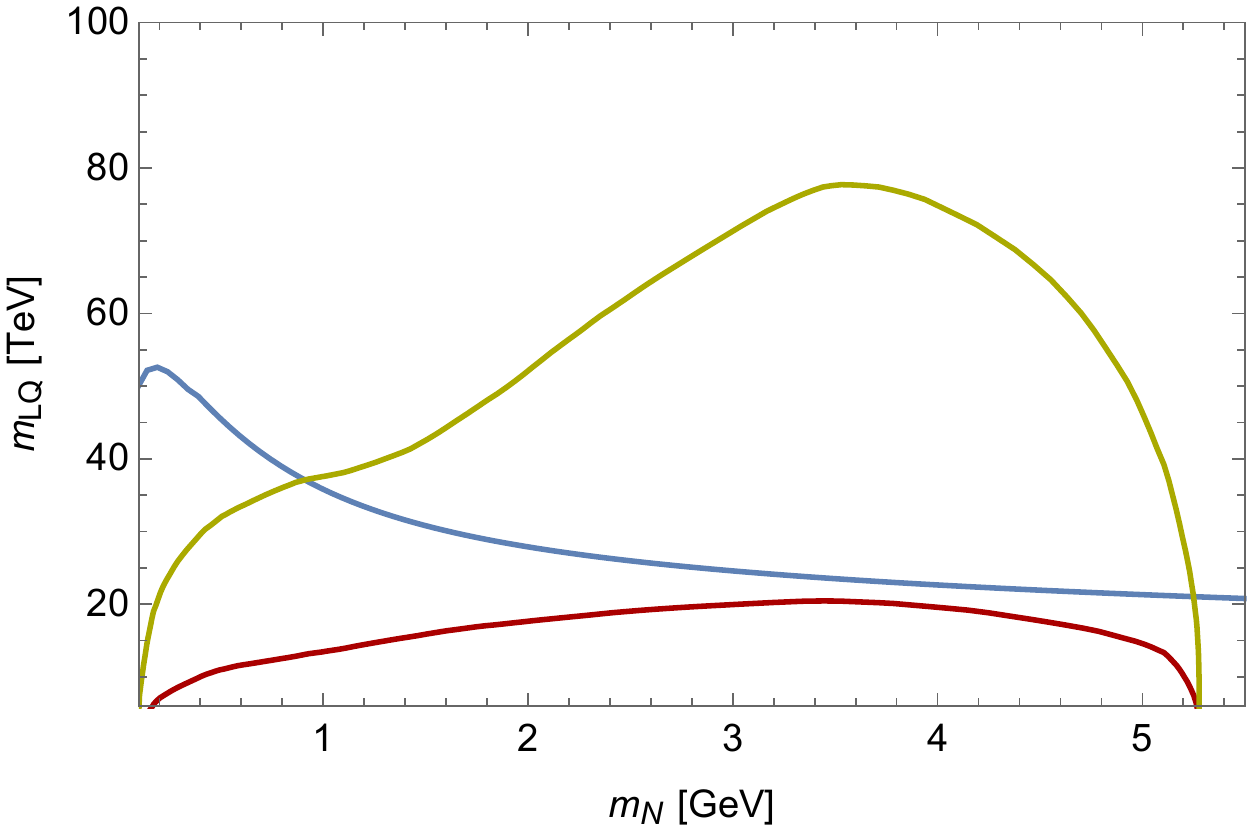}
\caption{Comparison between constraints from neutrinoless double beta decay of ${}^{136}$Xe (blue) \cite{KamLAND-Zen:2016pfg} and projected 
sensitivity of \texttt{FASER2} (red) and \texttt{MATHUSLA} (yellow) for the leptoquark scenario. In the left (right) panel we turned on LQ couplings 
corresponding to flavor benchmark 1 (3). See text for more details.
	}
\label{lqcontour}
\end{figure}

For these choices of couplings, we can calculate $0\nu\beta\beta$ decay rates and determine the LHC and \texttt{SHiP} sensitivity curves as a function of $m_N$ and $m_{\rm LQ}$ ($0\nu\beta\beta$ rates have a very small dependence on phases appearing in the $3+1$ neutrino mixing matrix and we neglect this dependence here for simplicity).
The results are shown in Fig.~\ref{lqcontour}. We stress that the uncertainties associated with hadronic and nuclear matrix elements for $0\nu\beta\beta$ decay rates are sizable and not included in the plot, for details we refer to Refs.~\cite{Cirigliano:2018yza,Dekens:2020ttz}. 
For flavor benchmark 1 (left panel of Fig.~\ref{lqcontour}), the limits from $0\nu\beta\beta$ are somewhat stronger than the prospected sensitivity of 
\texttt{FASER2} and \texttt{MATHUSLA}, chosen as representative experiments, in the relevant mass range. For flavor benchmark $3$ the prospected 
\texttt{MATHUSLA} overtake current $0\nu\beta\beta$ limits for masses between $1$ and $5$ GeV. 

We stress that the bounds from $0\nu\beta\beta$ decay experiments are only valid for Majorana neutrinos and final-state electrons. However, the 
sensitivity curves for the various LHC experiments discussed here are (roughly) valid for (pseudo-)Dirac neutrinos, and in the appropriate mass range also for 
couplings to muons and, to lesser extent, taus instead of electrons, and in general to a broader range of quark flavors. They are thus more 
general than $0\nu\beta\beta$ limits albeit in a much small sterile neutrino mass range.

\section{Conclusions}\label{sec:conclusion}

The possibility of sterile neutrinos provides one of the main motivations for the search for long-lived particles (LLPs). 
Sterile neutrinos provide compelling solutions to major problems in particle physics and cosmology, such as active 
neutrino masses and the baryon asymmetry of the Universe. Sterile neutrinos are in fact predicted in a variety of 
theoretical models, ranging from the minimal scenario where they interact with Standard Model (SM) particles through 
minimal mixing with active neutrinos, to more exotic scenarios involving new fields with masses well above the 
electroweak scale such as left-right symmetric models or grand unified theories.

In this work, we focused on relatively light sterile neutrinos with masses at the GeV scale, down to about 100\,MeV.
This mass range is interesting, as it is linked to low-scale leptogenesis and opens up the possibility of efficiently producing sterile neutrinos through the decays of pseudoscalar mesons, which are copiously 
produced in collider experiments.
In particular at CERN, besides the \texttt{ATLAS} and \texttt{CMS} LHC collaborations, various proposed experiments are presently under discussion specifically targeting the detection of LLPs, such as the fixed-target experiment \texttt{SHiP} and a number of so-called far detectors at various $pp$-collision interaction points \textit{e.g.} \texttt{FASER} and \texttt{MATHUSLA}.
A large number of mesons are expected to be produced at the interaction points of these experiments, which in turn can decay to sterile neutrinos. 
We focused on sterile neutrinos which can be produced from bottom and charm meson decays in hadronic collisions, and investigated the sensitivity reach of present and future LHC experiments, and \texttt{SHiP} to detect sterile neutrinos.

To avoid theoretical bias we approached this problem in the framework of the neutrino-extended SM effective field theory ($\nu$SMEFT).
This framework allows for a light gauge-singlet fermion, a sterile neutrino or heavy neutral lepton, and assumes other BSM fields to have masses $M\gg v$, the SM Higgs vacuum expectation value. This framework describes effective local interactions between sterile neutrinos and SM fields in a systematic expansion. 
We considered dimension-6 operators that allow for sterile neutrino production via mesonic decays at tree level. Our framework is general, but for concreteness we considered specific scenarios where a subset, or only a single, EFT operator is turned on at the same time.
These scenarios are motivated by UV completions like leptoquark or left-right-symmetric models. Other EFT scenarios or specific UV-complete models can be straightforwardly investigated by using the extensive formulae given in the appendix. 
To benchmark our calculations with the literature, we also considered the minimal scenarios where higher-dimensional operators are turned off.

We performed Monte-Carlo simulations to evaluate the sensitivity reach of the considered experiments: 
\texttt{ATLAS}, \texttt{CODEX-b}, \texttt{FASER}, \texttt{MATHUSLA}, \texttt{AL3X}, \texttt{ANUBIS}, \texttt{MoEDAL-MAPP}, 
and \texttt{SHiP}. For the minimal scenario, the obtained sensitivity curves are in agreement with existing results 
with minimal discrepancies, while we obtained sensitivity curves for the two \texttt{MoEDAL-MAPP} experiments for the first time.
For the EFT scenarios we consider active-sterile neutrino mixing, the EFT operators, and their interference terms simultaneously.
For each EFT scenario, we considered a series of different flavor benchmarks, where the EFT operators induce either $D$- or $B$-meson 
decays into sterile neutrinos, and the sterile neutrinos decay to an electron and various mesons, $N\to e+M_{ij}$, \textit{cf.} 
Tables~\ref{tab:scenario1}-\ref{tab:scenario5}. For the $D$-meson benchmarks, we found that \texttt{SHiP} and \texttt{MATHUSLA} have 
the most extensive sensitivity reach. They are sensitive to dimensionless Wilson coefficients at the $10^{-5}$ level, for most of the 
kinematically allowed sterile neutrino mass range. For such values of couplings, the production and decay of sterile neutrinos is dominated 
by the higher-dimensional operators with minimal sterile-active mixing playing a subleading role. Apart from dimensionless couplings and 
potential loop suppressions, the dimensionless Wilson coefficients scale as $v^2/\Lambda^2$, where $\Lambda$ is the high-energy scale 
where the $\nu$SMEFT operators are generated. The sensitivity drops at the edges of the allowed mass range, but does not disappear 
completely, even for sterile neutrinos with masses above $m_D$, because of the contributions from SM weak interactions and active-sterile 
mixing. Assuming a $v^2/\Lambda^2$ scaling, \texttt{SHiP} and \texttt{MATHUSLA} could probe scales around $80$ TeV.
This scale is lowered to $8$ TeV for \texttt{FASER} and $25$ TeV for \texttt{FASER2}, which are much smaller experiments. 

For our $B$-meson benchmarks, because of its much smaller $B$-meson production rates and a weaker acceptance, \texttt{SHiP} does 
not show the best performance. Instead, we found that \texttt{MATHUSLA} and \texttt{ANUBIS} would be sensitive to Wilson coefficients at the $5\cdot 10^{-6}$ level.
The sensitivity curves appear to be fairly independent of the Lorentz structure and the exact flavor configuration of the EFT operators, as long as sterile neutrinos can be produced through $D$- or $B$-meson decays and sterile neutrinos can decay semi-leptonically. 

For our \texttt{ATLAS} study we applied an overall flat efficiency factor of $10^{-3}$.  Of all the experiments we have studied here only
\texttt{ATLAS} is operational. Thus we strongly encourage our colleagues at \texttt{ATLAS} and \texttt{CMS} collaborations to perform a proper
full analysis of the scenarios we have presented and investigated here. This should put our approximations on a much firmer footing.

We compared our projected sensitivities with (projected) constraints obtained from other probes of sterile neutrinos with effective 
interactions such as light pseudoscalar meson decays, tau decays, coherent neutrino-nucleus scattering, LHC searches for missing 
transverse energy, and neutrinoless double beta decay. Our results are very competitive and complementary. We conclude that 
searches for displaced vertices of long-lived sterile neutrinos at the LHC and \texttt{SHiP} are an excellent probe of 
$\nu$SMEFT operators and of sterile neutrinos in general. 

A straightforward extension of our work is to final state muons instead of electrons. Here we expect basically the same results,
except at the very lowest end of the sterile neutrino mass we have considered. A more involved extension would be to also consider the
production of sterile neutrinos from the decay of the light $K$ and $\pi$ mesons. Furthermore in a future project, we shall consider
the case of final state $\tau$'s. This will restrict the kinematically viable regions, but would also require a proper investigation of
the detection efficiencies. Finally, in this work we have neglected direct sterile neutrino production through parton collisions which are subleading with respect to rare mesonic decays for sterile neutrinos masses below, roughly, $5$ GeV. It would be interesting to include partonic processes to investigate the sensitivity of various experiments to heavier sterile neutrinos.

\bigskip
\section*{Acknowledgements}

\bigskip
We thank Philip Bechtle, Haolin Li, and Vasiliki Mitsou for discussions. JdV is supported by the RHIC Physics Fellow Program of the RIKEN BNL Research Center.
Z.~S.~W. is supported partly by the Ministry of Science and Technology (MoST) of Taiwan with grant number MoST-109-2811-M-007-509, and partly by the Ministry of Science, ICT \& Future Planning of Korea, the Pohang City Government, and the Gyeongsangbuk-do Provincial Government through the Young Scientist Training Asia Pacific Economic Cooperation program of the Asia Pacific Center for Theoretical Physics.
The work of H.\,K.\,D. is supported through BMBF grant 05H18PDCA1.

\bigskip

\appendix
\section{Two-body Sterile Neutrino Production and Decay Processes}\label{sec:appendix}
\subsection{Charged Currents}
Sterile neutrinos can decay into a charged meson and a charged lepton by the weak interaction or EFT operators. Pseudoscalar mesons 
can decay into a two-body final state for pseudoscalar and axial-vector currents, while vector mesons can decay into a two-body final state 
via vector or tensor currents. For pseudoscalar mesons containing an anti-quark $\bar{q}_i$ and a quark $q_j$, we define meson decay 
constants via
\be\label{AM}
\langle 0|\bar{q}_i\gamma^\mu \gamma^5 q_j|M(q)\rangle \equiv iq^\mu f_{M}\,,
\ee
where $|M(q)\rangle$ is a pseudoscalar meson with momentum $q$. Current-algebra or leading-order chiral perturbation theory  gives
\begin{equation}\label{PSM}
\langle0|\bar{q}_i \gamma^5 q_j|M(q)\rangle =i\frac{m^2_{M}}{m_{q_i}+m_{q_j}}f_{M}\equiv if^S_{M}\
\end{equation}
for the pseudoscalar current. The vector and tensor currents only induce two-body final states for vector mesons. We define
\begin{equation}\label{VM*}
\begin{aligned}
\langle0|\bar{q}_i\gamma^\mu q_j|M^*(q,\epsilon)\rangle&\equiv i f^V_{M} m_{M^*}\epsilon^\mu\,,\\
\langle0|\bar{q}_i\sigma^{\mu\nu} q_j|M^*(q,\epsilon)\rangle&\equiv -f^T_{M}(q^\mu\epsilon^\nu-q^\nu \epsilon^\mu)\,,\\
\end{aligned}
\end{equation}
where $|M^*(q,\epsilon)\rangle$ denotes a vector meson $M^*$ with mass $m_{M^*}$, momentum $q$ and polarization $\epsilon^\mu$. Heavy-quark 
symmetry relates $ f^T_{M} \simeq f^V_{M} $. All decay constants are given below in Appendix~\ref{app:physical-parameters}.

Armed with these decay constants, we calculate the production and decay rates of neutrino mass eigenstates starting from Eq.~\eqref{final67}. We 
begin with neutrino production via the decay of pseudoscalar mesons $M^-_{ij} \rightarrow l_k^- + \nu_l$, and the corresponding decay $\nu_l 
\rightarrow M^+_{ij} + l_k^-$ where $ij$ denotes the generation of quark flavors that make up the meson (we drop these indices below for notational convenience), $k$ the charged lepton generation, and $l = \{1,\dots,\bar n\}$ the neutrino mass eigenstate. For the decay of the neutrino mass eigenstate we also include the decay to the charge-conjugate final state which is equally likely due to the Majorana nature of $\nu_l$. 

We obtain for the summed-over-spins squared amplitudes for sterile neutrino ($N$) production 
\begin{eqnarray}\label{am_s}
\sum_{\rm spins}|\mathcal M(M^- \rightarrow N + l_k^-)|^2&=& \frac{G_F^2}{2}\bigg\{ f_{M}^2 \left[ \left |C^{(6)}_{\rm VLL}\right |^2  +  \left|C^{(6)}_{\rm VLR}-C^{(6)}_{\rm VRR}\right |^2\right]f_{VV,1} \nn\\ 
&&\qquad + f_{M}^2 \mathrm{Re}\left[ C^{(6)}_{\rm VLL} (C^{(6)}_{\rm VLR}-C^{(6)}_{\rm VRR})^* \right]f_{VV,2}\nn\\
&&\qquad + (f^S_{M})^2 \left|C^{(6)}_{\rm SLR}-C^{(6)}_{\rm SRR}\right |^2 f_{SS} \nn\\ 
&&\qquad + f_{M}f^S_{M_{ij}} \mathrm{Re}\left[ C^{(6)}_{\rm VLL} (C^{(6)}_{\rm SLR}-C^{(6)}_{\rm SRR})^* \right]f_{VS,1}\\
&&\qquad + f_{M}f^S_{M_{ij}} \mathrm{Re}\left[ (C^{(6)}_{\rm VLR}-C^{(6)}_{\rm VRR})(C^{(6)}_{\rm SLR}-C^{(6)}_{\rm SRR})^* \right]f_{VS,2}\bigg\}\,,\nn 
\end{eqnarray}
where all Wilson coefficients carry flavor indices $ijkl$ and we defined the functions
\begin{eqnarray}
f_{VV,1} &\equiv& m_{M}^2(m_k^2 +m_N^2) -(m_k^2 -m_N^2)^2\,,\nn\\
f_{VV,2} &\equiv& - 4 m_{M}^2 m_k m_N\,,\nn\\
f_{SS} &\equiv& m_{M}^2-m_k^2 -m_N^2\,,\nn\\
f_{VS,1} &\equiv& -2 m_N(m_{M}^2+m_k^2 -m_N^2)\,,\nn\\
f_{VS,2} &\equiv& 2 m_k(m_{M}^2-m_k^2  + m_N^2).
\end{eqnarray}

For sterile neutrino decay,  we include the charge-conjugated final states leading to an additional factor $2$ compensating the additional 
$1/2$ from initial-spin averaging. We then obtain
\begin{eqnarray}
2\times\frac{1}{2}\sum_{\rm spins}|\mathcal M( N  \rightarrow M+ l_k)|^2&=& \sum_{\rm spins}|\mathcal M(M \rightarrow N + l_k)|^2 \bigg|_{f_{ab,i} \rightarrow g_{ab,i}}\,,
\end{eqnarray}
where $ab = \{VV,SS,VS\}$, $i=\{1,2\}$, and $g_{ab,i} = - f_{ab,i}$. 

Similarly, for processes involving vector mesons we find 
\begin{eqnarray}
\frac{1}{3}\sum_{\rm spins}|\mathcal M(M^{*\,-} \rightarrow N + l_k^-)|^2&=& \frac{G_F^2}{6}\bigg\{ (f^V_{M^*})^2 \left[ \left |C^{(6)}_{\rm VLL}\right |^2  +  \left|C^{(6)}_{\rm VLR}+C^{(6)}_{\rm VRR}\right |^2\right]f^*_{VV,1} \nn\\ 
&&\qquad +  (f^V_{M^*})^2 \mathrm{Re}\left[ C^{(6)}_{\rm VLL} (C^{(6)}_{\rm VLR}+C^{(6)}_{\rm VRR})^* \right]f^*_{VV,2}\nn\\
&&\qquad + (f^T_{M^*})^2 \left|C^{(6)}_{\rm TRR}\right |^2 f^*_{TT} \nn\\ 
&&\qquad + f^V_{M^*} f^T_{M^*} \mathrm{Re}\left[ C^{(6)}_{\rm VLL}C^{(6)\,*}_{\rm TRR}  \right]f^*_{VT,1}\\
&&\qquad + f^V_{M^*} f^T_{M^*} \mathrm{Re}\left[ (C^{(6)}_{\rm VLR}+C^{(6)}_{\rm VRR})C^{(6)\,*}_{\rm TRR}  \right]f^*_{VT,2}\bigg\}\,,\nn 
\end{eqnarray}
where
\begin{eqnarray}
f^*_{VV,1} &\equiv& 2 {m_{M^*}}^4-m_{M^*_{ij}}^2(m_k^2 +m_N^2) -(m_k^2 -m_N^2)^2\,,\nn\\
f^*_{VV,2} &\equiv& 12 m_{M^*}^2 m_k m_N\,,\nn\\
f^*_{TT} &\equiv& 16 \left(m_{M^*}^4 + m_{M^*_{ij}}^2(m_k^2 +m_N^2) - 2 (m_k^2 -m_N^2)^2\right)\,,\nn\\
f^*_{VT,1} &\equiv& -24 m_N(m_{M^*}^2+m_k^2 -m_N^2)\,,\nn\\
f^*_{VT,2} &\equiv& -24 m_k(m_{M^*}^2-m_k^2  + m_N^2)\,.
\end{eqnarray}
For sterile neutrino decay 
\begin{eqnarray}
2\times\frac{1}{2}\sum_{\rm spins}|\mathcal M( N  \rightarrow M^{*}+ l_k)|^2&=&\sum_{\rm spins}|\mathcal M(M^{*} \rightarrow N + l_k)|^2  \bigg|_{f^*_{ab,i} \rightarrow g^*_{ab,i}}\,,
\end{eqnarray}
where $ab = \{VV,TT,VT\}$, $i=\{1,2\}$, and $g^*_{VV,i} = - f^*_{VV,i}$, $g^*_{TT} = - f^*_{TT}$, $g^*_{VT,i}  = f^*_{VT,i} $. 

The expression for the decay rates is given by
\begin{equation}\label{2bodydecay}
\Gamma=\frac{\sqrt{\lambda(m_1^2,m_2^2,m_3^2)}}{16\pi m_1^3}\frac{1}{n}\sum_{\rm spins} |\mathcal M|^2
\end{equation}
where $n$ is the appropriate spin-averaging factor, $m_1$ is the mass of the decaying particle, and $m_2$ and $m_3$ are the 
masses of the final-state particles. The phase space function is defined as
\begin{equation}
\lambda(m_1^2,m_2^2,m_3^2) \equiv m_1^4+m_2^4+m_3^4-2m^2_1m^2_2-2m^2_1m^2_3-2m^2_2m^2_3\,.
\end{equation}

\subsection{Sterile Neutrino Decays into Neutral Pseudoscalar Mesons}

We follow Ref.~\cite{Bondarenko:2018ptm} and write the neutral weak axial current as
\begin{equation}\label{zweak}
J^A_{Z,\mu} =-\frac{1}{\sqrt{2}}(j^A_{3,\mu}+\frac{1}{\sqrt{3}}j^A_{8,\mu}-\frac{1}{\sqrt{6}}j^A_{0,\mu}+\frac{1}{\sqrt{2}}j^A_{\eta_c,\mu}+\cdots)\,,
\end{equation}   
 where 
 \begin{equation}
 \begin{aligned}
 j^A_{3,\mu}&=\frac{1}{\sqrt{2}}(\bar{u}\gamma_\mu\gamma_5 u-\bar{d}\gamma_\mu\gamma_5 d)\,,\\
 j^A_{8,\mu}&=\frac{1}{\sqrt{6}}(\bar{u}\gamma_\mu\gamma_5 u+\bar{d}\gamma_\mu\gamma_5 d-2\bar{s}\gamma_\mu\gamma_5 s)\,,\\
 j^A_{0,\mu}&=\frac{1}{\sqrt{3}}(\bar{u}\gamma_\mu\gamma_5 u+\bar{d}\gamma_\mu\gamma_5 d+\bar{s}\gamma_\mu\gamma_5 s)\,\\
 j^A_{\eta_c,\mu}&=\bar{c}\gamma_\mu\gamma_5 c\,.
 \end{aligned}
 \end{equation}
 We define 
 \begin{equation}
  \langle 0|j^A_{Z,\mu} |M^0_{a,P} (q)\rangle \equiv-i \frac{f_M}{\sqrt{2}} q_\mu\,,\qquad  \langle 0|j^A_{a,\mu} |M^0_{a,P} (q)\rangle \equiv i f^a_M q_\mu\,,  
 \end{equation}
 for $a=\{0,3,8,\eta_c\}$ and the $M$ subscript labels the final-state pseudoscalar meson. Clearly we have
  \begin{equation}
 f_M=f^3_M+\frac{1}{\sqrt{3}}f^8_M-\frac{1}{\sqrt{6}}f^0_M+\frac{1}{\sqrt{2}}f^{\eta_c}_M\,.
 \end{equation}
To clarify the notation: for neutral pions $f^0_{\pi^0} = f^8_{\pi^0} =f^{\eta_c}_{\pi^0}=0$ and $ f_{\pi^0} = f^3_{\pi^0}=  f_{\pi^\pm}$. 
For $\eta$ and $\eta'$ mesons, we have $f^3_{\eta^{(\prime)}} =f^{\eta_c}_{\eta^{(\prime)}}=0$ (neglecting isospin breaking), but 
we do need to take into account $\eta$-$\eta'$ mixing. We use the phenomenological parametrization in terms of two mixing angles
\bea
  \bma f^8_{\eta} &f^0_{\eta}\\[2mm] f^8_{\eta'}&f^0_{\eta'} \ema = \bma f_{8} \cos \theta_8 &-f_{0}\sin \theta_0\\[2mm] f_{8}\sin \theta_8&f_{0}\cos \theta_0 \ema\,,
\eea 
 where the values of $\theta_{0,8}$ and $f_{0,8}$ are given in Table \ref{eta}. We then obtain
 \begin{equation}
f_\eta=\frac{ f_{8} \cos \theta_8}{\sqrt{3}}+\frac{ f_{0} \sin \theta_0}{\sqrt{6}}\,,\qquad
f_{\eta'}=\frac{ f_{8} \sin \theta_8}{\sqrt{3}}-\frac{ f_{0} \cos \theta_0}{\sqrt{6}}\,.
 \end{equation} 
  
The decay width of $N\rightarrow \nu_e +M^0_P$ can now be written as
\begin{equation}
\Gamma (N\rightarrow \nu_e M^0_P)=2\times \frac{G^2_F f^2 m_N^3 |U_{e4}|^2}{32 \pi} (1-\frac{{m^0_P}^2}{m_N^2})^2\,,
\end{equation}
where we add a 2 to account for the Majorana nature of sterile neutrinos,
$f$  is given by $f_\eta$, $f_{\eta'}$, $f_{\eta_c}/    \sqrt{2}$ or $f_{\pi^\pm}$, and $m^0_P$ is the mass of $M^0_P$.

\subsection{Sterile Neutrino Decays into Neutral Vector Mesons}

We write the vector component of the neutral weak current as
\begin{equation}
j^V_{Z,\mu}=\left(\frac{1}{2}-\frac{4}{3}\sin^2\theta_w\right)\left(\bar{u}\gamma_\mu u+\bar{c}\gamma_\mu c\right) + 
\left(-\frac{1}{2}+\frac{2}{3}\sin^2\theta_w\right)\left(\bar{d}\gamma_\mu d+\bar{s}\gamma_\mu s\right)\,,
\end{equation}
where $\theta_w$ is the Weinberg angle. We define the currents 
\begin{equation}
\begin{aligned}
j^V_{\rho^0,\mu}&=\frac{1}{\sqrt{2}}(\bar{u}\gamma_\mu u-\bar{d}\gamma_\mu d)\,,\\
j^V_{\omega,\mu}&=\frac{1}{\sqrt{2}}(\bar{u}\gamma_\mu u+\bar{d}\gamma_\mu d)\,,\\
j^V_{\phi,\mu}&=\bar{s}\gamma_\mu s\,,\\
j^V_{\text{J},\mu}&=\bar{c}\gamma_\mu c\,,\\
\end{aligned}
\end{equation}
which correspond to the neutral vector mesons $\rho^0$, $\omega$, $\phi$, and $\textrm{J}/\Psi$, respectively.
We rewrite 
\begin{equation}\label{ZV}
j^V_{Z,\mu}=\frac{1}{\sqrt{2}}(1-2\sin^2\theta_w)j^V_{\rho^0,\mu}-\frac{\sqrt{2}}{3}\sin^2 \theta_w j^V_{\omega,\mu}
+\!\left(-\frac{1}{2}+\frac{2}{3}\sin^2\theta_w\right)\!j^V_{\phi,\mu}+(\frac{1}{2}-\frac{4}{3}\sin^2\theta_w)j^V_{\textrm{J}/\Psi,\mu}
\,.
\end{equation}
The hadronic matrix elements are defined as
\begin{equation}
\begin{aligned}
\langle 0|j^V_{a,\mu} |M^0_{a,V} (q,\epsilon)\rangle &=i f_a \epsilon_\mu\,,  \\ 
\end{aligned}
\end{equation}
where $a=\rho^0, \omega , \phi, \textrm{J}/\Psi$. The decay width becomes
\begin{equation}
\Gamma(N\rightarrow \nu_e M^0_{a,V})=2\times \frac{G_F^2 f_a^2 g_a^2 |U_{e4}|^2 m_N^3}{32\pi {m_a}^2}
\left(1+2 \frac{{m_a}^2}{m_N^2}\right)\left(1-\frac{{m_a}^2}{m_N^2}\right)^2\,,
\end{equation}
where $f_a$ and $g_a$ are listed in Table \ref{constantV} and $m_a$ is the mass of $M^0_{a,V}$.

\section{Sterile Neutrino Production in Three-body Decays}

\subsection{Automizing Three-body Phase Space Integral Calculations}

This work involves a large amount of three-body phase-space computations, which are straightforward but tedious 
to evaluate. We briefly discuss here our approach to automize these computations using Mathematica. 
In the rest-frame of the decaying pseudoscalar meson, the decay rate is given by 
\begin{eqnarray}
\Gamma &=& \frac{1}{2 M} \int \frac{d^3 p'}{2 {p'}^0 (2\pi)^3}\int \frac{d^3 p_l}{2 p_l^0 (2\pi)^3}\int \frac{d^3 p_N}
{2 p_N^0 (2\pi)^3} \sum \abs{\mathcal{M}}^2  (2\pi)^4 \delta^4(p-p'-p_l-p_N)\,,
\end{eqnarray}
where $p$, $p'$, $p_l$, and $p_N$ denote the momentum of the decaying meson, outgoing meson, SM lepton, and sterile neutrino, respectively, and $M$
is the mass of the decaying pseudoscalar meson. $\sum \abs{\mathcal{M}}^2$ denotes the spin-averaged product of the 
leptonic and hadronic matrix element squared. It 
can be decomposed into a hadronic form factor, which only depends on $ q^2$, where $q\equiv p-p'$, and a function of 
four-momentum invariant scalar products. The 
form factors are defined below and   the spin-averaged matrix elements are calculated with 
standard techniques and checked with PackageX \cite{Patel:2015tea}.  We convert to four-dimensional integrals and write
\begin{eqnarray}
\Gamma &=& \frac{1}{2 M} \frac{1}{(2\pi)^5} \int d^4 p' \int d^4 p_l \int d^4 p_N\, \delta({p'}^2-m^2) \delta(p^2_l-m_l^2) 
\delta(p^2_N-m_N^2)\nonumber\\&&\times   \sum \abs{M}^2  (2\pi)^4 \delta^4(p-p'-p_l-p_N)\,,
\end{eqnarray}
where for notational convenience we have omitted three Heaviside step functions.  We perform the $p_N$ integral by 
setting $p_N = q- p_l$ and introduce the variable $a$ via a factor of $1=\int da\, \delta(a-q^2)$ to obtain
\begin{eqnarray}
\Gamma &=& \frac{1}{2 M} \frac{1}{(2\pi)^5}\int da\int d^4 p' \int d^4 p_l \, \delta(a-q^2) \delta({p'}^2-m^2) \delta(p^2_l-m_l^2) 
\delta((q-p_l)^2-m_N^2)\nn\\
&& \times \sum \abs{\mathcal{M}}^2   \bigg |_{p_N = q-p_l}\,.
\end{eqnarray}
Here $m,m_l,m_N$ denote the masses of the outgoing meson, lepton, and sterile neutrino, respectively. The hadronic form 
factor contained in $ \sum |M|^2$ only depends on $a$ and, together with three of the six scalar 
products 
 \begin{eqnarray}\label{set1}
 p' \cdot p = \frac{1}{2}\left(M^2 -a + m^2 \right)\,,\qquad
 p' \cdot q = \frac{1}{2}\left(M^2 -a - m^2 \right)\,,\qquad
 p \cdot q =  \frac{1}{2}\left(M^2  +a - m^2 \right)\,,
 \end{eqnarray}
 can be taken out of the $p'$ and $p_l$ integrals. The last delta function gives the additional relations
  \begin{eqnarray}\label{set2}
  p_l \cdot q = \frac{1}{2}\left(a+m_l^2-m_N^2\right)\,,\qquad p_l \cdot p' =  p_l \cdot p -\frac{1}{2}\left(a+m_l^2-m_N^2\right)\,.
  \end{eqnarray}
 These relations imply that the spin-averaged matrix element squared can be written in the form
  \begin{equation}
  \sum \abs{M}^2 \bigg |_{p_N = q-p_l} = \sum_{n=0}^N c_n(a) (p_l \cdot p)^n\,,
  \end{equation}
where $c_n(a)$ are process-dependent functions of $a$ and particle masses,  and they also contain the hadronic form
 factors. For our calculations we have $N\leq 2$. The remaining integrals can be explicitly computed. We need
 \begin{equation}
I_P = \int d^4 p'\,  \delta({p'}^2-m^2) \delta\left[a-(p-p')^2\right] = \frac{\pi}{2} \frac{\lambda^{1/2}(a,m^2,M^2)}{M^2}\,,
\end{equation}
and
 \begin{equation}
I_n = \int d^4 p_l \, \delta(p^2_l-m_l^2) \delta\left[(q-p_l)^2-m_N^2\right](p_l \cdot p)^n \,,
\end{equation}
for $n=\{0,1,2\}$. A straightforward calculation gives
 \begin{eqnarray}
 I_0 &=&  \frac{\pi}{2} \frac{\lambda^{1/2}(a,m_l^2,m_N^2)}{a}\,,\nn\\
 I_1 &=&  \frac{ (p_l\cdot q)(p\cdot q)}{a} I_0\,\nn\\
 I_2 &=& -\frac{1}{3a}\left\{ \frac{(p \cdot q)^2}{a}\left[a m_l^2 - 4(p_l \cdot q)^2\right] -M^2 \left[a m_l^2 - (p_l \cdot q)^2\right] \right\}I_0\,,
 \end{eqnarray}
 where the scalar products appearing in $I_{1,2}$ should be evaluated via Eqs.~\eqref{set1} and \eqref{set2} and are 
 thus functions of $a$ only. The final decay rate requires one remaining integral over $a = q^2$ 
\begin{eqnarray}
\Gamma &=& \frac{1}{2 M} \frac{1}{(2\pi)^5}\int^{(M-m)^2}_{(m_l + m_N)^2} da\,  I_P\, \sum_{n=0}^N c_n(a) I_n\,.
\end{eqnarray}
We have automized the above procedure in a few lines of Mathematica code. In certain cases when no or just simple 
hadronic form factors appear, the integrals can be performed analytically. When this is possible we have checked our 
results with the literature. In most cases, however, the integrals are computed numerically.
 
\subsection{Definition of Three-body Decay  Form Factors}

A sterile neutrino $N$ can be produced via the decay of a pseudoscalar meson, $M_P$, with mass $M$
\begin{equation}
M_P \rightarrow M'_{P/V} +e^{\pm} +N\,,
\end{equation}
where $M'_{P/V}$ is a pseudoscalar or vector meson with mass $m$. For a final-state pseudoscalar meson, we require 
the following form factors:
\begin{equation}\label{pform}
\begin{aligned}
\langle M'_{P}(p')| \bar{q}_1 \gamma^\mu q_2 |M_P(p)\rangle&=f_+(q^2) \left[(p+p')^\mu-\frac{M^2-m^2}{q^2} q^\mu\right]+f_0(q^2)\frac{M^2-m^2}{q^2} q^\mu\,,\\
\langle M'_{P}(p')| \bar{q}_1 q_2 |M_P(p)\rangle&=f_S(q^2)\,,\\
\langle M'_{P}(p')| \bar{q}_1 \sigma^{\mu\nu} q_2 |M_P(p)\rangle&=\frac{2 i}{M+m} [p^\mu p'^{\nu}-p^\nu p'^{\mu}]f_T (q^2)\,,
\\
\end{aligned}
\end{equation}
where $q^\mu=p^\mu-p'^\mu$. Applying the equations of motion, the scalar form factor becomes
\begin{equation}
f_S(q^2) =f_0 (q^2) \frac{M-m}{m_1-m_2}\,,
\end{equation}
where $m_{1}$ and $m_2$ denote the mass of $q_1$ and $q_2$, respectively. A similar trick for the tensor form factor gives
\begin{equation}
f_T(q^2) = \frac{(M+m)^2}{q^2} \left[f_+(q^2)-f_0(q^2)\right]\,,
\end{equation}
which agrees fairly well with lattice computations of Ref.~\cite{Lubicz:2018rfs} in cases where a comparison is possible.

When a vector meson is produced additional form factors are required
 \begin{equation}\label{vform}
 \begin{aligned}
 \langle M'_{V}(p',\epsilon)| \bar{q}_1 \gamma^\mu q_2 |M_P(p)\rangle&=ig(q^2) \epsilon^{\mu\nu\alpha\beta}\epsilon^*_\nu P_\alpha q_\beta
 \,,\\[1.2mm]
 \langle M'_{V}(p',\epsilon)| \bar{q}_1\gamma^\mu \gamma^5 q_2 |M_P(p)\rangle&=f(q^2)\epsilon^{*\mu}+a_+(q^2)P^\mu\epsilon^*\cdot p  +a_-(q^2) q^\mu\epsilon^*\cdot p 
 \,,\\[1.2mm]
 \langle M'_{V}(p',\epsilon)| \bar{q}_1 \sigma^{\mu\nu} q_2 |M_P(p)\rangle&=g_+(q^2)\epsilon^{\mu\nu\alpha\beta}\epsilon^*_\alpha P_\beta+ g_-(q^2)\epsilon^{\mu\nu\alpha\beta}\epsilon^*_\alpha q_\beta +g_0(q^2)\epsilon^{\mu\nu\alpha\beta} p_\alpha p'_{\beta} p\cdot \epsilon^*  
 \,,
 \\[1.2mm]
  \langle M'_{V}(p',\epsilon)| \bar{q}_1 \gamma^5 q_2 |M_P(p)\rangle&=f_{PS} \epsilon^*\cdot p\,,
 \end{aligned}
 \end{equation}
 where $\epsilon^{*\mu}$ is the polarization vector of the vector meson, and  $P^\mu=p^\mu+p'^\mu$. The pseudo-scalar form factor  is given by
  \begin{equation}\label{fPS}
  f_{PS}=\frac{1}{m_1+m_2} \left[f(q^2)+a_+(q^2)(M^2-m^2)+a_-(q^2)q^2\right]\,.
 \end{equation}
 
\section{Physical Parameters, Decay Constants, and Form Factors}
\label{app:physical-parameters}

We list all  parameters we use in this paper in this section. The values of relevant CKM matrix elements are extracted from Ref.~\cite{Tanabashi:2018oca}
\begin{eqnarray}
|V_{cd}|=0.218, &|V_{ud}|=0.974, &|V_{us}|=0.224,\nonumber\\
|V_{cs}|=0.997, &|V_{ub}|=0.00394, &|V_{cb}|=0.0422.
\end{eqnarray}
We use the quark masses at a renormalization scale of $\mu =2 $ GeV in $\overline{\mathrm{MS}}$
\begin{eqnarray}
m_u = 2.2\,\mathrm{MeV}\,, &m_d = 4.7\,\mathrm{MeV}\,, &m_s = 93\,\mathrm{MeV}\,,\nonumber\\
m_c = 1.27\,\mathrm{GeV}\,, &m_b = 4.18\,\mathrm{GeV}\,.
\end{eqnarray}

Decay constants for pseudoscalar and vector mesons are given in Tables \ref{table:psmesons}. Parameters to calculate decay constants for the neutral pseudoscalar and vector mesons are given in Tables~\ref{eta} and \ref{constantV}, respectively.

\begin{table}[t]
	\centering
	\begin{tabular}{||c | c | c |c ||  }
	         \hline
		meson $M_P$ &  $f_M$ [MeV] & meson $M_V$ &  $f^V_M$ [MeV]\\
		\hline 
		$D^\pm$      &    212 \cite{Rosner:2015wva}	& $D^{*\pm}$	 & 266 \cite{Dhiman:2017urn}   \\
		$D_s^\pm$	  & 249 \cite{Rosner:2015wva}	& $D_s^{*\pm}$   &308 \cite{Dhiman:2017urn}	\\
		$B^\pm$	   &  187 \cite{Rosner:2015wva}		& $K^{*\pm}$     & 230 \cite{Dreiner:2006gu}	\\
		$B_c^\pm$  &  434 \cite{Colquhoun:2015oha}	& $\rho^\pm$   &209 \cite{Ebert:2006hj}	\\
		$K^\pm$         & 155.6 \cite{Rosner:2015wva}	&&	\\
		$\pi^\pm$	     &  130.2 \cite{Rosner:2015wva}.  &&\\
	\hline
	\end{tabular}
	\caption{Decay constants for charged pseudoscalar and vector mesons. }
	\label{table:psmesons}
\end{table}

{\renewcommand{\arraystretch}{1.3}\begin{table}[t!]\small
		\center
		\begin{tabular}{||c|c||}
			\hline
			$f_0$ & $0.148 $ GeV  \cite{Escribano:2015yup} \\ \hline
			$f_8$ & $0.165 $ GeV  \cite{Escribano:2015yup} \\ \hline
			$f_{\eta_c}$ & $0.335 $ GeV \cite{Edwards:2000bb}  \\ \hline
			$\theta_0$ & -6.9\textdegree   \cite{Escribano:2015yup} \\ \hline
			$\theta_8$ & -21.2\textdegree  \cite{Escribano:2015yup}  \\ \hline
		\end{tabular}
		\caption{Decay constants and angles for $\eta$, $\eta'$ and $\eta_c$.
		} \label{eta}
\end{table}}

\begin{table}[t]
	\centering
	\begin{tabular}{||c | c  c||}
	\hline
		meson $M$ & $f_a$ [Ge$\text{V}^2$] & $g_a$ \\
		\hline 
		$\rho^0$  \cite{Coloma:2020lgy}    &  0.171 &  $1-2\sin^2\theta_w$ \\
		$\omega$	\cite{Coloma:2020lgy}   & 0.155  & -$\frac{2}{3}\sin^2\theta_w$ \\
		$\phi$	\cite{Coloma:2020lgy}   & 0.232  & $\sqrt{2}(-\frac{1}{2}+\frac{2}{3}\sin^2\theta_w)$ \\
		$\text{J} $\cite{Becirevic:2013bsa}	   &  1.29 &   $\sqrt{2}(\frac{1}{2}-\frac{4}{3}\sin^2\theta_w)$ \\
		\hline
	\end{tabular}
	\caption{Decay constants and $g_a$ of neutral vector mesons.  }
	\label{constantV}
\end{table}

\subsection{Form Factors for $B_{(s)}\rightarrow M_P^\prime$}

We apply the Bourrely-Caprini-Lellouch (BCL) method \cite{Bourrely:2008za,Bondarenko:2018ptm} to parameterize the form factors,
\begin{equation}
\begin{aligned}
f_{+} (q^2)&=\frac{1}{1-q^2/m^2_{\rm{pole}}} \sum_{k=0}^{K-1} b^+_k [(z(q^2))^k-(-1)^{k-K}\frac{k}{K}z(q^2)^K]\,,\\[1.1mm]
f_{0} (q^2)&= \frac{1}{1-q^2/m^2_{\rm{pole}}} \sum_{k=0}^{K-1} b^0_k z(q^2)^k\,,
\end{aligned}
\end{equation}
where $z(q^2)$ is the function 
\begin{equation}
z(q^2)=\frac{\sqrt{t_+-q^2}-\sqrt{t_+-t_0}}{\sqrt{t_+-q^2}+\sqrt{t_+-t_0}}\,,
\end{equation}
with $t_+ = (M+m)^2$ and $t_0=(M+m)(\sqrt{M}-\sqrt{m})^2$.
We set $K=3$ and $f_+(0)=f_0(0)$. This determines $b^0_2$ through
\begin{equation}
b^0_2 =\frac{f_+(0)-b^0_0-b^0_1 z(0)}{z(0)^2}\,.
\end{equation}  
The  best-fit parameter values are given in Table \ref{fBP}.

\subsection{Form Factors for $D\rightarrow M_P^\prime$}

For $D\rightarrow \pi$ and $D\rightarrow K$ transitions, we use the methods of Ref\cite{Lubicz:2017syv}. We write
\begin{equation}
f_{+/0}=\frac{f(0) +c \left[z(q^2)-z(0)\right]\left[1+\frac{z(q^2)+z(0)}{2}\right]}{(1-P q^2)}\,,
\end{equation}
and list the best-fit parameter values in Table \ref{fdp}.

{\renewcommand{\arraystretch}{1.3}\begin{table}[t!]\small
		\center
		\begin{tabular}{||c|c|c|c|c||}
			\hline $f$ & $m_{\rm{pole}}$ [GeV]  &  $b_0$ &  $b_1$ & $b_2$ \\
			\hline
			$f_+^{B_{(s)}\rightarrow D_{(s)}}$ & $\infty$ & $0.909$ & $-7.11$ &$66$  \\ \hline
			 $f_0^{B_{(s)}\rightarrow D_{(s)}}$&  $\infty$ &  $0.794$ & $-2.45$ & $\null$  \\ \hline 
			 	$f_+^{B_s\rightarrow K}$ & $m_{B^*}=5.325$ & $0.360$ & $-0.828$ &$1.1$  \\ \hline
			 $f_0^{B_s\rightarrow K}$&  $m_{B^*(0^+)}=5.65$ &  $0.233$ & $0.197$ & $\null$  \\ \hline 
			$f_+^{B\rightarrow \pi}$ & $m_{B^*}=5.325$ & $0.404$ & $-0.68$ &$-0.86$  \\ \hline
		$f_0^{B\rightarrow \pi}$&  $\infty$ &  $0.49$ & $-1.61$ & $\null$  \\ \hline 
		\end{tabular}
		\caption{Best fit parameters values for the form factors in $B\rightarrow D$ and $B\rightarrow \pi$ transitions from Ref.~\cite{Aoki:2019cca}.
		} \label{fBP}
\end{table}}

{\renewcommand{\arraystretch}{1.3}\begin{table}[t!]\small
		\center
		\begin{tabular}{||c|c|c|c||}
			\hline $f$ & $f(0)$  &  $c$ &  $P$ [Ge$\text{V}^2$] \\
			\hline
			$f_+^{D\rightarrow K}$ & $0.7647$ & $-0.066$ & $0.224$  \\ \hline
			$f_0^{D\rightarrow K}$&  $0.7647$ &  $-2.084$ & $0$   \\ \hline 
			$f_+^{D\rightarrow \pi}$ & $0.6117$ & $-1.985$ & $0.1314$   \\ \hline
			$f_0^{D\rightarrow \pi}$&  $0.6117$ &  $-1.188$ & $0.0342$  \\ \hline 
		\end{tabular}
		\caption{Best fit parameters for the form factors in $D\rightarrow K$ and $D\rightarrow \pi$ transitions from Refs.~\cite{Lubicz:2017syv,Bondarenko:2018ptm}.
		} \label{fdp}
\end{table}}

\subsection{Form Factors for $B_{(s)}$ and $D$ Decaying into $M_V^\prime$}\label{BtoV}

Eq. \eqref{vform} contains seven form factors which must be determined. The pseudoscalar form factor is related to $f(q^2)$ 
and $a_\pm(q^2)$ through Eq.~\eqref{fPS}.  To better present these form factors and follow the conventions of 
Ref.~\cite{Melikhov:2000yu}, we define the following dimensionless combinations
\begin{equation}
\begin{aligned}
V(q^2)&=(M+m)g(q^2)\,, \qquad  A_1(q^2)= \frac{f(q^2)}{M+m}\,, \qquad A_2(q^2)=-(M+m)a_+(q^2)\,,\\ A_0&=\frac{1}{2 m}\left[f(q^2)+a_-(q^2) q^2 +a_+(q^2) P\cdot q\right]
\,, \qquad T_1 =-g_+(q^2)\,,\\
T_2(q^2) &=-g_+(q^2)-\frac{q^2}{P\cdot q}g_-(q^2)\,,\qquad
T_3(q^2)=g_-(q^2)-\frac{P\cdot q}{2} g_0(q^2)\,.
\end{aligned}
\end{equation}
and choose the following three-parameter formula
\begin{equation}\label{1stformv}
f(q^2)=\frac{f(0)}{(1-q^2/M_{\mathrm{pole}}^2)(1-\sigma_1 q^2/M_{\mathrm{pole}}^2+\sigma_2 q^4/M_{\mathrm{pole}}^4)}\,,
\end{equation}
to describe $V$, $T_1$ and $A_0$. $M$ is the pole mass, which is $M_P$($0^-$) for $A_0$ and $M_V$($1^-$) for $V$ and $T_1$.  For the remaining form factors $A_1$, $A_2$, $T_2$ and $T_3$, we use the simpler form \cite{Melikhov:2000yu}
\begin{equation}\label{2ndformv}
f(q^2)=\frac{f(0)}{1-\sigma_1 q^2/M_V^2+\sigma_2 q^4/M_V^4}\,.
\end{equation}
 In all scenarios we consider, the transition $B_s\rightarrow D_s^*$ is only induced by SM weak interactions and we do not list form factors associated to BSM currents. The values of the best fit parameters are given in Table \ref{1stfptov} to Table \ref{3rdfptov}.
{\renewcommand{\arraystretch}{1.3}\begin{table}[t!]\small
		\center
		\begin{tabular}{||c|c|c|c|c|c|c|c||}
			\hline $f(0)$& $f_V(0)$& $f_{A_0}(0)$ & $f_{A_1}(0)$ &$f_{A_2}(0)$ & $f_{T_1}(0)$ &$f_{T_2}(0)$ &$f_{T_3}(0)$ \\
			\hline $D\rightarrow K^*$& $1.03$& $0.76$ & $0.66$ &$0.49$ & $0.78$ &$0.78$ &$0.45$ \\
			\hline $D\rightarrow \rho$& $0.9$& $0.66$ & $0.59$ &$0.49$ & $0.66$ &$0.66$ &$0.31$ \\
			\hline $B\rightarrow D^*$& $0.76$& $0.69$ & $0.66$ &$0.62$ & $0.68$ &$0.68$ &$0.33$ \\
			\hline $B\rightarrow \rho$& $0.31$& $0.30$ & $0.26$ &$0.24$ & $0.27$ &$0.27$ &$0.19$ \\
			\hline $B_{s}\rightarrow D_s^*$& $0.95$& $0.67$ & $0.70$ &$0.75$ & $\null$ &$$ &$$ \\
			\hline $B_{s}\rightarrow K^*$& $0.38$& $0.37$ & $0.29$ &$0.26$ & $0.32$ &$0.32$ &$0.23$ \\
			\hline
		\end{tabular}
		\caption{Part I: Best-fit parameters values for Eqs.~\eqref{1stformv}-\eqref{2ndformv} from Refs.~\cite{Melikhov:2000yu,Bondarenko:2018ptm}.
		} \label{1stfptov}
\end{table}}

{\renewcommand{\arraystretch}{1.3}\begin{table}[t!]\small
		\center
		\begin{tabular}{||c|c|c|c|c|c|c|c||}
			\hline $\sigma_1$& $\sigma_1(V)$& $\sigma_1(A_0)$ & $\sigma_1(A_1)$ &$\sigma_1(A_2)$ & $\sigma_1(T_1)$ &$\sigma_1(T_2)$ &$\sigma_1(T_3)$ \\
			\hline $D\rightarrow K^*$& $0.27$& $0.17$ & $0.30$ &$0.67$ & $0.25$ &$0.02$ &$1.23$ \\
			\hline $D\rightarrow \rho$& $0.46$& $0.36$ & $0.50$ &$0.89$ & $0.44$ &$0.38$ &$1.10$ \\
			\hline $B\rightarrow D^*$& $0.57$& $0.59$ & $0.78$ &$1.40$ & $0.57$ &$0.64$ &$1.46$ \\
			\hline $B\rightarrow \rho$& $0.59$& $0.54$ & $0.73$ &$1.40$ & $0.60$ &$0.74$ &$1.42$ \\
			\hline $B_{s}\rightarrow D_s^*$& $0.372$& $0.350$ & $0.463$ &$1.04$ & $\null$ &$$ &$$ \\
			\hline $B_{s}\rightarrow K^*$& $0.66$& $0.60$ & $0.86$ &$1.32$ & $0.66$ &$0.98$ &$1.42$ \\
						\hline
		\end{tabular}
		\caption{Part II: Best-fit parameters values for Eqs.~\eqref{1stformv}-\eqref{2ndformv} from Refs.~\cite{Melikhov:2000yu,Bondarenko:2018ptm}.
		} \label{2ndfptov}
\end{table}}

{\renewcommand{\arraystretch}{1.3}\begin{table}[t!]\small
		\center
		\begin{tabular}{||c|c|c|c|c|c|c|c|c|c||}
			\hline $\sigma_2,m_{\rm{pole}}$& $\sigma_2(V)$& $\sigma_2(A_0)$ & $\sigma_2(A_1)$ &$\sigma_2(A_2)$ & $\sigma_2(T_1)$ &$\sigma_2(T_2)$ &$\sigma_2(T_3)$ & $M_P$(GeV)& $M_V$(GeV)\\
			\hline $D\rightarrow K^*$&  $0$& $0$ & $0.20$ &$0.16$ & $0$ &$1.80$ &$0.34$ & $m_{D_s}= 1.968$& $m_{D^*_s}=2.112$ \\
			\hline $D\rightarrow \rho$&  $0$& $0$ & $0$ &$0$ & $0$ &$0.50$ &$0.17$ & $m_D= 1.87$& $m_{D^*}=2.01$ \\
			\hline $B\rightarrow D^*$ & $0$& $0$ & $0$ &$0.41$ & $0$ &$0$ &$0.50$ & $m_{B_c}= 6.275$& $m_{B_c^*}=6.331$ \\
			\hline $B\rightarrow \rho$& $0$& $0$ & $0.10$ &$0.50$ & $0$ &$0.19$ &$0.51$ & $m_B= 5.279$& $m_{B^*}=5.325$ \\
			\hline $B_{s}\rightarrow D_s^*$& $0.561$& $0.600$ & $0.510$ &$0.070$ & $\null$ &$$ &$$ & $m_{B_c}=6.275$ & $m_{B^*_c}=6.331$\\
			\hline $B_{s}\rightarrow K^*$& $0.30$& $0.16$ & $0.60$ &$0.54$ & $0.31$ &$0.90$ &$0.62$ &$m_{B_s}=5.367$ &$m_{B^*_s}=5.415$ \\
						\hline
		\end{tabular}
		\caption{Part III: Best-fit parameters for Eqs.~\eqref{1stformv}-\eqref{2ndformv} from Refs.~\cite{Melikhov:2000yu,Bondarenko:2018ptm}.
		} \label{3rdfptov}
\end{table}}

\subsection{The Semi-leptonic Decay of $D_s$}

Through the semi-leptonic decay of $D_s$ meson,  $\eta$, $\eta'$, $K^0$, $K^*_0$ and $\phi$ can be produced. The pseudoscalar mesons $\eta$ and $\eta'$ mix with each other and we can consider them as the mixtures of $\eta_n\equiv \frac{\bar{u}u+\bar{d}d}{\sqrt{2}}$ and $\eta_s\equiv \bar{s}s$ \cite{Melikhov:2000yu}
\begin{equation}
\eta=\cos (\psi)\eta_n-\sin (\psi) \eta_s\,, \qquad \eta'=\sin (\psi)\eta_n +\cos (\psi) \eta_s\,,
\end{equation}
where the mixing angle $\psi$ is around $40$\textdegree \cite{Anisovich:1996hh,Feldmann:1998vh,Melikhov:2000yu}. The decay rates are \cite{Melikhov:2000yu}
\begin{equation}
\begin{aligned}
\Gamma(D_s \rightarrow \eta+e+\nu_e)& = \sin^2(\psi)\Gamma(D_s \rightarrow \eta_s(m_\eta)+e+\nu_e) \,,\\
\Gamma(D_s \rightarrow \eta'+e+\nu_e) &= \cos^2(\psi)\Gamma(D_s \rightarrow \eta_s(m_{\eta'})+e+\nu_e)\,,
\end{aligned}
\end{equation}
where $\eta_s (m_{\eta^{(')}})$ means we consider the mass of $\eta_s$ as $m_{\eta^{(')}}$ when calculating the decay width. For the decay $D_s \rightarrow \eta_s/K^0$, we use Eq.~\eqref{1stformv}  to parameterize the form factors $f_+$ and  $f_T$ with $M_{\rm pole}=m_{D_s^*}/m_{D^*}$, and use Eq.~\eqref{2ndformv} to parameterize $f_0$ with $M_V=m_{D_s^*}/m_{D^*}$. For the remaining decay channels, we can use the same method as that in \ref{BtoV} with $M_P=m_D$, $M_V=m_{D^*}$ for $D_s \rightarrow   K_0^*$, and $M_P=m_{D_s}$, $M_V=m_{D_s^*}$ for $D_s \rightarrow \phi$.  All the related best fit values are given in Tables \ref{formDs1} and \ref{formDs2}.
 \begin{table}[t!]
 	\renewcommand{\arraystretch}{1.2}
 	\center\small
 	\begin{tabular}{|c|c|c|c|c|c|c||c|c|c|c|c|c|c|}
 		\hline
 		$\null$ & \multicolumn{3}{|c|}{ $D_s\rightarrow \eta_s(m_\eta)$ }&\multicolumn{3}{|c||}{ $D_s \rightarrow \eta_s(m_{\eta'})$ } &  \multicolumn{7}{c|}{$Ds  \rightarrow \phi$} \\
 		\hline
 		$\null$ & $f_+$ & $f_0$ & $f_T$     & $f_+$ & $f_0$   		& $f_T $             & $V$ & $A_0$& $A_1$ & $A_2$& $T_1$& $T_2$& $T_3$ \\
 \hline
 $f(0)$ & $0.78$ & $0.78$ & $0.80$   &0.78   & $0.78$   		& $0.94 $             & $1.10$ & $0.73$& $0.64$ & $0.47$& $0.77$& $0.77$& $0.46$\\

 $\sigma_1$ & $0.23$ & $0.33$ & $0.24$     &0.23 & $0.21$   		& $0.24 $             & $0.26$ & $0.10$& $0.29$ & $0.63$& $0.25$& $0.02$& $1.34$\\
 
$\sigma_2$ & $0$ & $0.38$ & $0$   &0   & $0.76$   		& $0 $             & $0$ & $0$& $0$ & $0$& $0$& $2.01$& $0.45$\\
\hline
 	\end{tabular}
 	\caption{ Part I: Best-fit parameters for $D_s$ decays from Ref.~\cite{Melikhov:2000yu}.}  
 	\label{formDs1}
 \end{table}

\begin{table}[t!]
	\renewcommand{\arraystretch}{1.2}
	\center\small
	\begin{tabular}{|c|c|c|c||c|c|c|c|c|c|c|}
		\hline
		$\null$ & \multicolumn{3}{|c|}{ $D_s\rightarrow K$ } &  \multicolumn{7}{c|}{$Ds  \rightarrow K^*$} \\
		\hline
		$\null$ & $f_+$ & $f_0$ & $f_T$     & $V$ & $A_0$& $A_1$ & $A_2$& $T_1$& $T_2$& $T_3$ \\
		\hline
		$f(0)$ & $0.72$ & $0.72$ & $0.77$   &1.04   & $0.67$   		& $0.57 $             & $0.42$ & $0.71$& $0.71$ & $0.45$\\
		
		$\sigma_1$ & $0.20$ & $0.41$ & $0.24$     &0.24 & $0.20$   		& $0.29 $             & $0.58$ & $0.22$& $-0.06$ & $1.08$\\
		
		$\sigma_2$ & $0$ & $0.7$ & $0$   &0   & $0$   		& $0.42 $             & $0$ & $0$& $0.44$ & $0.68$\\
		\hline
	\end{tabular}
	 	\caption{ Part II: Best-fit parameters for $D_s$ decays from Ref.~\cite{Melikhov:2000yu}.}  
	\label{formDs2}
\end{table}

\medskip

\bibliographystyle{utphysmod}

\providecommand{\href}[2]{#2}\begingroup\raggedright\endgroup

\end{document}